\documentclass[fleqn,usenatbib]{mnras}

\usepackage[T1]{fontenc}

\DeclareRobustCommand{\VAN}[3]{#2}
\let\VANthebibliography\thebibliography
\def\thebibliography{\DeclareRobustCommand{\VAN}[3]{##3}\VANthebibliography}

\usepackage{graphicx}	
\usepackage{amsmath}	
\usepackage{amssymb}	
\usepackage{physics}
\usepackage{bm}

\title[Planet-Driven Scatterings of Planetesimals Into a Star]{Planet-Driven Scatterings of Planetesimals Into a Star: Probability, Timescale and Applications}

\author[Laetitia Rodet \& Dong Lai]{
Laetitia Rodet $^{1}$,
Dong Lai $^{1,2}$\thanks{E-mail: dong@astro.cornell.edu}
\\
$^{1}$Cornell Center for Astrophysics and Planetary Science, Department of Astronomy, Cornell University, Ithaca, NY 14853, USA\\
$^{2}$ Tsung-Dao Lee Institute, Shanghai Jiao-Tong University, Shanghai, 520 Shengrong Road, 201210, China 
}

\date{Accepted XXX. Received YYY; in original form ZZZ}

\pubyear{2021}

\begin{document}
\label{firstpage}
\pagerange{\pageref{firstpage}--\pageref{lastpage}}
\maketitle

\begin{abstract}
A planetary system can undergo multiple episodes of intense dynamical activities throughout its life, resulting in the production of star-grazing planetesimals (or exocomets) and pollution of the host star. Such activity is especially pronounced when giant planets interact with other small bodies during the system's evolution. However, due to the chaotic nature of the dynamics, it is difficult to determine the properties of the perturbing planet(s) from the observed planetesimal-disruption activities. In this study, we examine the outcomes of planetesimal-planet scatterings in a general setting. We focus on one-planet systems, and determine the likelihood and timescale of planetesimal disruption by the host star as a function of the planet properties. 
We obtain a new analytical expression for the minimum distance a scattering body can reach, extending previous results by considering finite planet eccentricity and non-zero planetesimal mass.
Through $N$-body simulations, we derive the distribution of minimum distances and the likelihood and timescales of three possible outcomes of planetesimal-planet scatterings: collision with the planet, ejection, and disruption by the star. 
For planetesimals with negligible mass, we identify four defining dimensionless parameters (the planet eccentricity, planet-to-star mass ratio, planet radius to semi-major axis ratio, and the stellar disruption radius to planet semi-major axis ratio) that enable us to scale the problem and generalize our findings to a wide range of orbital configurations. Using these results, we explore three applications: falling evaporating bodies in the $\beta$ Pictoris system, white dwarf pollution due to planetesimal disruption and planet engulfment by main-sequence stars.
\end{abstract}




\section{Introduction}

A planetary system can experience many episodes of dynamical disruptions throughout its life. At the very beginning of the system's life, planetesimals are accreted or scattered by the growing and migrating protoplanets. A few million years later, when the protoplanetary disc disperses, the planets undergo a phase of giant impacts that change the orbital architecture of the system and further excite its dynamics \citep[see][and references therein]{helledPlanetFormation2021,raymondPlanetFormationKey2022}. As the star evolves on the main-sequence, orbital evolution can occur on a secular timescale, with processes such as high-eccentricity migration or secular chaos, sometimes leading to new violent planet scatterings \citep[e.g.;][]{mustillDestructionInnerPlanetary2015,Teyssandier2019}. Stellar flybys can occasionally contribute to the disruption, exciting in particular the planetary bodies at the outskirts of the system \citep[e.g.,][]{malmbergEffectsFlybysPlanetary2011,Rodet2021,Rodet2022}. Finally, as the star leaves the main sequence, its expansion and mass loss may completely reshape the remaining of the planetary system \citep[e.g.,][]{bonsorDynamicalEffectsStellar2011,verasPostmainsequencePlanetarySystem2016,O'connor2022}. The pollution in heavy elements observed in cool white dwarfs is evidence of a sustained population of scattered planetesimals \citep{zuckermanMetalLinesWhite2003,zuckermanAncientPlanetarySystems2010}.

When a small planet or planetesimal encounters a giant planet, it can be destroyed or removed from the system in three different ways: ejection from the system, collision with the giant planet, or disruption by the star. Previous studies mainly focused on the first two possibilities, which are by far the most common outcomes when the two scattering bodies are initially on near-circular orbits \citep[e.g.;][]{andersonInSituScatteringWarm2020,liGiantPlanetScatterings2021}. However, understanding the processes by which the host star might accrete/disrupt planetesimals and planets has become a critical issue for a variety of problems, from the link between exoplanets and exocomets \citep[e.g.][]{beustFallingEvaporatingBodies2000} to white dwarf pollution \citep[e.g.][]{steckloffHowSublimationDelays2021}, to planet engulfment by main-sequence and post-main-sequence stars \citep{spinaChemicalEvidencePlanetary2021,Behmard2023,De2023}.

This topic has been mostly studied in the context of white dwarf metal pollution. The origin of this pollution is linked to the fate of planetary systems when their host stars leave the main-sequence \citep{verasPostmainsequencePlanetarySystem2016}. Numerous studies have explored the dynamics of planet-planetesimal interactions around white dwarfs, under various conditions, including the effects of resonances \citep[e.g.,][]{smallwoodRoleResonancesPolluting2021,verasHighresolutionResonantPortraits2023}, secular chaos \citep[e.g.,][]{oconnorSecularChaosWhite2022} and exomoons \citep[e.g.,][]{doyleIcyExomoonsEvidenced2021,verasSmallestPlanetaryDrivers2023}, etc. \citep[for a recent review, see][]{verasPlanetarySystemsWhite2021}. \cite{bonsorDynamicalEffectsStellar2011} investigate how the decrease in stellar mass during the post-main sequence phase can destabilize a debris belt. The authors start from a Neptune + Kuiper belt equivalent, and use $N$-body simulations to compute the amount of material sent into the inner planetary system. 
Rather than examining the probability of a planetesimal entering a star-grazing orbit, they focus on calculating the likelihood of it ending up in orbits inwards of the planet ($0.8$--$0.9~a_{\rm p}$, where $a_{\rm p}$ is the semi-major axis of the planet). They hypothesize that the planets will continue to perturb the planetesimals, until the latters get disrupted by the star. In fact, we will demonstrate in this paper that (for example) a lone circular Neptune-mass planet at 30 au is unable to send planetesimals inwards of roughly 20 au. In a follow-up paper, \cite{bonsorScatteringSmallBodies2012} study analytically the minimum possible periastron distance of planetesimals scattered by a circular planet using the Tisserand parameter. This approach becomes invalid when the planet has a finite eccentricity or/and when the planetesimal has a non-negligible mass, as the Jacobi integral is no longer a constant of motion.
In this paper, we generalize the analysis to eccentric planets and to finite-mass  planetesimals using energy and angular momentum conservation laws. Finally, \cite{bonsorScatteringSmallBodies2012} investigate how multiple planets can increase the likelihood of sending planetesimals into the star. The full evolution of multiple planets around a white dwarf and their disruption of neighbouring planetesimals have also been studied by \cite{mustillUnstableLowmassPlanetary2018}, who quantify the rate of planetesimal accretion onto the white dwarf throughout the stellar evolution, for three different sets of planetary masses and semi-major axes. 


\cite{frewenEccentricPlanetsStellar2014} carry out a detailed study of the fate of planetesimals scattered by a giant planet. 
Using $N$-body simulations, they determine the branching ratios, i.e. the probability of the different scattering outcomes, as well as the associated timescales, as a function of the planet mass and eccentricity. However, their studies focus on a specific planet semi-major axis ($a_{\rm p} = 4$ au), radius ($R_{\rm p} = 0.6~\mathrm{R_J}$), and disruption radius ($r_{\rm dis} = 5\times 10^{-3}$ au). In this work, we seek to obtain the outcomes for planet-planetesimal scatterings under a wide range of conditions -- such general results would be valuable for many different applications.  In particular, we reproduce the \cite{frewenEccentricPlanetsStellar2014} numerical results and extend them to different ratios $r_{\rm dis}/a_{\rm p}$ and $R_{\rm p}/a_{\rm p}$. Moreover, we aim to connect the numerical simulations to the analytical results mentioned above.

Our paper is organized as follows.
In Section~\ref{sec:qmin}, we derive the minimum periastron distance that a scattered planetesimal (or a small planet) can reach as a function of the giant planet properties. We present $N$-body simulations in Section~\ref{sec:fate}, which aim to validate the analytical results as well as to derive the branching ratio and timescale associated with the scattering process. We compare our results to the \cite{frewenEccentricPlanetsStellar2014} study, and discuss how our more general results can be used to predict the flux of scattered planetesimals under different conditions. In Section~\ref{sec:2planet}, we briefly study the case of planetesimals  scattered by two planets and compare the results to the single planet case. Finally, in Section~\ref{sec:discussion}, we discuss our results in relation to three different applications: exocomets in the $\beta$ Pictoris system, white dwarf pollution by planetesimal disruption and planet engulfment by main-sequence stars. We summarize our study in Section~\ref{sec:conclusion}.

\section{Star-grazing orbits by scatterings: analytical Results}
\label{sec:qmin}

We consider a planet of mass $m_{\rm p}$ orbiting a star of mass $M_\star$, with a semi-major axis $a_{\rm p}$ and eccentricity $e_{\rm p}$. The planet encounters a low-mass object ($m < m_{\rm p}$), initially orbiting on a nearby orbit of semi-major axis $a_0$, eccentricity $e_0$ and inclination $i_0$ with respect to the planet's orbital plane. Throughout the paper, we shall call the low-mass object ``planetesimal'', although some of our analytical results (Section~\ref{sec:conservation}) apply even when $m/m_{\rm p}$ is not negligible. The goal of our paper is to determine the probability for the planetesimal $m$ to be scattered inward and disrupted by the star. Close encounters are generally chaotic, but the conservations of energy and angular momentum limit the possible outcomes. In particular, they allow us to derive the minimum possible periastron distance that $m$ can attain as a function of the planet mass and orbital properties. We will demonstrate that some initial configurations cannot lead to stellar disruption or consumption of planetesimals.

\subsection{Restricted three-body problem with a circular planet}
\label{sec:tisserand}

Let us assume that the planetesimal is a test mass ($m = 0$) and the planet is on a circular orbit ($e_{\rm p} = 0$). In this case, the Tisserand parameter of the test mass is conserved
\citep[e.g.][]{murraySolarSystemDynamics2000}, i.e.
\begin{align}
	T ={}&  \frac{a_{\rm p}}{a} + 2\sqrt{\frac{a}{a_{\rm p}}(1-e^2)}\cos i  \nonumber\\
	={}& \frac{a_{\rm p}}{a_0} + 2\sqrt{\frac{a_0}{a_{\rm p}}(1-e_0^2)}\cos i_0, \label{eq:T}
\end{align}
where $a$, $e$ and $i$ are the orbital elements of the test mass after close encounters. For small $e_0$, $i_0$ and for $a\simeq a_{\rm p}$, we have
\begin{equation}
	T = 3 + \frac{3}{4}\left(\frac{a_0-a_{\rm p}}{a_{\rm p}}\right)^2 - e_0^2 - i_0^2 + O\left[(a_0-a_{\rm p})^j e_0^k i_0^l\right] \label{eq:Tapprox},
\end{equation}
with $j+k+l \geq 3$. \cite{bonsorScatteringSmallBodies2012} have used the Tisserand parameter to predict the minimum possible periastron distance $q = a(1-e)$ for test masses with $T < 3$. Their results do not apply to test masses on initially circular and coplanar orbits, and/or in the extended chaotic zone around the planet; these particles can reach $T$ values of $3$--$4$. Here we adopt a different approach to extend their results to all $T$'s. 

\begin{figure}
	\centering
	\includegraphics[width=\linewidth]{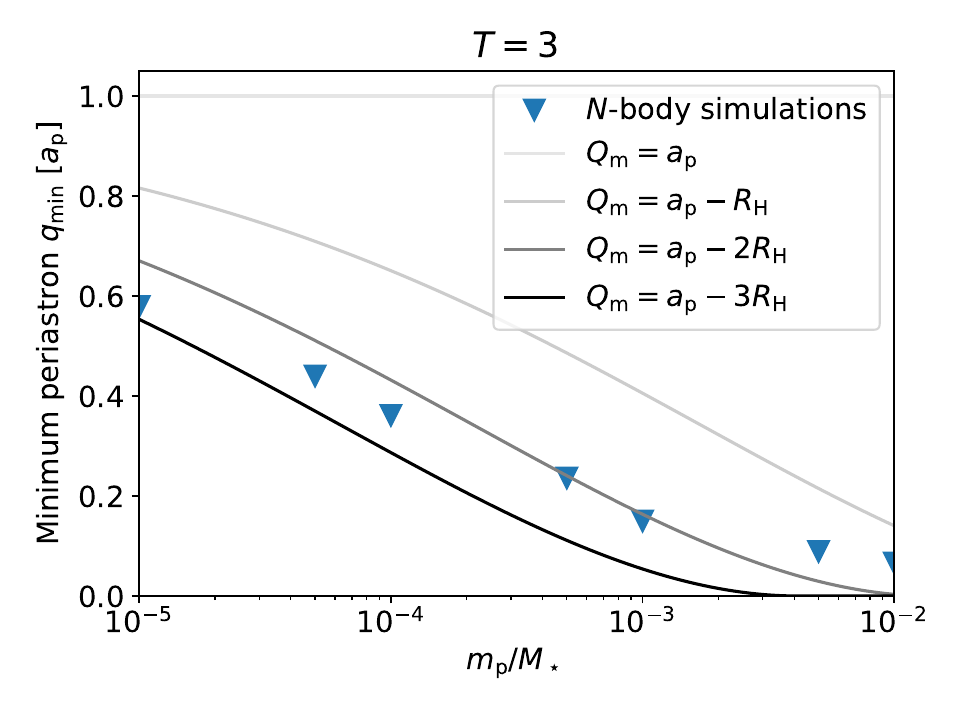}
	\caption{Minimum periastron $q_{\rm min}$ of a test mass scattered by a circular planetary perturber orbiting at $a_{\rm p}$, as a function of the planet-to-star mass ratio $m_{\rm p}/M_\star$. The blue triangles represent the minimum periastron reached by 1000 initially nearly coplanar and circular test particles in $N$-body simulations (see Section~\ref{sec:fate} for details). The grey curves are the theoretical minima (equation~\ref{eq:qmin}) corresponding to $T = 3$ and critical apoastrons $Q_{\rm m} = \tilde{Q}_{\rm m} a_{\rm p}$ ranging from $a_{\rm p}-3 R_{\rm H}$ to $a_{\rm p}$.}\label{fig:qminvsmu}
\end{figure}

Let us rewrite $T$ as a function of the dimensionless periastron $\tilde{q} \equiv q/a_{\rm p}$ and apoastron $\tilde{Q} \equiv Q/a_{\rm p} = a(1+e)/a_{\rm p}$:
\begin{equation}
	T = \frac{2}{\tilde{q}+\tilde{Q}} + 2\sqrt{\frac{2\tilde{Q}\tilde{q}}{\tilde{q}+\tilde{Q}}}\cos i. \label{eq:teq}
\end{equation}
We focus on the coplanar case ($i = 0$), which corresponds to the minimum reachable periastron. Theoretically, according to equation~\eqref{eq:teq}, for a given $T$, the minimum periastron $q$ can reach zero if the apocenter distance can reach $2/T$. However, there is an additional constraint to take into account for the post-close encounter orbit of the test mass. Due to the time symmetry of the dynamics, a close encounter cannot end in two stable bound orbits. This means that the orbit of the test mass cannot be completely detached from the orbit of the planet: either the orbits cross, or the periastron or apoastron of the test mass remains  close to $a_{\rm p}$ so that the test mass can be perturbed again (i.e. the test mass orbit intersects with the so-called chaotic zone around the planet). 
The stability between two bodies is related to the mutual Hill radius, $R_{\rm H} = r_{\rm H} (a+a_{\rm p})/2$, where
\begin{equation}
	r_{\rm H} = \left(\frac{m + m_{\rm p}}{3M_\star}\right)^\frac{1}{3},
\end{equation}
\citep{gladmanDynamicsSystemsTwo1993,chambersStabilityMultiPlanetSystems1996}. Thus, we add the following condition to the post-encounter orbit for the test mass:
\begin{equation}
	\tilde{Q} \ge 1 - \beta r_{\rm H} \equiv \tilde{Q}_{\rm m}, \label{eq:condition}
\end{equation}
where $\beta$ typically equal $2$--$3$. This condition ensures that near the apocenter, the test mass remains in the planet's chaotic zone. The prescription $\beta = 3$ is very similar (within 10\%) to the chaotic zone estimate derived empirically by  \cite{frewenEccentricPlanetsStellar2014}. However, we find (see Fig.~\ref{fig:qminvsmu}) that $\beta = 2$ is a better fit to $\tilde{Q}_{\rm m}$, since the post-encounter orbit of the test mass may not cover the entire extent of the chaotic zone around the planet.

The minimum periastron distance depends on both $T$ and $\tilde{Q}_{\rm m}$. From equation (\ref{eq:teq}), we see that if $2/T \geq \tilde{Q}_{\rm m}$, then a zero periastron is allowed: the test mass can theoretically be sent into the star, whatever the stellar radius. This condition requires $T \le 2/(1-\beta r_{\rm H})$, corresponding to an initially eccentric and/or inclined orbit for the test mass. On the other hand, if $2/T < \tilde{Q}_{\rm m}$, then the minimum periastron has a finite value.
We rewrite equation~\eqref{eq:teq} to obtain $\tilde{q}$ as a function of $T$ and $\tilde{Q}$:
\begin{equation}
	\tilde{q} = \frac{4\tilde{Q}^2 + 2 T - \tilde{Q} T^2 - 4\sqrt{ \tilde{Q}(2+\tilde{Q}^3-T\tilde{Q})}}{T^2-8\tilde{Q}}. \label{eq:q}
\end{equation}
We study the variation of $\tilde{q}(\tilde{Q})$ in  Appendix~\ref{sec:appendixtisserand}, with $\tilde{Q} > \tilde{Q}_{\rm m}$. If $2/T < \tilde{Q}_{\rm m}$, the function $\tilde{q}(\tilde{Q})$ is minimal at $\tilde{Q}=\tilde{Q}_{\rm m}$; the minimum $\tilde{q}$ is simply obtained by replacing $\tilde{Q}$ by $\tilde{Q}_{\rm m}$ in equation~\eqref{eq:q}: 
\begin{equation}
	\tilde{q}_{\rm min} = \frac{4\tilde{Q}_{\rm m}^2 + 2 T - \tilde{Q}_{\rm m} T^2 - 4\sqrt{ \tilde{Q}_{\rm m}(2+\tilde{Q}_{\rm m}^3-T\tilde{Q}_{\rm m})}}{T^2-8\tilde{Q}_{\rm m}}.\label{eq:qmin}
\end{equation}
If $Q_{\rm m} < 2/T$, then $\tilde{q}_{\rm min} = 0$.

In Section~\ref{sec:fate}, we use $N$-body simulations to determine the minimum distance from the star that a belt of planetesimals scattered by a giant planet can reach. We compare these simulation results to equation~(\ref{eq:qmin}) in Figure~\ref{fig:qminvsmu} for different planet-to-star mass ratios. The simulations agree reasonably well with the theory, and suggest that $\beta=2$-$3$ is a good approximation of the chaotic zone accessible to the planetesimals after close encounters with the planet.

Our result for $q_{\rm min}$ differs from that of \cite{bonsorScatteringSmallBodies2012}. In that  paper, the authors restrict to the test mass population with $T < 3$ and assume $\beta = 0$. This approximation significantly overestimates the minimum periastron distance for the parameters we consider (see Figures~\ref{fig:qminvsmu} or \ref{fig:qminvsT}).

\subsection{Eccentric planet and finite planetesimal mass}
\label{sec:conservation}

\begin{figure}
	\centering
	\includegraphics[width=\linewidth]{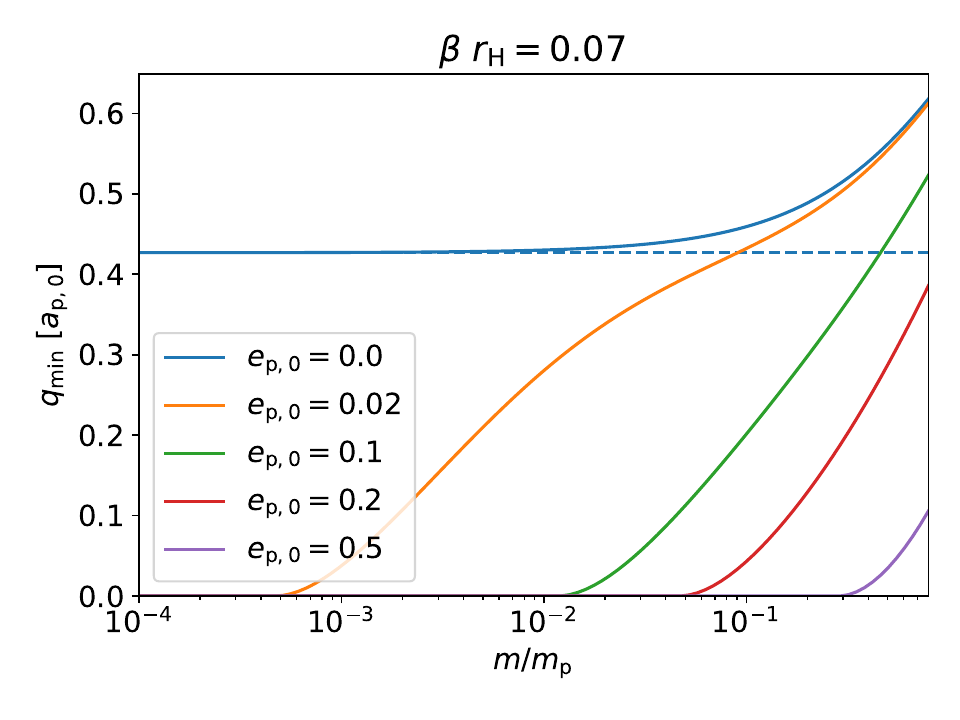}
	\includegraphics[width=\linewidth]{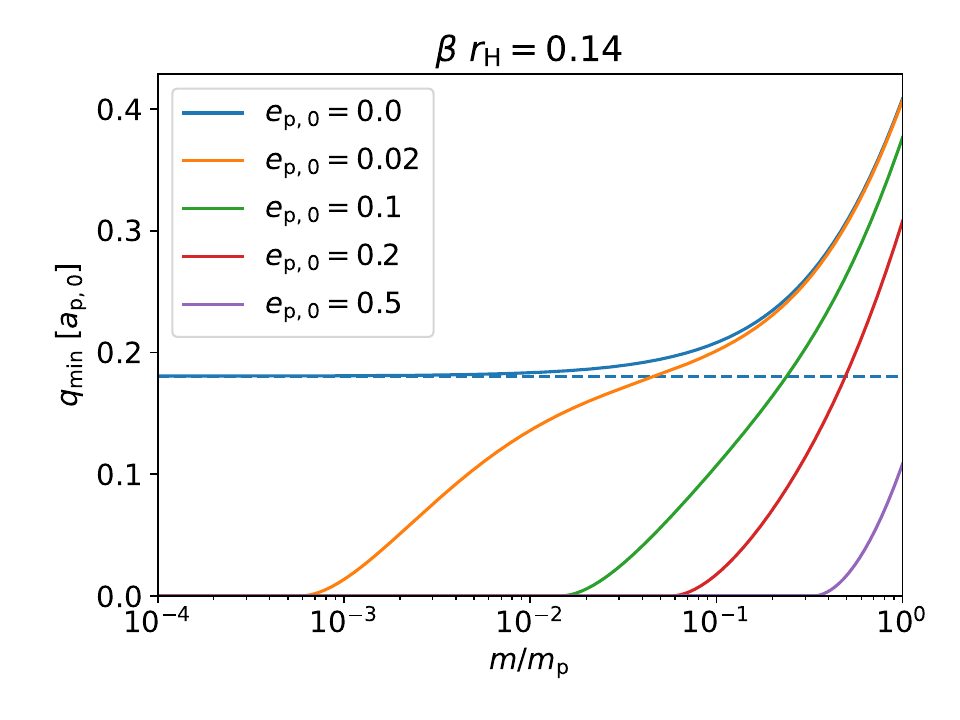}
	\includegraphics[width=\linewidth]{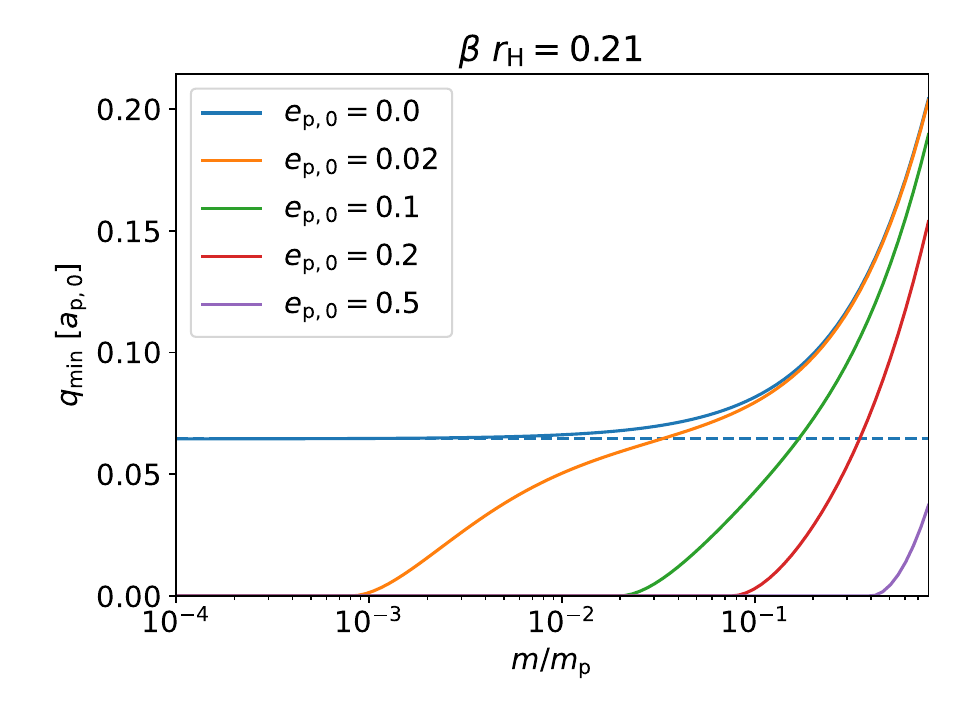}
	\caption{Minimum periastron $q_{\rm min}$ of a planetesimal scattered by a planetary perturber initially at $a_{\rm p, 0}$, as a function of the planetesimal-to-planet mass ratio $m/m_{\rm p}$. The planetesimal has initially zero eccentricity and has a semi-major axis of $a_{0} = a_{\rm p, 0} + \beta R_{\rm H}$, corresponding to $T = 3.01$ (the results do not depend strongly on $a_0$ and $e_0$ as long as $T\simeq 3$.). Each different colour corresponds to a fixed initial planetary eccentricity $e_{\rm p,0}$. The curves are obtained semi-analytically, by finding the minimum $q$ that satisfies equations \eqref{eq:E}, \eqref{eq:h} and \eqref{eq:q1} for different $\beta r_{\rm H}$ values ($\beta r_{\rm H} = 0.14$ corresponds to $\beta =2$ and $m_{\rm p}/M_\star = 10^{-3}$). The dashed blue line corresponds to the analytical result of the restricted three-body problem (equation~\ref{eq:qmin}).}\label{fig:qmin_vs_m} 
\end{figure}

When the planet is eccentric or the planetesimal has a finite mass, the result of Section~\ref{sec:tisserand} cannot apply, because the Tisserand parameter is not conserved. However, during the scattering process, both the total energy and angular momentum are conserved. In the coplanar case, we have
\begin{align}
	E ={}& \frac{m_{\rm p}}{a_{\rm p}} + \frac{m}{a} = \frac{m_{\rm p}}{a_{\rm p,0}} + \frac{m}{a_{0}}, \label{eq:E}\\
	h ={}& m_{\rm p}\sqrt{a_{\rm p}(1-e_{\rm p}^2)} + m\sqrt{a(1-e^2)} \nonumber \\
	={}& m_{\rm p}\sqrt{a_{\rm p,0}(1-e_{\rm p,0}^2)} + m\sqrt{a_{0}(1-e_{0}^2)},\label{eq:h}
\end{align}
where we set the gravitational constant $G=1$.
Moreover, as previously stated, the two orbits must stay close to one another so that they remain unstable. This can be translated to the following conditions:
\begin{align}
	\frac{q}{Q_{\rm p}} &\le 1+\beta~r_{\rm H}, \label{eq:rh}\\
	\frac{Q}{q_{\rm p}} &\ge 1-\beta~r_{\rm H}.  \label{eq:rh2}
\end{align}
Exploring the allowed parameter space, we notice that $q$ is minimum when $Q$ is minimum, similarly to the test mass case described above. We thus set
\begin{align}
	\frac{Q}{q_{\rm p}} = 1-\beta~r_{\rm H}.  \label{eq:q1}
\end{align}
The minimum periastron $q_{\rm min}$ can then be determined semi-analytically, by finding the  parameters $(a, q_{\rm p}, a_{\rm p})$ that minimize $q$ while satisfying equations~\eqref{eq:E}, \eqref{eq:h}, and \eqref{eq:q1}. The resulting $q_{\rm min}$ is plotted in Figure~\ref{fig:qmin_vs_m} as a function of the mass ratio $m/m_{\rm p}$,
and the details of the derivation can be found in Appendix~\ref{sec:appendixconservation}. 

When $e_{\rm p,0} = 0$ and $m \to 0$, $q_{\rm p} \to a_{\rm p}$. Using equations \eqref{eq:E}, \eqref{eq:h}, and \eqref{eq:q1}, we can derive analytically the minimum $q$ in the test mass limit: we retrieve the result of the previous section (equation~\ref{eq:qmin}). The derivation is detailed in Appendix~\ref{sec:appendixconservation}.

\begin{figure}
	\centering
	\includegraphics[width=\linewidth]{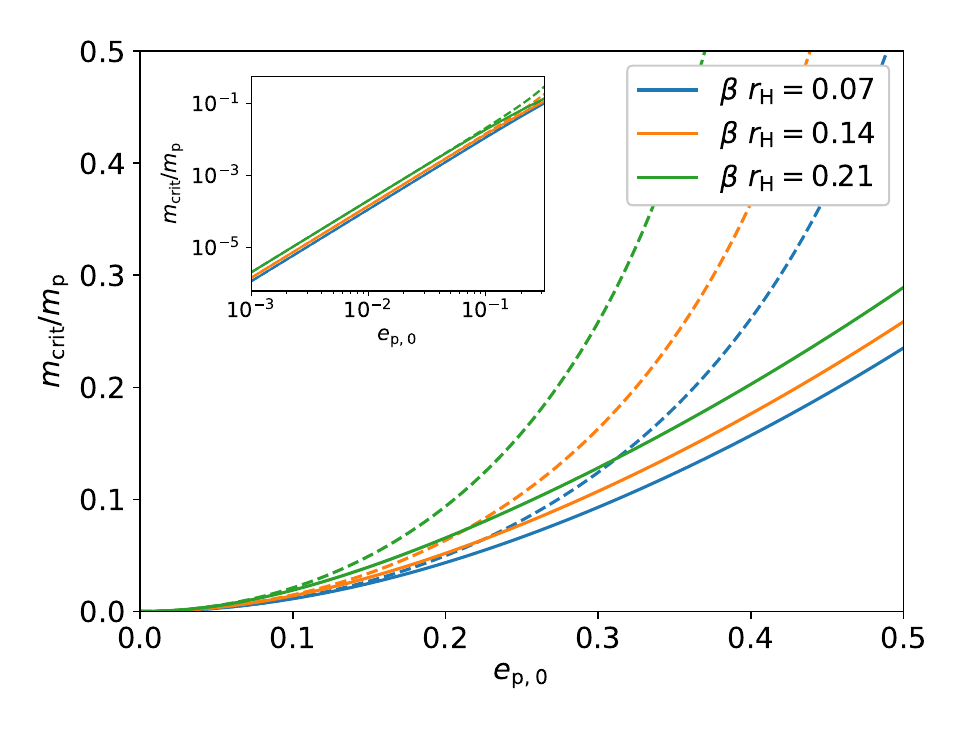}
	\caption{Critical planetesimal-to-planet mass ratio below which there is no theoretical minimum to the periastron distance of the planetesimal (i.e. $q_{\rm min} = 0$), for different $\beta r_{\rm H}$ values ($\beta r_{\rm H} = 0.14$ corresponds to $\beta =2$ and $m_{\rm p}/M_\star = 10^{-3}$). The dashed lines correspond to the small-mass approximation (equation~\ref{eq:ecrit}), the solid lines to the exact result (equation~\ref{eq:condition_finitem}). The planetesimal has initially zero eccentricity and has a semi-major axis of $a_{0} = a_{\rm p, 0} + 2 R_{\rm H}$, corresponding to $T = 3.01$. The upper left subplot is a log-log close-up of the small eccentricity part of the main plot.}\label{fig:mcrit}
\end{figure}

We can also derive analytically the threshold between the configurations with $q_{\rm min} = 0$ and with $q_{\rm min} > 0$ in the low-mass limit. Equations~\eqref{eq:E}, \eqref{eq:h} and \eqref{eq:rh2} give the following constraints:
\begin{align}
	&h = m_{\rm p}\sqrt{q_{\rm p}\left(2-\frac{q_{\rm p}}{a_{\rm p}}\right)}, \label{eq:h_q=0}\\
	&q_{\rm p} \le \frac{2a}{1-\beta r_{\rm H}} = \frac{2m}{(1-\beta r_{\rm H})\left(E-\frac{m_{\rm p}}{a_{\rm p}}\right)}.\label{eq:qpsolmax}
\end{align}
Using equation~\eqref{eq:h_q=0}, we can express $q_{\rm p}$ as a function of $h$ and $a_{\rm p}$, and inject this expression into equation~\eqref{eq:qpsolmax}. After some algebra (see Appendix~\ref{sec:appendixconservation}), we find a sufficient condition for $q_{\rm min} = 0$:
\begin{align}
	\frac{h^2E}{m_{\rm p}^3} \le \frac{2m}{m_{\rm p}(1-\beta r_{\rm H})} + 1. \label{eq:condition_finitem}
\end{align}
The left-hand side of the equation is a quantity that appears in other works related to the three-body problem \citep[e.g. $c^2 h / {M^*}^3$ in][]{marchalHillRegionsGeneral1975}, but, to our knowledge, the condition derived above  is new. This equation becomes a sufficient and necessary condition for $m/m_p\ll 1$, where it reduces to
\begin{equation}
	m \le m_{\rm  crit} = m_{\rm p} \frac{e_{\rm p,0}^2}{T\sqrt{1-e_{\rm p, 0}^2}-\frac{2}{1-\beta r_{\rm H}}}.\label{eq:ecrit}
\end{equation}
Thus, for an eccentric planet, there is a critical planetesimal mass $m_{\rm  crit}$ below which the minimum pericenter distance to the star can reach zero. This critical mass is shown in Figure~\ref{fig:mcrit} as a function of the planet initial eccentricity. Although equation~\eqref{eq:ecrit} is strictly valid only for $m/m_{\rm p}\ll 1$, it still provides a reasonable approximation of the critical mass for $m \lesssim m_{\rm p}$. Moreover, equation~\eqref{eq:ecrit} implies that in the test mass regime, $q_{\rm min} = 0$ as long as $e_{\rm p} > 0$. The different behaviors between the circular and eccentric planet cases have been observed numerically, for example in \cite{antoniadouLinkingLongtermPlanetary2016} or \cite{verasEntryGeometryVelocity2021} in the context of white dwarf pollution.

\section{Fate of the scattered planetesimals: Simulation Results}
\label{sec:fate}

\begin{figure*}
	\centering
	\includegraphics[width=\linewidth]{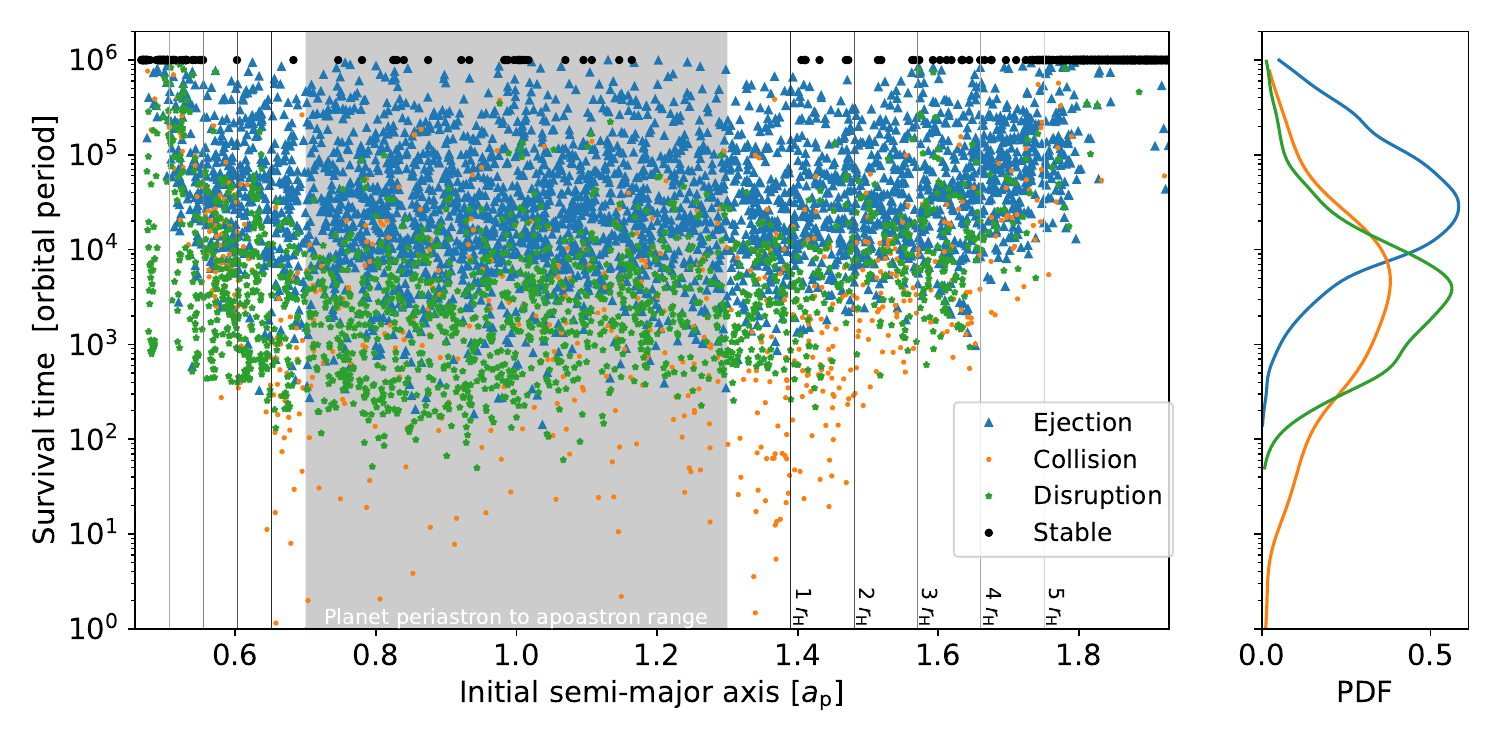}
	\caption{Survival time (in unit of the planet orbital period) with respect to the initial semi major axis (in unit of the planet semi-major axis), for $5,000$ test particles randomly populating the instability zone \citep[as computed by][ equations~\ref{eq:Petrovich1} and \ref{eq:Petrovich2}]{petrovichStabilityFatesHierarchical2015} around a planet with eccentricity $e_{\rm p} = 0.3$, planet-to-star mass ratio $10^{-3}$ and radius $R_{\rm p} = 5\times 10^{-4} a_{\rm p}$. The shape and colour coding of the dots correspond to the fate of the particles within the $10^6$ orbital periods. We chose a star disruption radius of $0.1 a_{\rm p}$. The details of the $N$-body simulation are described in Section~\ref{sec:fate}. The shaded region corresponds to $0.7a_{\rm p}<a_0<1.3a_{\rm p}$ (planet periastron to apoastron range).  The thin vertical lines correspond to one, two, three, four and five Hill radii distances from the planet orbit, where $r_{\rm H}=1.3a_{\rm p} (m_{\rm  p}/3M_\star)^{1/3}$ for $a_0>1.3a_{\rm p}$, and $r_{\rm H}=0.7a_{\rm p} (m_{\rm  p}/3M_\star)^{1/3}$ for $a_0<0.7a_{\rm p}$.
 } \label{fig:timescale_vs_ci}
\end{figure*}

In the previous section, we discussed the theoretical constraints on the outcomes of planet-planetesimal scatterings, and derived analytical and semi-analytical results for the minimum stellar approach $q_{\rm min}$ of the planetesimal. We showed the importance of the planet initial eccentricity and of the planet-to-star mass ratio. In particular, we demonstrated that there is no theoretical barrier that prevents a test particle to plunge into the star after being scattered by a planet, as long as this planet has a finite eccentricity (equation~\ref{eq:ecrit}). However, the theory does not give any information on the proportion of scatterings that lead to a disruption/consumption of the planetesimal by the star, nor on the timescale needed to achieve it. In this section, we perform $N$-body simulations of planetesimals (test particles) initially orbiting close to a giant planet, and monitor their fate.

The simulations are performed using the \texttt{IAS15} integrator in the \texttt{Rebound} Python package. They are initialized as follows: The planet of radius $R_{\rm p}$ is set on an orbit with semi-major axis $a_{\rm p}$ with eccentricity $e_{\rm p}$. A the beginning of the simulations, the test particles have eccentricities randomly distributed between $0$ and $0.05$, inclinations between $0$ and $2^\circ$ (following a cosine prior) with respect to the planet orbit, with the argument of periastron, longitude of the node and mean anomaly uniformly distributed between $0$ and $360^\circ$. Their initial semi-major axes are sampled uniformly within the instability range computed by \cite{petrovichStabilityFatesHierarchical2015}: the inner and outer edges of the instability region $a_{\rm in}$ and $a_{\rm out}$ are the solutions of the following equations:
\begin{align}
	&\frac{a_{\rm p}(1-e_{\rm p})}{a_{\rm in}} = 2.4 \left(\frac{m_{\rm p}}{M_* + m_{\rm p}}\right)^\frac{1}{3} \left(\frac{a_{\rm p}}{a_{\rm in}}\right)^\frac{1}{2} + 1.15\label{eq:Petrovich1},\\
	&\frac{a_{\rm out}}{a_{\rm p}(1-e_{\rm p})} = 2.4 \left(\frac{m_{\rm p}}{M_* + m_{\rm p}}\right)^\frac{1}{3} \left(\frac{a_{\rm out}}{a_{\rm p}}\right)^\frac{1}{2} + 1.15. \label{eq:Petrovich2}
\end{align}
 The resulting semi-major axis range encompasses the eccentric chaotic zone derived by \cite{frewenEccentricPlanetsStellar2014}, so that we are confident that our simulations cover the complete behaviors of the test particle instabilities. The simulations run for $10^6$ planet orbital periods. The test particles are removed only if they are ejected or if they collide with the planet, and their minimum distance to the star is monitored, as detailed below. 

The ejection outcome (``eject'') in the simulation is defined by $e > 1$ (the eccentricity is checked when the particle's distance to the star is larger than 100$a_{\rm p}$). The collision outcome (``coll'') is defined by the particle approaching the planet at a distance less than $R_{\rm p}$, the radius of the planet. We use the routine included in \texttt{Rebound} that draws a line between two positions at consecutive times to ensure it did not miss the collision. The stellar disruption outcome (``dis'') corresponds to the particle approaching the star at a distance less than a certain threshold $r_{\rm dis} < 0.1~a_{\rm p}$. However, for efficiency, we do not remove the particle when it reaches the threshold. Instead, we monitor each close encounter with the star (distance less than $0.1~a_{\rm p}$) by recording its time and the corresponding periastron of the particle's orbit. In addition, we monitor the minimum stellar distance reached by the test particle during the whole integration, using the periastron of the particle's orbit when its computation is reliable. This technique allows us to set the stellar disruption threshold after the simulation: we do not need to perform new simulations if we want to change this threshold.

An example of the simulation of $5,000$ test particles around an eccentric giant planet ($e_{\rm p} = 0.3$, $m_{\rm p} = 10^{-3}M_\star$) is shown in  Figure~\ref{fig:timescale_vs_ci}. The initial semi-major axes of the particles range from $0.5~a_{\rm p}$ to $1.9~a_{\rm p}$, following equations~\eqref{eq:Petrovich1} and \eqref{eq:Petrovich2}. The particles are colour-coded by their fates, among the four possible outcomes (Ejection, Collision, disruption by the Star, Stability). A correlation between the particles instability timescales and their fates can clearly be seen. Some particles remain stable for the entirety of the simulation run, even though they are located deep in the chaotic zone. This is because the particles trajectories are chaotic and their outcomes at $10^6$ orbital periods are probabilistic.

\begin{figure}
	\centering
	\includegraphics[width=\linewidth]{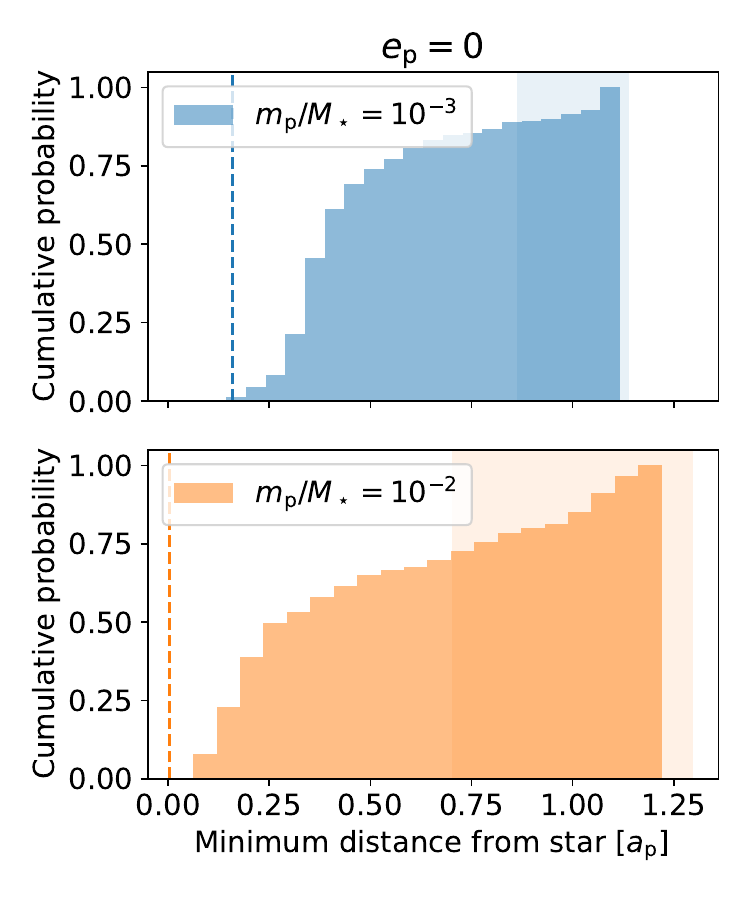}
	\caption{Distribution of the minimum distances to the star from the simulations of two samples of 1000 test particles scattered by a circular planetary body of mass ratio $m_{\rm p}/M_\star$ during $10^6$ orbital periods. The upper panel corresponds to a giant planet ($m_{\rm p}/M_\star = 10^{-3}$), and the lower panel to a brown dwarf ($m_{\rm p}/M_\star = 10^{-2}$). 
 The planet has a radius $R_{\rm p} = 5\times 10^{-4} a_{\rm p}$. The lightly shaded regions correspond to $(a_{\rm p}-2R_{\rm H}, a_{\rm p}+2R_{\rm H})$, where most of the interaction occurs. The dashed vertical lines correspond to the analytical predictions for the minimum periastron distance given by equation~\eqref{eq:qmin} with $\beta = 2$ and $T = 2.996$ (corresponding to $e = 0.05$ and $i = 2^\circ$). A higher planetary mass decreases the minimum possible periastron distance, without significantly changing the overall distribution of the closest approaches.} \label{fig:dmin}
\end{figure}

\begin{figure}
	\centering
	\includegraphics[width=\linewidth]{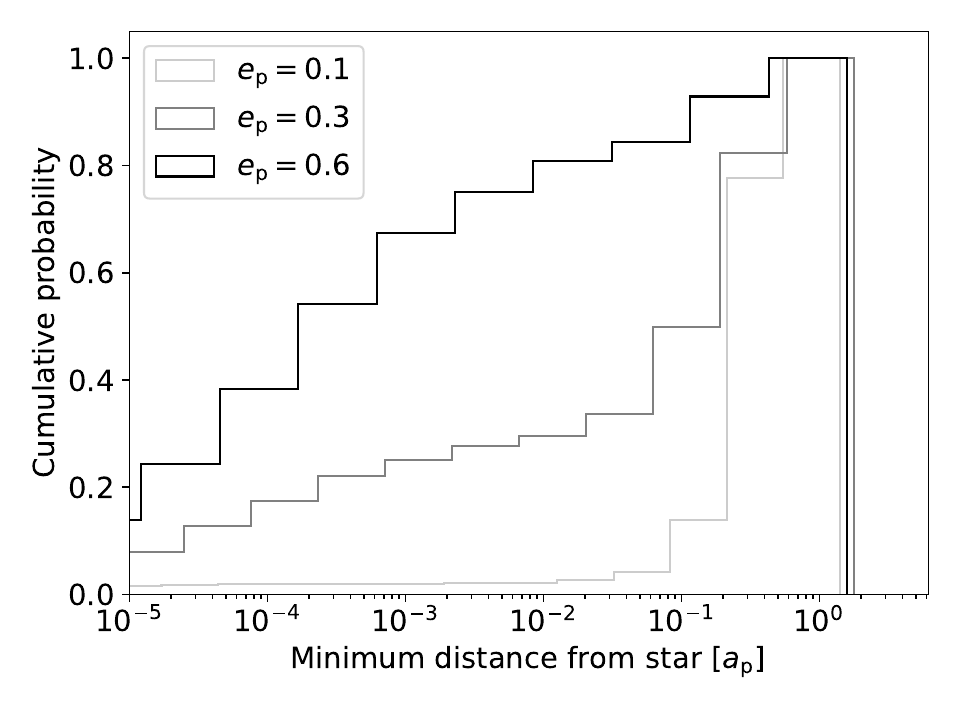}
	\caption{Same as Figure~\ref{fig:dmin}, for a mass ratio of $m_{\rm p}/M_\star = 10^{-3}$ and different planetary eccentricities. A finite eccentricity strongly increases the number of test particles that reach the close vicinity of the star.} \label{fig:dmin_log}
\end{figure}

\subsection{Key parameters}

The probability and timescale of each outcome for the whole population of unstable test particles depend on four parameters:
\begin{itemize}
	\item Planet eccentricity, $e_{\rm p}$;
	\item Planet to star mass ratio, $m_{\rm p}/M_\star$;
	\item Planet radius to semi-major axis ratio, $R_{\rm p}/a_{\rm p}$;
	\item Disruption radius $r_{\rm dis}$ of planetesimals by the star to planet semi-major axis ratio, $r_{\rm dis}/a_{\rm p}$.
\end{itemize}
The planetesimal disruption radius $r_{\rm dis}$ is the combination (maximum) of three quantities: (i) the stellar radius $R_\star$, (ii) the tidal radius $\sim (M_\star/\rho_{\rm ps})^{1/3} \simeq R_\odot (M_\star/M_\odot)^{1/3} (\rho_{\rm ps}/1~\mathrm{g.cm}^3)^{-1/3}$,  where $\rho_{\rm ps}$ is the bulk density of the planetesimal, (iii) the evaporation radius \citep[e.g.,][]{beustPictorisCircumstellarDisk1996}. Since these radii depend on the problem under study and on the stellar type,  we consider here a range of $r_{\rm dis}/a_{\rm p}$, from $10^{-4}$ to $10^{-1}$.
The above four quantities, along with the distribution of the initial conditions of the test particles, entirely determine the outcomes of scatterings. They allow us to study the problem in a scale-free way, and to adapt our results to a variety of situations. In the following, we will examine the impacts of varying each parameter using $N$-body simulations, building on our analytical results derived in Section~\ref{sec:qmin}.

\subsection{Distribution of minimum distances}

In Section~\ref{sec:qmin}, we computed analytically the minimum periastron distance that is theoretically reachable for a planetesimal scattered by a massive planet. For a circular planet, this limit is finite and varies with the planet-to-star mass ratio $m_{\rm p}/M_\star$ (see Figure~\ref{fig:qminvsmu}). For an eccentric planet, there is no such minimum. 

To determine the distribution of the periastron distances, we perform $N$-body simulations with $1000$ test particles orbiting close to a circular giant planet ($m_{\rm p}/M_\star = 10^{-3}$) and brown dwarf ($m_{\rm p}/M_\star = 10^{-2}$). We then monitor the closest encounter each planetesimal has with the star before being ejected or colliding with the planet ($R_{\rm p} = R_{\rm J}$). The cumulative distribution of the minimum distances is shown in Figure~\ref{fig:dmin}. We do not consider stellar disruption in generating this plot.
When $m_{\rm p}/M_\star = 10^{-3}$, Section~\ref{sec:qmin} predicts a minimum periastron $q_{\rm min} \simeq 0.18~a_{\rm p}$.
On the other hand, when $m_{\rm p}/M_\star = 10^{-2}$, the minimum distance $q_{\rm min} \simeq 0.07~a_{\rm p}$ goes below the $0.1~a_{\rm p}$ that we typically set as stellar disruption radius. In addition to changing the theoretical minimum, we see that the planet-to-star mass ratio also modifies the shape of the distribution of the closest approaches, with more weight given to smaller distances as $m_{\rm p}/M_\star$ increases.

In both cases depicted in Fig.~\ref{fig:dmin}, the proportion of planetesimals that get disrupted by the star is negligible if we assume a disruption radius $r_{\rm dis} < 0.1 a_{\rm p}$. Inspired by the eccentricity dependence suggested by Figures~\ref{fig:qmin_vs_m} and \ref{fig:mcrit},
we perform additional simulations with $e_{\rm p} = 0.1$, $e_{\rm p} = 0.3$ and $e_{\rm p} = 0.6$, and show the cumulative histograms of the minimum distances in Figure~\ref{fig:dmin_log}. The number of stellar disruptions increases dramatically with higher eccentricities: 5\% for $e_{\rm p} = 0.1$, 40\% for $e_{\rm p} = 0.3$ and 80\% for $e_{\rm p} = 0.6$. This trend is true also if we strengthen the requirement for stellar disruption, down to a stellar disruption radius of $10^{-4}~a_{\rm p}$ (as will be discussed below).

Finally, the distribution of minimum distances does not significantly vary with the planet radius to semi-major axis ratio $R_{\rm p}/a_{\rm p}$, as long as collisions do not represent the majority of the outcome. When they do, however, the overall lifetime of the particle is reduced, which leads to far less particles reaching small stellar distances.

\subsection{Branching Ratio}
\label{sec:branchingratio}

\begin{figure}
	\centering
	\includegraphics[width=\linewidth]{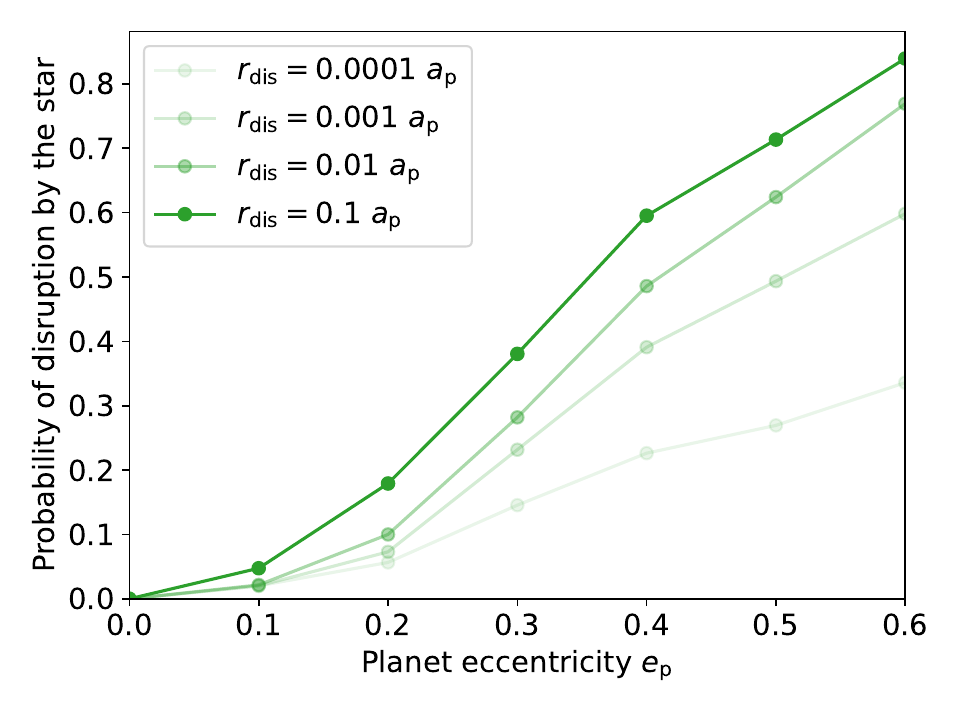}
	\caption{Probability for a scattered test particle to reach the stellar disruption radius $r_\mathrm{dis}$ as a function of the perturbing planet eccentricity $e_{\rm p}$. These results are computed from the $N$-body simulations of 1000 test particles (see details in  Section~\ref{sec:fate}) with a planet-to-star mass ratio of $m_{\rm p}/M_\star = 10^{-3}$ and a radius of $R_{\rm p} = 5\times 10^{-4} a_{\rm p}$. Following our analytical results (Section~\ref{sec:qmin}), the probability is zero for a circular planet and $r_{\rm dis} < 0.1~a_{\rm p}$; it increases strongly with the planet eccentricity regardless of the disruption radius.} \label{fig:proba_vs_e}
\end{figure}

From Figures~\ref{fig:dmin} and \ref{fig:dmin_log}, we see that the probability for the scattered particles to be disrupted by the star depends strongly on the planet eccentricity, and moderately on the exact value of the disruption radius $r_{\rm dis}$. We show these dependencies in Figure~\ref{fig:proba_vs_e}. For $r_{\rm dis} = 0.1~a_{\rm p}$, more than half of the unstable test particles are disrupted by the star if the planet eccentricity is greater than $0.4$. By assuming a much smaller disruption radius or a much wider planetary orbit ($r_{\rm dis} = 0.0001~a_{\rm p}$), more than $20\%$ of the test particles are still disrupted by the star for the same planet eccentricity.

\begin{figure}
	\centering
	\includegraphics[width=\linewidth]{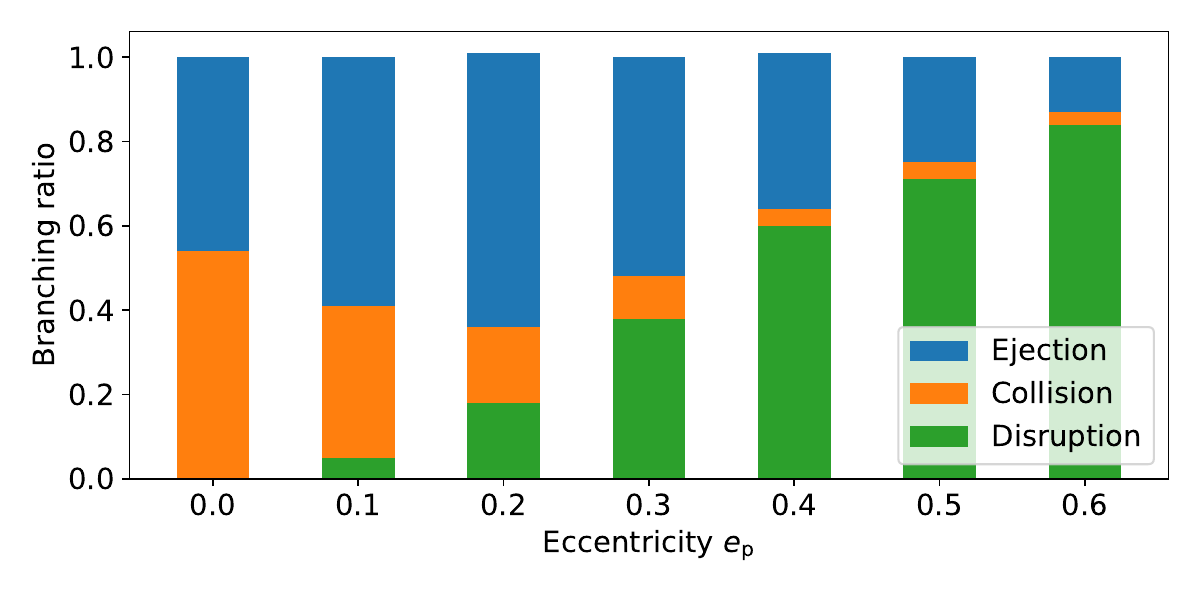}
	\caption{Branching ratios (i.e. probabilities of the different outcomes, i.e., ejection, collision with the planet or disruption by the star) of a scattered test particle as a function of the eccentricity of the planetary perturber, for a stellar disruption radius of $r_\mathrm{dis} = 0.1~a_{\rm p}$. The planet has a mass $m_{\rm p} = 10^{-3}~M_\star$ and a radius of $R_{\rm p} = 5\times 10^{-4} a_{\rm p}$. These results are computed from the $N$-body simulations of 1000 test particles (see details in  Section~\ref{sec:fate}). }\label{fig:branchingratio}
\end{figure}

Apart from the disruption by the star, the probability of ejection or collision with the planet also varies with the planet eccentricity. In Figure~\ref{fig:branchingratio}, we show the branching ratios (i.e. probabilities of different outcomes) of unstable particles as a function of the planet eccentricity (assuming a fixed disruption radius $r_{\rm dis} = 0.1~a_{\rm p}$). Both outcomes become much rarer as the eccentricity increases and the probability of disruption by the star skyrockets.

\begin{figure*}
	\centering
	\includegraphics[width=\linewidth]{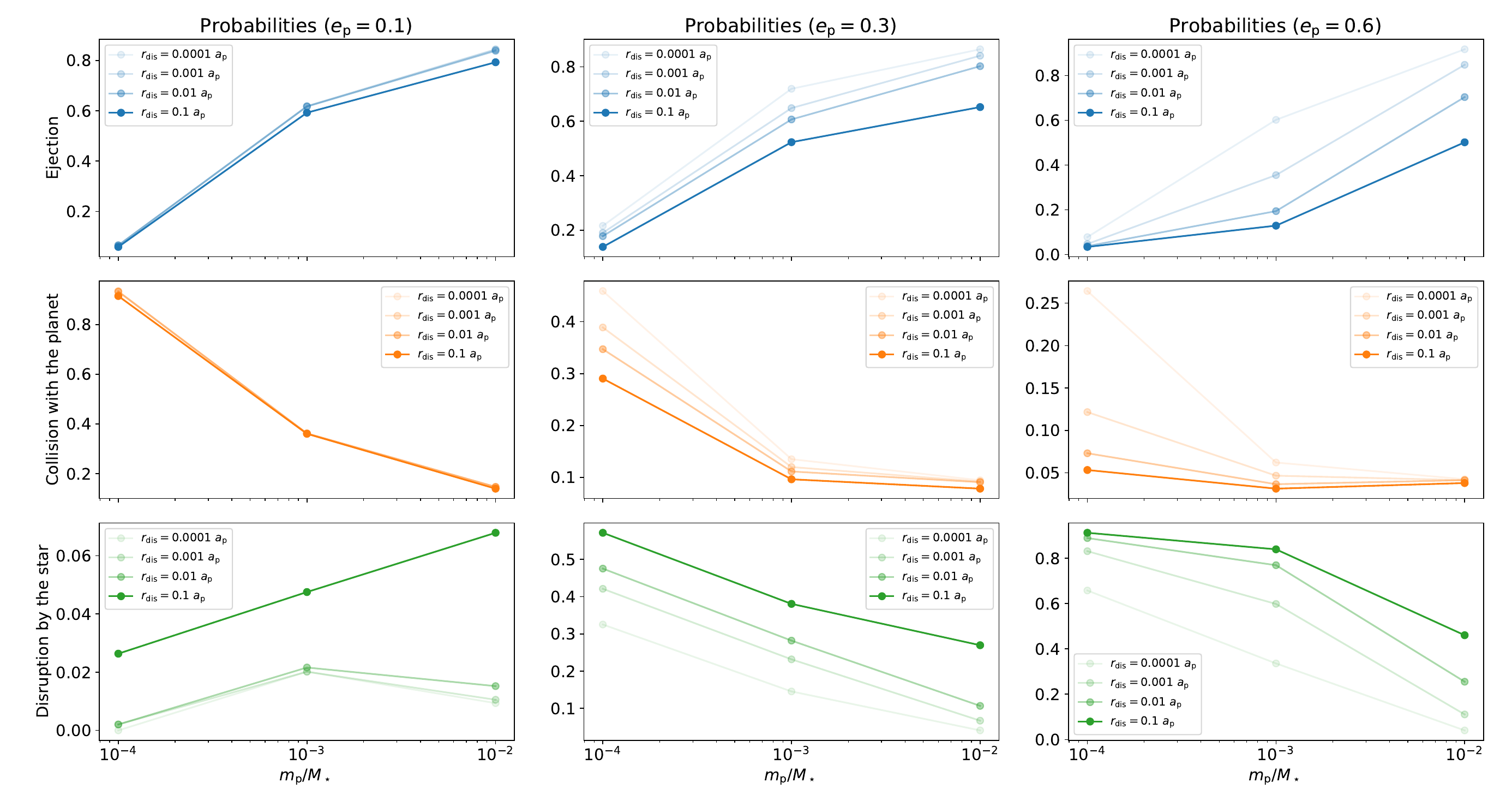}
	\includegraphics[width=\linewidth]{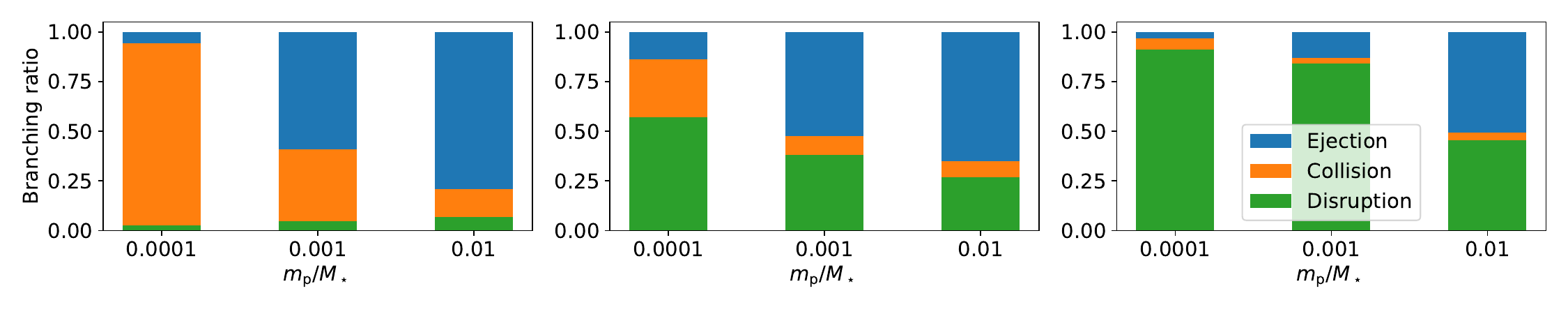}
	\caption{Probabilities for a scattered test particle to be ejected, collide with the planet or be disrupted by the star as a function of the planet to star mass ratio $m_{\rm p}/M_\star$ and stellar disruption radius to semi-major axis ratio $r_{\rm dis}/a_{\rm p}$. These results are computed from the $N$-body simulations of 1000 particles (see details in Section~\ref{sec:fate}), assuming the planet radius $R_{\rm p} = 5\times 10^{-4} a_{\rm p}$ and three different eccentricities (left column $e_{\rm p} = 0.1$, middle column $e_{\rm p} = 0.3$ and right column $e_{\rm p} = 0.6$). The branching ratios in the bottom row correspond to  $r_{\rm dis}/a_{\rm p} = 0.1$. 
} \label{fig:proba_vs_mp}
\end{figure*}

The probabilities of different outcomes also depend on the planet-to-star mass ratio. Figure~\ref{fig:proba_vs_mp} explores this dependence for giant planets, with different values of the eccentricity and stellar disruption radius. We fix the planet radius $R_{\rm p}$ to semi-major axis $a_{\rm p}$ ratio at $5\times 10^{-4}$, corresponding to $1~R_{\rm J}$ at $1$ au. When the mass ratio $m_{\rm p}/M_\star$ increases, the ejection outcome becomes more likely, while the probabilities of planet collision and stellar disruption decrease. When the stellar disruption radius increases, less particles end up being disrupted by the star, and the particles that were previously tagged as `disruption' are then ejected from the system or collide with the planet.  

\begin{figure*}
	\centering
	\includegraphics[width=\linewidth]{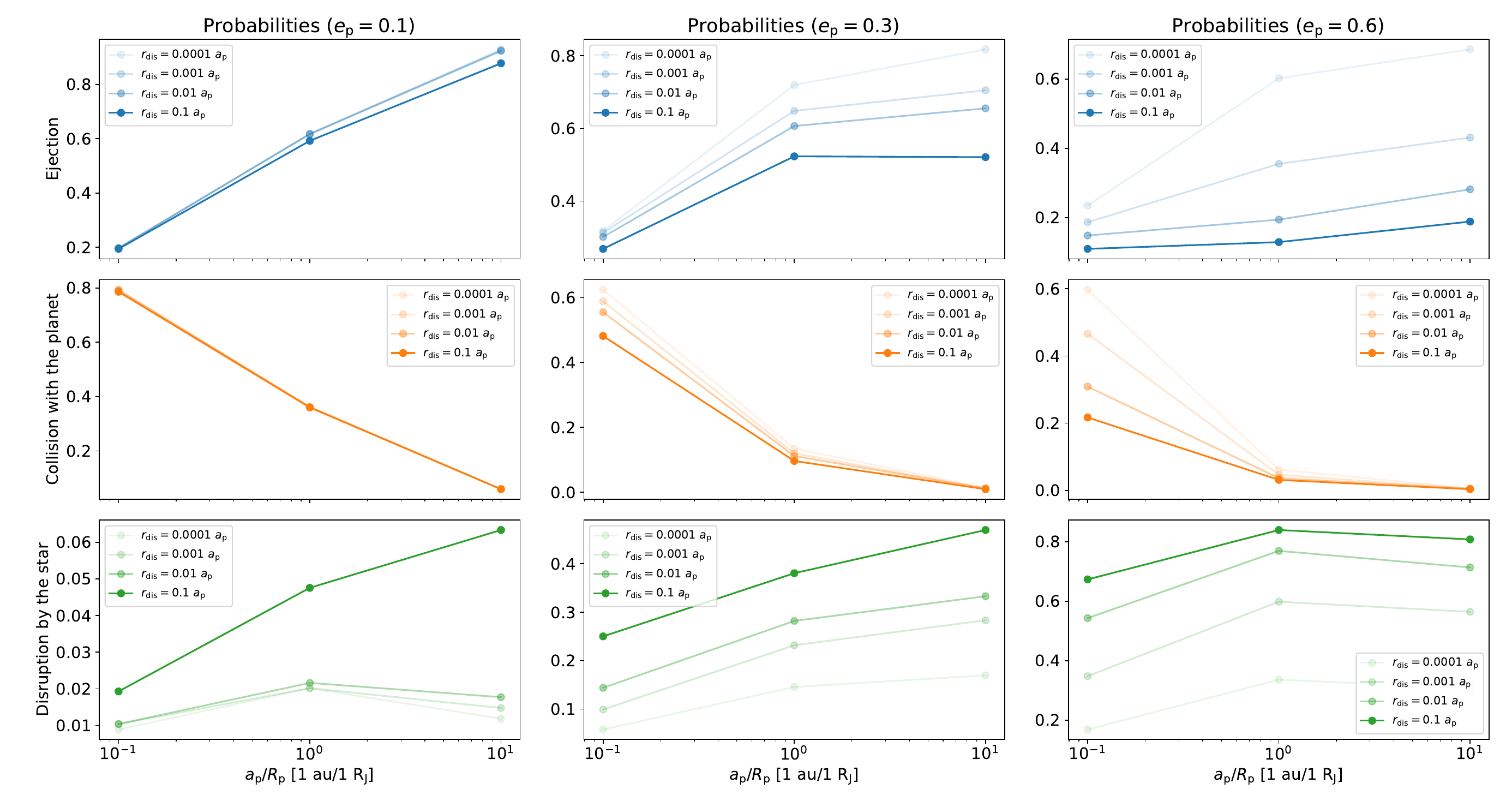}
	\includegraphics[width=\linewidth]{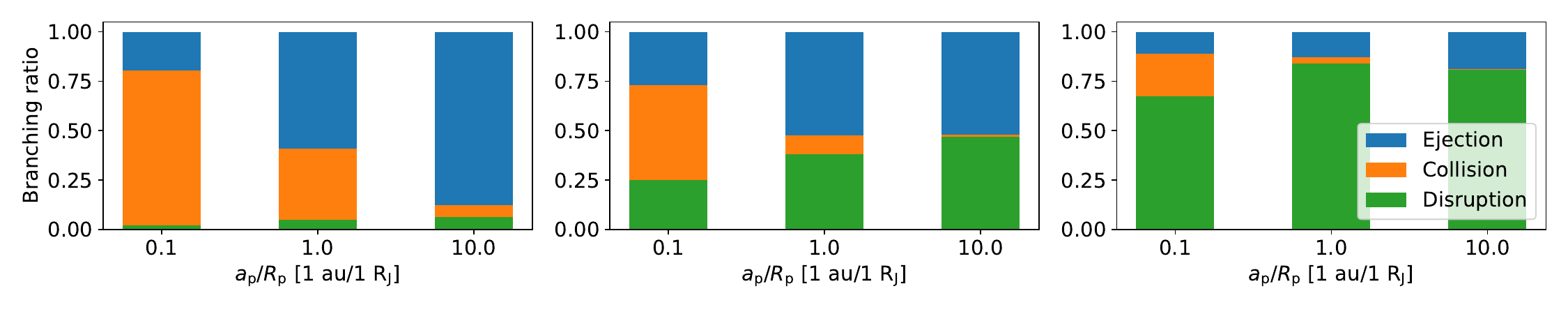}
	\caption{Same as Figure~\ref{fig:proba_vs_mp}, for a fixed planet to star mass ratio of $m_{\rm p}/M_\star = 10^{-3}$, varying the ratio between the perturbing planet's semi-major axis and radius, $a_{\rm p}/R_{\rm p}$.} \label{fig:proba_vs_ap}
\end{figure*}

Finally, Figure~\ref{fig:proba_vs_ap} explores the dependence on the semi-major axis of the planet, or equivalently on the planet radius to semi-major axis ratio, for a giant planet with fixed mass ratio ($m_{\rm p}/M_\star = 10^{-3}$) and different values of the planet eccentricity and stellar disruption radius. As expected, the probability of ejection increases with the semi-major axis and the probability of collision increases with the planet radius. Moreover, for fixed $r_{\rm dis}/a_{\rm p}$, the probability of disruption by the star hardly depends on $a_{\rm p}/R_{\rm p}$. This can be understood by looking at the distribution of minimum distances in Figure~\ref{fig:dmin_log}.

Previous works have highlighted the importance of the Safronov number in determining the outcomes of planetary scatterings \citep{safronovEvolutionProtoplanetaryCloud1972,fordDynamicalInstabilitiesExtrasolar2001,andersonInSituScatteringWarm2020,liGiantPlanetScatterings2021}. This number is defined as:
\begin{equation}  
	S = 2 \frac{m_{\rm p}}{M_\star} \frac{a_{\rm p}}{R_p}. \label{eq:safronov}
\end{equation}
The ejection-to-collision ratio is proportional to the Safronov number when there is no disruption by the star, for a fixed planet eccentricity (see Figure~\ref{fig:safronov}). This dependence becomes more complex when accounting for stellar disruption, especially for large planetary eccentricities, where most of the particles are disrupted before having the chance of ejection or collision with the planet.

\subsection{Timescales and Rates}
\label{sec:timescales}

\begin{figure}
	\centering
	\includegraphics[width=\linewidth]{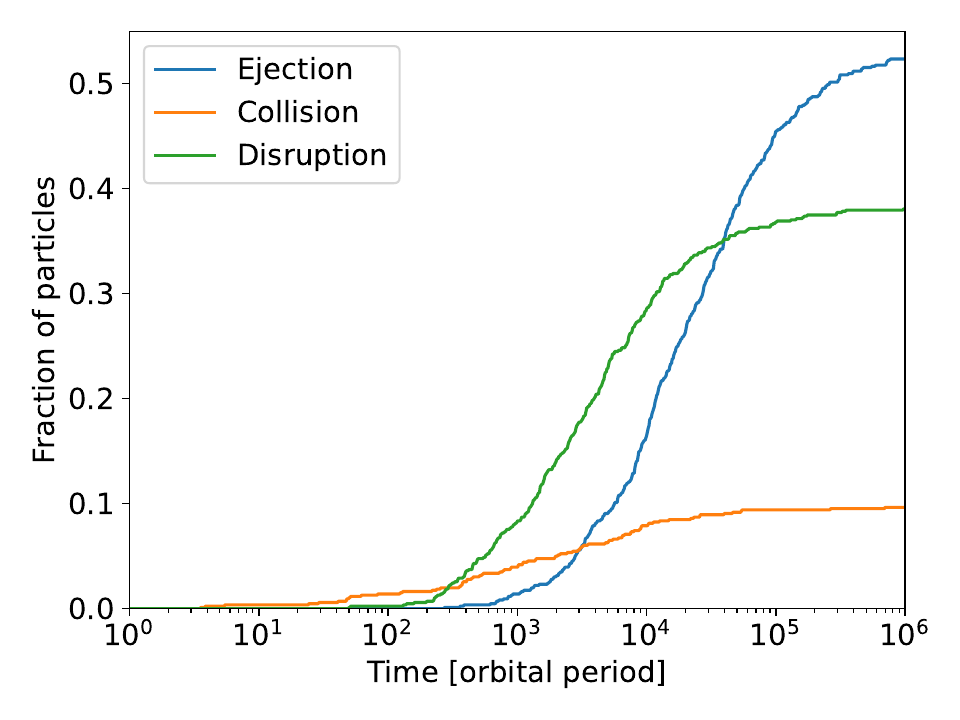}
	\includegraphics[width=\linewidth]{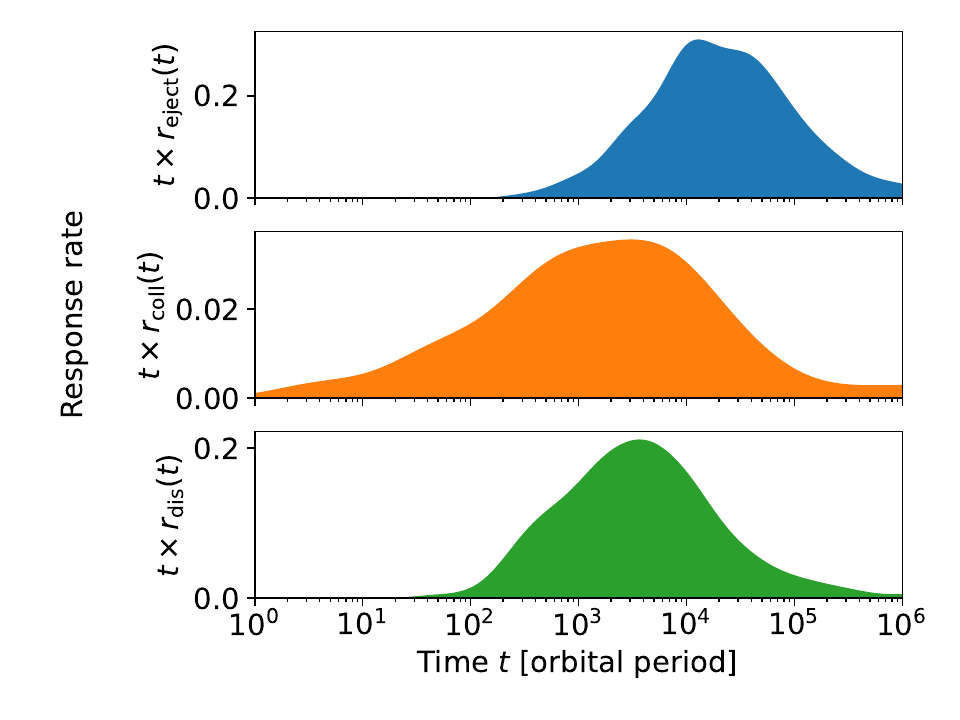}
	\caption{Fractions (top) and rates (bottom) of ejection, collision with the planet and disruption by the star within a set of $1000$ test particles scattered by a planet perturber, computed  from $N$-body simulations (see details in Section~\ref{sec:fate}). The time unit is the planet orbital period. The planet has an eccentricity $e_{\rm p} = 0.3$, mass ratio $m_{\rm p}/M_\star = 10^{-3}$, radius $R_{\rm p} = 5\times 10^{-4} a_{\rm p}$, and the criterion for disruption by the star is set to $r_{\rm dis} = 0.1~a_{\rm p}$. The rates are defined per unstable particle and are adapted to the logarithmic scale (see equations~\ref{eq:r} and \ref{eq:rlog}).} \label{fig:f-r}
\end{figure}

Now that we have explored the probabilities of the different outcomes, let us consider their typical timescales (in unit of the planet orbital period). Figure~\ref{fig:timescale_vs_ci} suggests that ejection takes longer on average than disruption by the star or collision with the planet. We plot the fraction of particles that have experienced one of these three outcomes as a function of time in Figure~\ref{fig:f-r} for a moderately eccentric giant planet. Each probability evolves differently with time, with a peak of efficiency between $10^3$ and $10^5$ orbital periods.

To quantify more accurately the timescale profiles, we define the rate of each outcome in the following way. Assuming a one-time injection of unstable test particles at $t=0$ of the form $\mathcal{\dot N}_i(t) = \mathcal{N}_{i,0}~\delta(t)$ (such as in our simulations), the rates are defined as:
\begin{equation}
	r_{\rm eject}(t) \equiv \frac{\dot{\mathcal{N}}_{\rm eject}(t)}{\mathcal{N}_{i,0}}, \quad r_{\rm coll}(t) \equiv \frac{\dot{\mathcal{N}}_{\rm coll}(t)}{\mathcal{N}_{i,0}},\quad
	r_{\rm dis}(t) \equiv \frac{\dot{\mathcal{N}}_{\rm dis}(t)}{\mathcal{N}_{i,0}}, \label{eq:r}
\end{equation}
where $\mathcal{N}_{\rm eject}(t)$, $\mathcal{N}_{\rm coll}(t)$ and $\mathcal{N}_{\rm dis}(t)$ are the numbers of particles that are ejected, collide with the planet or are disrupted by the star, respectively. To compute the rate numerically, we perform a kernel density estimate (using the \texttt{PyQt} Python module) of the distribution of the instability times, and renormalize the product by the probability of each outcome among the unstable particles, so that:
\begin{equation}
	\int_0^{+\infty} \left[r_{\rm eject}(t) + r_{\rm coll}(t) + r_{\rm dis}(t)\right] \dd{t} = 1.
\end{equation}
The time-dependent fractions of particles and corresponding rates are plotted in Figure~\ref{fig:f-r}. For better readability, we plot them with respect to the logarithm of the instability times:
\begin{equation}
	\frac{1}{\mathcal{N}_{i,0}} \dv{\mathcal{N}_j}{\ln(t)} = r_{\rm j}(t) \times t,\label{eq:rlog}
\end{equation}
where $j=$"eject", "coll" and "dis".
As can be seen in Figure~\ref{fig:f-r}, collisions with the planet tend to happen before the other outcomes, but peak at around the same time as the ejections or disruptions by the star (between $10^3$ and $10^4$ orbits) for our fiducial case. Most of the events happen before $10^6$ orbital periods. All outcome timescales span several orders of magnitude.

\begin{figure*}
	\centering
	\includegraphics[width=\linewidth]{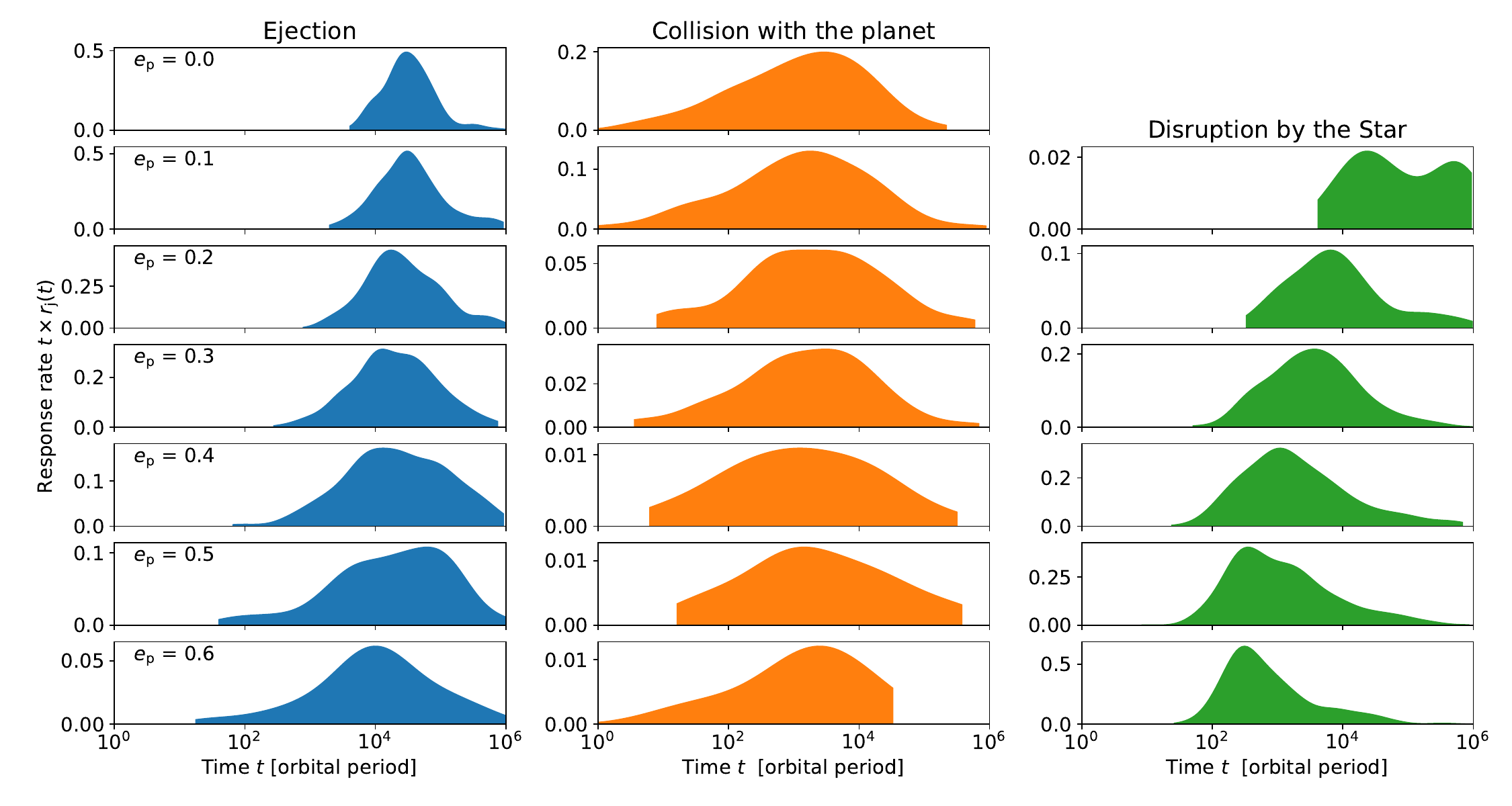}
	\caption{Same as the bottom panel of Figure~\ref{fig:f-r}, for different planet eccentricities (from $e_{\rm p} = 0$ to $0.6$; each row corresponds to a different eccentricity).}\label{fig:timescales}
\end{figure*}

Of course, the rates depend strongly on the property of the planet perturber. in Figure~\ref{fig:timescales}, we show their dependence on the planet eccentricity. The peak times of the response rates for ejection and planet collision do not depend sensitively on $e_{\rm p}$, but both processes begin sooner as $e_{\rm p}$ increases. On the other hand, the particles that get disrupted by the star are removed more quickly as $e_{\rm p}$ increases. The bimodal shape of the rate of disruption by the star for $e_{\rm p} = 0.1$ is likely due to the $2:1$ mean-motion resonance being at the outskirts of the instability zone. 
Overall, a high planetary eccentricity tends to increase the likelihood of events happening close to the injection time.

\begin{figure*}
	\centering
	\includegraphics[width=\linewidth]{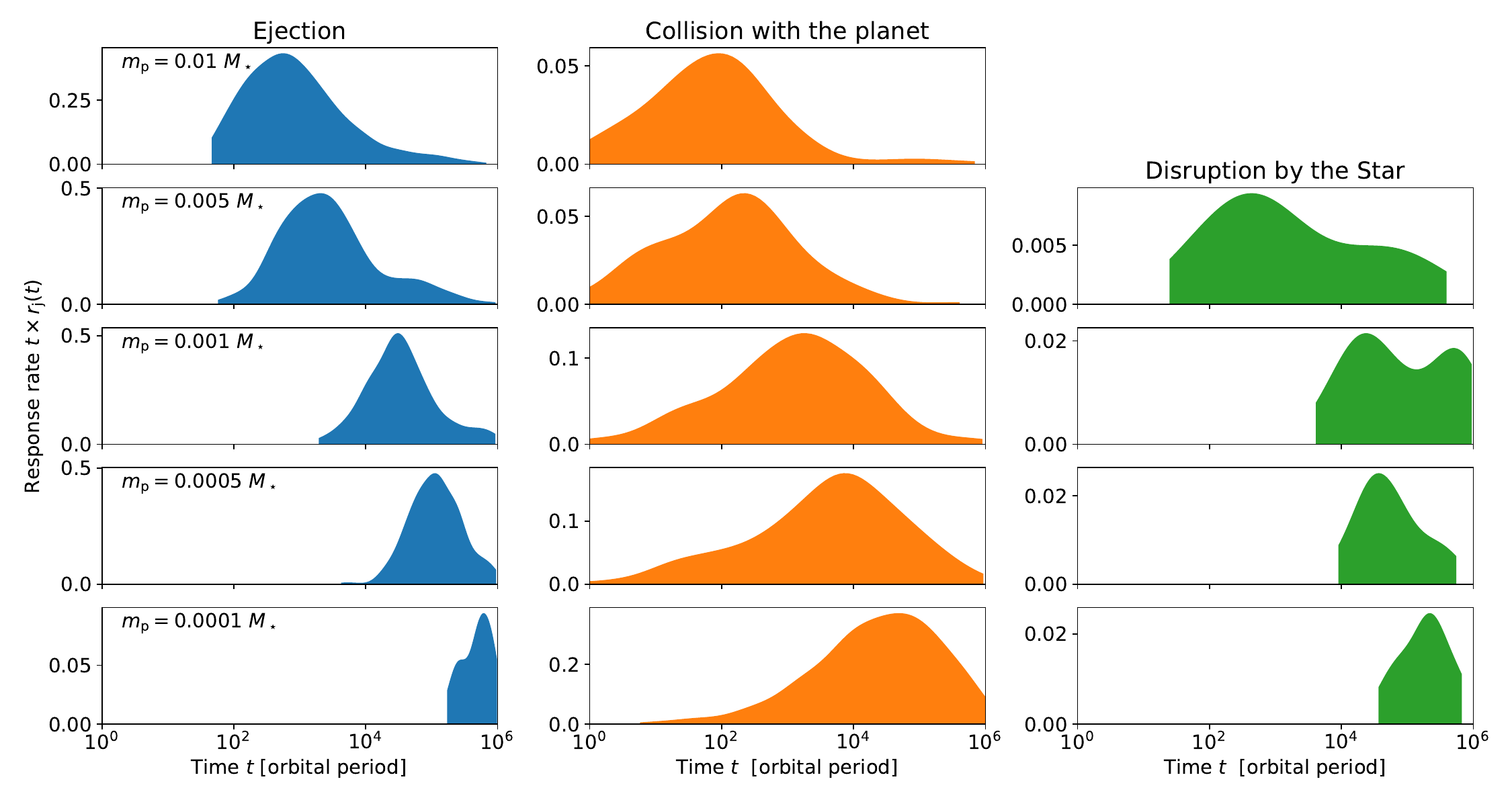}
	\caption{Same as Figure~\ref{fig:timescales} for a planet of eccentricity $e_{\rm p} = 0.1$ and different planet-to-star mass ratio $m_{\rm p}/M_\star$ between $10^{-4}$ and $10^{-2}$.}\label{fig:timescales_vs_mu}
\end{figure*}

\begin{figure*}
	\centering
	\includegraphics[width=\linewidth]{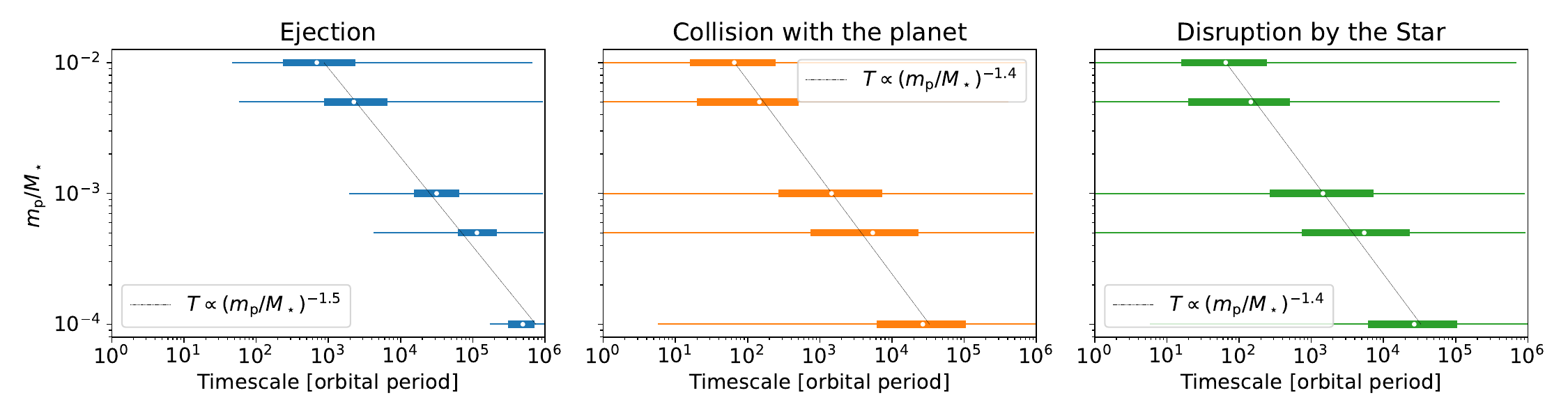}
	\caption{Timescales of ejection, collision with the planet and disruption by the star computed from the $N$-body simulations of 1000 particles (see details in Section~\ref{sec:fate}). The planet has a radius $R_{\rm p} = 5\times 10^{-4} a_{\rm p}$, eccentricity $e_{\rm p} = 0.1$ and varying mass ratio $m_{\rm p}/M_\star$  between $10^{-4}$ and $10^{-2}$. The  disruption radius is set to $r_{\rm dis} = 0.1~a_{\rm p}$. The horizontal lines span the entire range of timescales observed in the simulations for each mass ratio, and the thick lines highlight 50\% of the time distribution, between the 25\% and 75\% quartiles. The thin curves are fits to the medians (white dots) assuming timescales proportional to $(m_{\rm p}/M_\star)^\alpha$ (we assume an individual $\alpha$ for each outcome), following the scaling derived empirically by \citet{puStrongScatteringsCold2021}.}\label{fig:timescales_vs_mu_fit}
\end{figure*}

In Figures~\ref{fig:timescales_vs_mu} and \ref{fig:timescales_vs_mu_fit}, we vary the planet-to-star mass ratio for $e_{\rm p} = 0.1$ and find that all timescales are approximately proportional to $(m_{\rm p} / M_\star)^{-1.5}$. This agrees with the empirical finding of \cite{puStrongScatteringsCold2021} (in the context of planet-planet scatterings) or \cite{verasSmallestPlanetaryDrivers2023} (in the context of white-dwarf pollution), and is  consistent with the rough analytical estimate of the energy diffusion timescale [$t \propto (m_{\rm p} / M_\star)^{-2}$] for test particles undergoing scatterings with a giant planet 
\citep{tremaineDistributionCometsStars1993}.

\begin{figure}
	\centering
	\includegraphics[width=\linewidth]{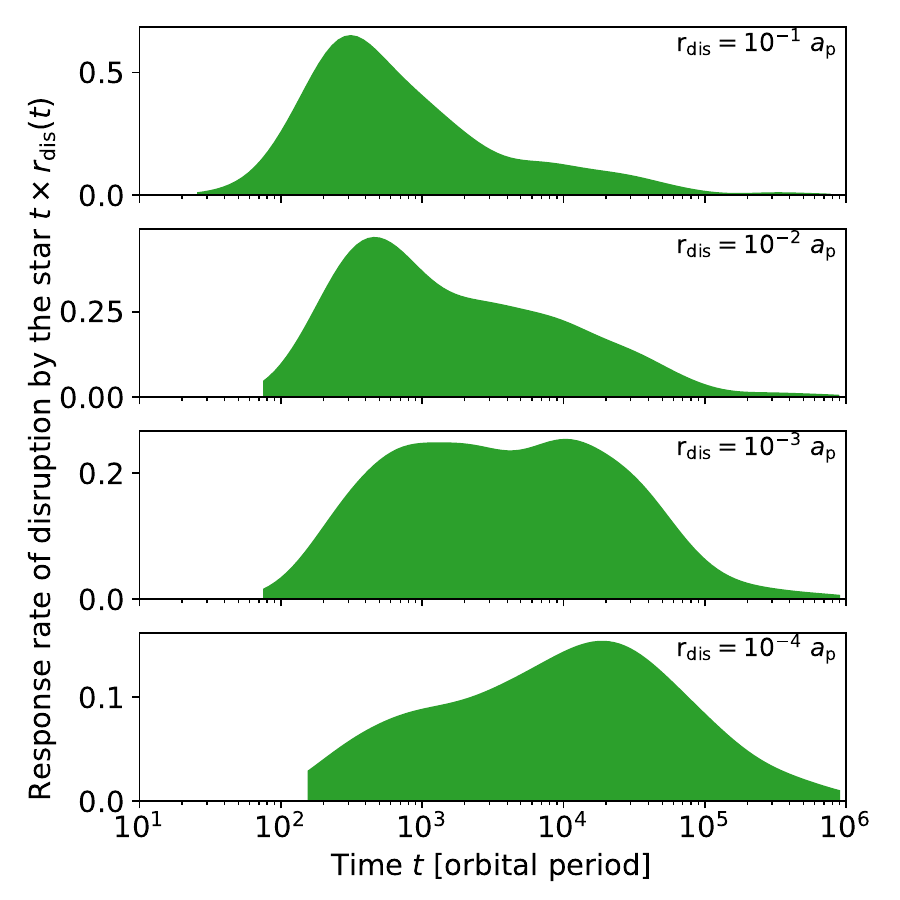}
	\caption{Same as Figure~\ref{fig:timescales}, but for $e_{\rm p} = 0.6$ and different stellar disruption radii $r_{\rm dis}$.
 }\label{fig:timescalesthreshold}
\end{figure}

In Figure~\ref{fig:timescalesthreshold}, we vary the stellar disruption radius $r_{\rm dis}$ for an eccentric planet ($e_{\rm p} = 0.6$, a case with a large fraction of stellar disruptions). As expected, the smaller $r_{\rm dis}$, the longer the average time needed for particles to get disrupted. 

\begin{figure}
	\centering
	\includegraphics[width=\linewidth]{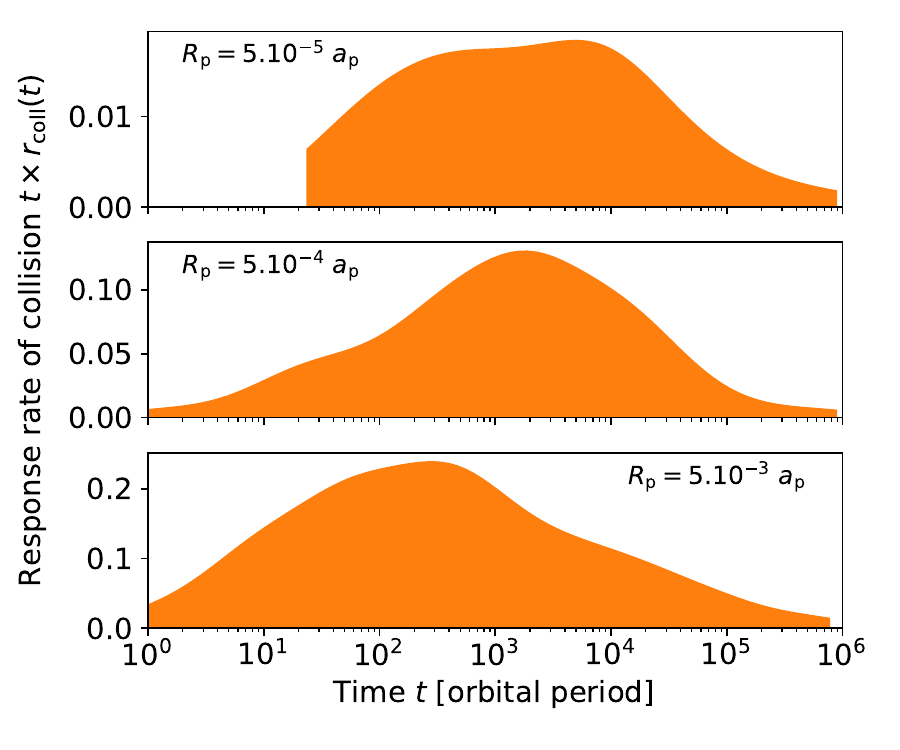}
	\caption{Same as Figure~\ref{fig:timescales}, but for $e_{\rm p} = 0.1$ and different planet radii $R_{\rm p}$. }\label{fig:timescales_vs_rp}
\end{figure}

Finally, the timescale of collision with the planet depends on the radius of the planet in unit of $a_{\rm p}$, in a similar fashion that the star disruption timescale depends on $r_{\rm dis}$. We vary $R_{\rm p}/a_{\rm p}$ by two orders of magnitude in Figure~\ref{fig:timescales_vs_rp}. As expected, the larger the planet size, the longer the average collision time. However, since the collision timescales span so many orders of magnitude, the variation in timescale is not as striking as the variation in the branching ratio when we vary $R_{\rm p}/a_{\rm p}$ (see Figure~\ref{fig:proba_vs_ap}).

\subsection{Continuous Injection of Test Particles}

So far we assumed a one-time injection of test particles in the unstable (chaotic) zone around a planet, i.e. the injection rate $\mathcal{\dot N}_i(t) = \mathcal{N}_{i,0}~\delta(t)$. In the general case where the injection of test particles is not localized in time, the rate of a process $j$ (="eject", "coll" and "dis") is given by:
\begin{equation}
	\dot{\mathcal{N}}_j(t) = \int_{-\infty}^{t} \dot{\mathcal{N}}_i(t') r_j(t-t') dt'.
\end{equation}
If we assume that $\dot{\mathcal{N}}_i$ is constant in time, we get:
\begin{equation}
	\dot{\mathcal{N}}_j = \dot{\mathcal{N}}_i \int_{0}^{\infty} r_j(t') dt'.
\end{equation}
If the process has an overall probability of $p_j$, then
\begin{equation}
	\dot{\mathcal{N}}_j = \dot{\mathcal{N}}_i~p_j.
\end{equation}
We can imagine that such a constant injection profile may take place at the beginning of a system's life, when the planets have not yet cleared the debris around their orbits, or else when the star enters the giant branch, through the radiative breakup of asteroids \citep[e.g.,][]{verasPostmainsequenceDebrisRotationinduced2014,verasPostmainsequenceDebrisRotationinduced2020}. However, it is unlikely that it will be sustained at a significant rate for a long period of time, given the limited mass reservoir of debris discs.

\begin{figure}
	\centering
	\includegraphics[width=\linewidth]{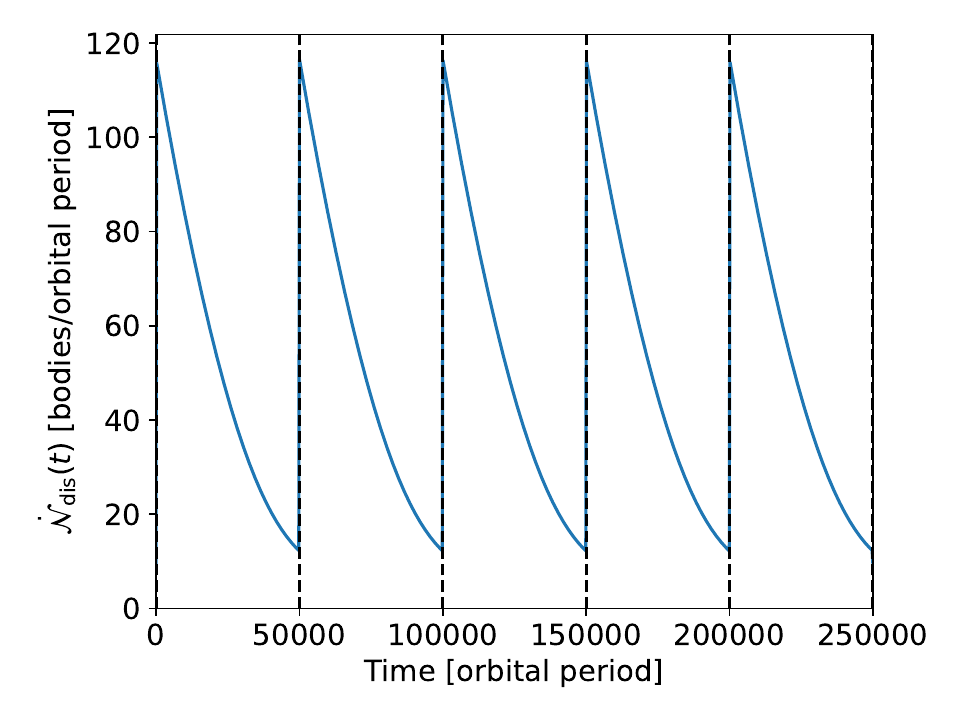}
	\caption{Frequency of particles disrupted by the star $\dot{\mathcal{N}}_s$ assuming the periodic injection (equation~\ref{eq:spiked}) of $10^7$ planetesimals every 50,000 orbital periods close to a mildly eccentric planet ($e_{\rm p} = 0.3$), with mass ratio $10^{-3}$ and radius $R_{\rm p} = 5.10^{-4} a_{\rm p}$. The criterion for disruption by the star is $r_{\rm dis} = 0.1~a_{\rm p}$. We used the rate $r_{\rm s}$ computed from our simulations (see Section~\ref{sec:fate}) and represented in Figure~\ref{fig:timescales}.} \label{fig:spiked}
\end{figure}

At later time ($\gtrsim 10$ Myr), instead, the injection may consist of a collection of transient influx of material, that is
\begin{equation}
	\dot{\mathcal{N}}_i(t) = \sum_k \delta(t-t_k) \mathcal{N}_{i,k}.
\end{equation}
The rates of the different processes are then given by
\begin{equation}
	\dot{\mathcal{N}}_j(t) = \sum_{k\text{ such that }t_k < t} \mathcal{N}_{i,k} ~ r_j(t-t_k).\label{eq:spiked}
\end{equation}
This equation gives the expected number of particle event rate  using our previously determined results for $r_j(t)$.
An example of such rate profile is shown in Figure~\ref{fig:spiked}, and could account for the observation of high comet activity that is unlikely to be sustained for a long period of time. This injection pattern could be induced by occasional collisions between neighbouring minor bodies, leading to the initiation of a massive collisional cascade and a temporary filling of the instability (chaotic) zone around the planet.


\subsection{Comparison with previous works}

\cite{frewenEccentricPlanetsStellar2014} previously studied the outcomes and timescales of planet-planetesimal scatterings using $N$-body simulations. They varied the planet-to-star mass ratio and planet eccentricity, but assumed a fixed semi-major axis ($a_{\rm p} = 4$ au), planet radius ($0.6~\mathrm{R_J} \simeq 7\times 10^{-5}~a_{\rm p}$) and stellar disruption radius ($r_{\rm dis} = 5\times 10^{-3}$ au $\simeq 10^{-3}~a_{\rm p}$).
Their results agree qualitatively with ours: As the planet eccentricity increases, the probability of collision with the planet decreases while the probability of disruption by the star increases (see their Fig. 6).  On the other hand, as the planet-to-star mass ratio increases, the probability of stellar disruption  decreases while the probability of ejection increases (see their Fig. 4). The timescales increase as the mass ratio decreases (see their Fig. 5).

Quantitative comparisons can be done for a planet with mass ratio $m_{\rm p}/M_\star = 10^{-3}$ using the results of our Figure~\ref{fig:proba_vs_ap}. Frewen \& Hansen's planet semi-major axis and radius values correspond to $a_{\rm p}/R_{\rm p}\simeq 7~\mathrm{au/R_J}$. Assuming the same stellar disruption radius of $\simeq 10^{-3}~a_{\rm p}$, we predict 90\% ejections and 10\% collisions with the planet for $e_{\rm p} = 0.1$, and 40\% ejections and 60\% disruptions by the star for $e_{\rm p} = 0.6$. Their Fig. 6 shows very similar values. Our paper can thus reproduce the previously published results and allow us to extrapolate to other values of planet and star 
radii.

\section{Planetesimal scatterings by two planets}
\label{sec:2planet}

\begin{figure*}
	\centering
	\includegraphics[width=\linewidth]{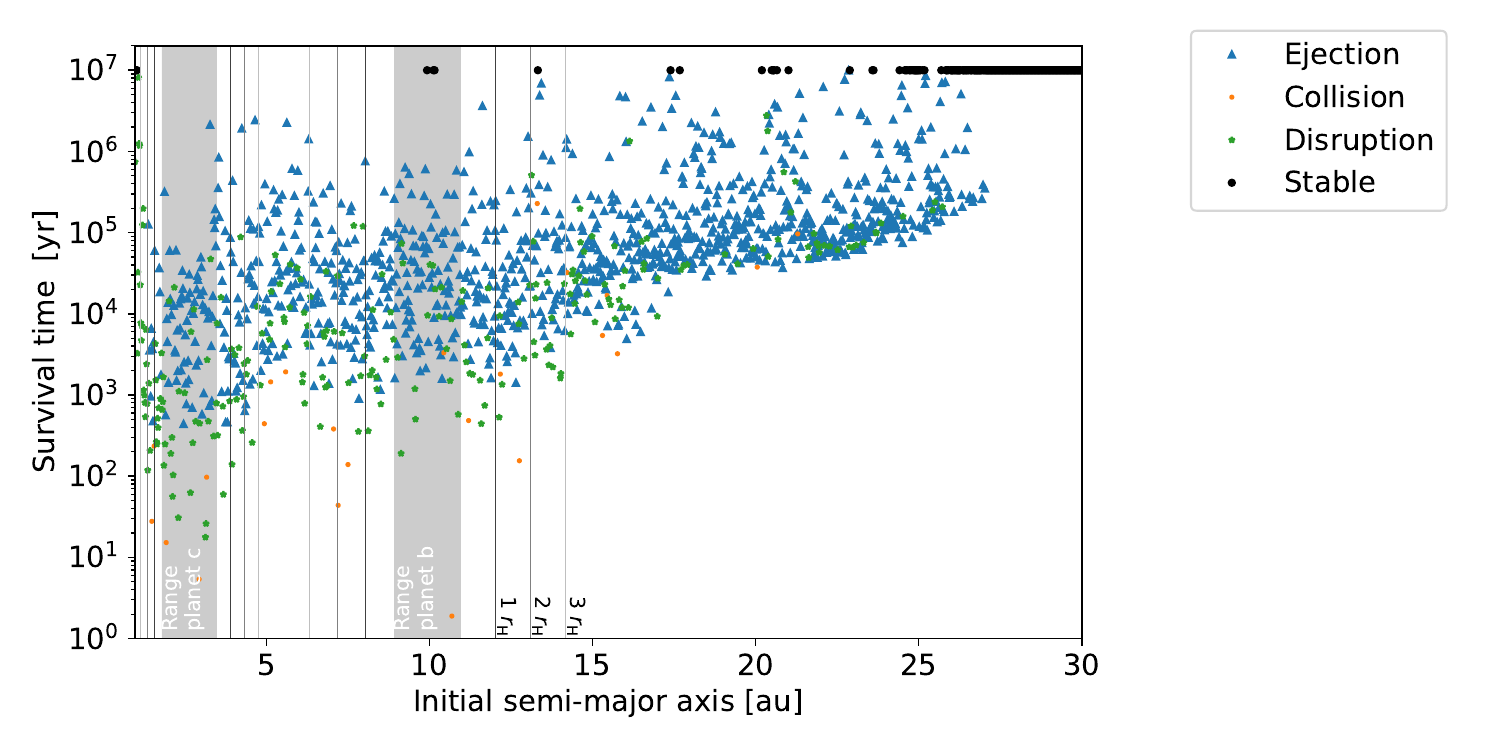}
	\caption{Survival time with respect to initial semi major axis, for 1641 test particles populating the Beta Pictoris two-planet system between $1$ and $30$ au. The orbital elements of the two planets are given in Section~\ref{sec:2planet}. The shape and colour coding of the dots correspond to the fate of the particles within the $10^7$ yr of evolution. The stellar disruption radius is set to $r_{\rm dis} = 0.3$ au. The grey area highlights the particle's semi-major axes that intersect with either planet's orbit. The thin vertical lines correspond to the one, two and three Hill radii distances from each planet's  orbit.}\label{fig:timescale_vs_ci_betapic}
\end{figure*}

\begin{figure*}
	\centering
	\includegraphics[width=\linewidth]{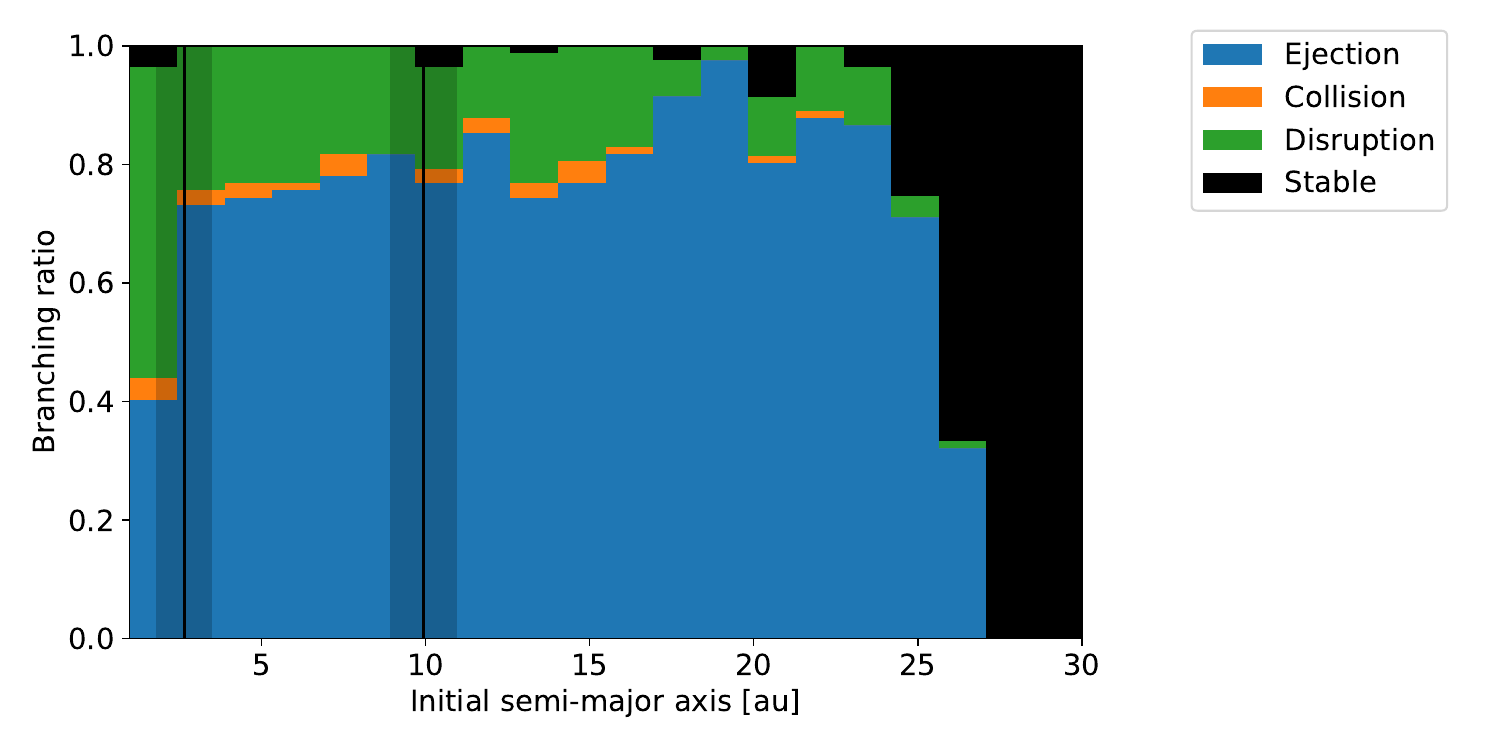}
	\caption{Same as Figure~\ref{fig:timescale_vs_ci_betapic}, depicting the branching ratios of the planetesimal population for the different bins of initial semi-major axes (each bin gathers about $82$ planetesimals).}\label{fig:branchingratio_vs_ci_betapic}
\end{figure*}

In the last two sections, we have thoroughly studied the outcomes of 
planetesimal scatterings by a single planet for a wide range of parameters.
\citet{bonsorScatteringSmallBodies2012} argued that the probability of 
inward scatterings of planetesimals can be  
significantly enhanced by the presence of multiple planets. 
Exploring the entire range of parameters characterizing the scatterings with more than one planet is beyond the scope of this paper. Here we carry out a limited study (using the parameters similar to those of the $\beta$ pictoris system), with the aim of understanding qualitatively how the results of the last section change when there are two planets.

In our simulations, we add a second planet to the setup introduced in Section~\ref{sec:fate}. We model the two planets similar to the two giant planets found in the system $\beta$ Pictoris (see Section~\ref{sec:betapic})
as constrained in \cite{lacourMassPictorisPictoris2021} (Table 2, two-planet model). The inner planet $\beta$ Pic c (hereby ``planet 1'') has a semi-major axis of $a_{\rm c} = 2.64$ au, an eccentricity $e_{\rm c} = 0.32$ and a mass ratio $m_{\rm c}/M_\star = 6 \times 10^{-3}$. The stellar mass is $1.7~M_\odot$ and the disruption radius is set to $r_{\rm dis} = 0.1~a_{\rm c} = 0.3$ au. The outer planet $\beta$ Pic b (``planet 2'') has a semi-major axis of $a_{\rm b} = 9.94$ au, an eccentricity $e_{\rm b} = 0.103$ and a mass ratio $m_{\rm b}/M_\star = 8.3\times 10^{-3}$. We adopt a radius of $R = 1.3~\mathrm{R_J}$ for both planets \citep{nowakDirectConfirmationRadialvelocity2020}, and simulate the $10^7$ yr evolution of test particles with semi-major axis initially ranging from $1$ to $30$ au. Figure~\ref{fig:timescale_vs_ci_betapic} shows their fate and survival time as a function of their initial conditions. Figure~\ref{fig:branchingratio_vs_ci_betapic} shows the associated branching ratios.

Before analysing the results on the planetesimals, we note that while the planetary configuration in $\beta$-pic is 
dynamically stable \citep{lacourMassPictorisPictoris2021}, 
the two planets are undergoing significant periodic variations in eccentricities on secular timescales. The amplitudes of the variations are of order $0.05$, and the secular period is 
about $65,000$ yr \citep{murraySolarSystemDynamics2000}. Thus, our integration time of $10^7$ yr samples around $150$ secular periods, so that the final results take into account the secular effects.

The addition of a second planet has the effect of significantly enlarging the instability zone for the planetesimals. The overlap between the influences of the two planets gives every unstable planetesimal a non-negligible likelihood to reach the host star, even those originally on a wide orbit. According to Figure~\ref{fig:proba_vs_mp}, the percentage of planetesimals expected to reach the star assuming the system comprises only planet 2 is around 3\%, and around 30\% if it comprises only planet 1. The combination of both planets pushes an average of $\simeq 20\%$ of the unstable planetesimals towards the star from much farther away than the chaotic zone of planet 1 alone. As a rough estimate, we can thus consider that the inner planet controls the percentage of planetesimals being disrupted by the star for the entire range of initial conditions when 
using the single-planet results derived in the previous sections.

We note that the planetesimals have a low probability of colliding with either planets regardless of their initial conditions. From Figure~\ref{fig:proba_vs_mp}, we would indeed expect a low collision probability ($< 10\%$) with planet 1, given its large ($\simeq 0.3$) eccentricity. However, collisions with planet 2 would have been much more frequent ($\simeq 20\%$) if planet 1 did not exist. It appears that, thanks to its eccentricity and despite its lower mass, planet 1 changes the dynamics of planetesimal collision with its neighbour. 

As for the timescales, we note that the orbital periods of the two planets are $P_1 = 3.3$ and $P_2 = 24$ yr. While in the single-planet case, the plenetesimal instability/survival timescale does not depend strongly on the initial semi-major axis, there is a clear trend of increasing instability time with the initial semi-major axis in the two-planet case. In the neighbourhood of planet 1, the time required for planetesimal ejection is around $10^3-10^4~P_1$ and it is $10^2-10^4~P_1$ for disruption by the star. This is about one order of magnitude larger than the predictions from the single-planet results depicted in Figure~\ref{fig:timescales_vs_mu}. On the other hand, around planet 2, the time required for ejection is around $10^2-10^4~P_2$ and it is $10^2-10^3~P_2$ for disruption by the star, which better matches the predictions of Figure~\ref{fig:timescales_vs_mu}. In the outer regions ($a>20$ au), the time required for ejection and disruption is around $10^3-10^5~P_2$, larger than predicted for each planet of such masses. Overall, the presence of two planets seems to increase the instability time, especially in the outer region far from the chaotic zone of individual planet.

\section{Applications}
\label{sec:discussion}

In the previous sections, we have presented general results on the outcomes of planetesimal scatterings by planets.  In this section, we apply these results to three topics of current interest.

\subsection{Infalling planetesimals in $\beta$ Pictoris}
\label{sec:betapic} 

$\beta$ Pictoris is a unique system, comprising a resolved debris disc and a pair of exoplanets. The presence of planetary-mass object(s) was first inferred in 1990s following the detection of numerous falling evaporating bodies (FEBs) onto the host star  \citep{beustBetaPectorisProtoplanetary1991,beustMeanMotionResonancesSource1996,beustFallingEvaporatingBodies2000}, a decade before both planets were discovered \citep{lagrangeGiantPlanetImaged2010,lagrangeEvidenceAdditionalPlanet2019}. This infall of evaporating bodies has continued  since then \citep{pavlenkoNewExocometsPic2022}.

The properties of the $\beta$ Pictoris system are detailed in Section~\ref{sec:2planet}, since we used it as a test case for our two-planet scattering study. The rate of observed FEBs is around $\dot{\mathcal{N}}_s \simeq 10^3$ yr$^{-1}$ \citep{beustMeanMotionResonancesSource1996} for km-sized planetesimals falling at a distance of $\lesssim 0.27$ au $\simeq 30~R_\star$ to the star \citep{beustFallingEvaporatingBodies2000}. Since the inner planet $\beta$ Pic c has a semi-major axis of $2.7$ au, this stellar disruption distance corresponds to $0.1\, a_{\rm p}$, which is similar to the fiducial value used in this paper. The simulations we performed in Section~\ref{sec:2planet} give an average stellar disruption probability of $p_s \simeq 0.2$ among unstable planetesimals (Figure~\ref{fig:branchingratio_vs_ci_betapic}). It follows that the injection rate of planetesimals in the planet's unstable surrounding is
\begin{equation}
	\dot{\mathcal{N}}_i = \frac{\dot{\mathcal{N}}_s}{p_s} \simeq 5\times 10^3~\mathrm{yr}^{-1}.
\end{equation}
This injection could be sustained either periodically (see Figure~\ref{fig:spiked}) 
or incidentally through collisions between planetary bodies and/or migration of planetesimals.
We expect the planetesimal reservoir to be at $a_{\rm r} \gtrsim 25$ au (see Figure~\ref{fig:timescale_vs_ci_betapic}). The observed FEBs are thought to have a typical size of $\sim 30$ km \citep{beustFallingEvaporatingBodies2000}. Assuming the planetesimals have a density of $\sim 3$ g~cm$^{-3}$, then the injection rate corresponds to
a mass flux $\dot{M}_i$ of
order $4\times 10^{-5}~\mathrm{M}_{\earth}$ per year. 

Our simulations show that the typical FEB production timescale through scatterings is $5\times 10^5$ yr (see Figure~\ref{fig:timescale_vs_ci_betapic}). This timescale was correctly estimated early on, in the first study to investigate the planet scattering hypothesis for the system \citep{beustBetaPectorisProtoplanetary1991}. Although this timescale is much smaller than the system's age ($25$ Myr), the probability to witness a unique event of that duration is not negligible ($\simeq 2\%$), so that we cannot rule out the incidental origin. To account for the 
observed FEB rate of $10^3$ per year for $5\times 10^5$ yr,  
the number of planetesimals in the planets' surrounding unstable region 
is order $2.5\times 10^9$; this corresponds to 20 $M_{\earth}$, which is a significant fraction of the estimated mass of the entire debris disc in 
$\beta$ pic \citep[$140~M_{\earth}$, ][]{artymowiczBetaPictorisEarly1997}. 

Direct scatterings by planets is not the only way to produce FEBs. Another process that is usually invoked in this context is the extreme eccentricity excitation of planetesimals in mean-motion resonance (MMR) with an eccentric planet \citep{beustMeanMotionResonancesSource1996}. Most of the planetesimals trapped in the MMR will remain stable for an extended period of time, so that the replenishment rate can be reduced compared to the planetary scattering scenario. The MMR mechanism has the advantage of pontentially explaining the observed asymmetry in the influx of FEBs \citep[bias towards redshift,][]{chauvinOrbitalCharacterizationBeta2012}. However, the reservoir of planetesimals should be located close to the MMR in the inner part of the $\beta$ pic system, where the debris disc is strongly depleted \citep{thebaultDustProductionCollisions2003}. Moreover, the MMR scenario was developed before the discovery of two planets in the system. Although its predictions matched well with the properties of $\beta$ Pic b \citep{chauvinOrbitalCharacterizationBeta2012}, it must now be revised following the recent detection of $\beta$ Pic c  \citep{beustDynamicsBetaPictoris2023}.  

\subsection{White Dwarf pollution}
\label{sec:whitedwarf}

Between 25-50\% of white dwarfs (WDs) exhibit spectra ``polluted'' by heavy elements, indicating recent or ongoing accretion of 
rocky material onto the WD from a circumstellar debris disc \citep[e.g.,][]{zuckermanMetalLinesWhite2003,zuckermanAncientPlanetarySystems2010, barstowEvidenceExternalOrigin2014,koesterFrequencyPlanetaryDebris2014}. Such an influx of material is expected at the beginning of the WD phase, as the planetary system was just destabilized by the post-main sequence stellar radius inflation and mass loss. However, debris discs are found
in relatively cool (thus older, Gyrs old) WDs \citep[e.g.,][]{debesGyrWhiteDwarf2019,blouinNoEvidenceStrong2022}.
This is at odds with the understanding that the material composing debris discs have a small life expectancy \citep[typically Myrs,][]{girvenConstraintsLifetimesDisks2012,verasLifetimesPlanetaryDebris2020}) due to Poynting-Robertson drag and other dissipative processes \citep[e.g.,][]{farihiCircumstellarDebrisPollution2016}. 
Such discs thus need to be regularly replenished by planetesimals scattered in the "disruption zone"  from the outer region. Various mechanisms have been suggested to induce this scattering, such as mean-motion resonances with the remaining planets \citep{debesLinkPlanetarySystems2012} or planet-planet scatterings \citep{bonsorDynamicalEffectsStellar2011,bonsorScatteringSmallBodies2012}. However, the uncertainty surrounding the outer architectures of white dwarfs makes it hard for numerical studies to provide general results on the evolution and dependencies of white dwarf pollution. The theoretical insights and scalings provided in this paper could help identify the regimes of interest for future in-depth studies.

The dynamical fate of planetary systems when the host star becomes a white dwarf has been extensively studied in recent years, and the topic is reviewed in \cite{verasPostmainsequencePlanetarySystem2016}. For a Jovian planet to survive the post-main sequence phase, it must be located farther than $a_{\rm p} = 3-5$ au depending on the initial mass of the star. On the other hand, for planetesimals to become part of a close-in debris disc and eventually fall into the star, their periastron must reach $q_{\rm min} \simeq R_\odot \simeq 5\times 10^{-3}$ au. To compare these constraints with our results in Section \ref{sec:fate}, we set the stellar disruption radius to $r_{\rm dis} \simeq 10^{-3}~a_{\rm p}$. According to Figure~\ref{fig:proba_vs_e}, planet scattering could produce a non-negligible fraction of disrupting planetesimals, as long as the planet eccentricity $e_{\rm p}$ is modest ($\gtrsim 0.2$). Moreover, a planet at larger distance (e.g. $50$ au) can also induce white dwarf pollution if its eccentricity is larger. However, such a process would deplete the planetesimals in the planet's surroundings in about $10^5$ orbital periods (see Figure~\ref{fig:timescalesthreshold}), that is around $10^6$ yr for $a_{\rm p} = 5$ au and $4\times 10^7$ yr for $a_{\rm p} = 50$ au. This is much shorter than the observed timescales of WD pollution \citep[e.g.,][]{koesterFrequencyPlanetaryDebris2014,hollandsCoolDZWhite2018,blouinNoEvidenceStrong2022}. The presence of several outer planets could extend the scattering timescale (see Section~\ref{sec:2planet}), or else we need to consider alternative ways to replenish the debris disc on secular timescales (e.g. secular chaos, \citealt{oconnorSecularChaosWhite2022}; radiation mechanisms, \citealt{verasOrbitDecay21002022}; bombardments, \citealt{verasGeneratingMetalpollutingDebris2020}). 

\subsection{Planet engulfment by main-sequence stars}
\label{sec:engulfment}

Recent works suggest that some fraction of solar-type stars may have engulfed a planet during their lifetimes \citep{spinaChemicalEvidencePlanetary2021}
\citep[see also][]{Behmard2023a,Behmard2023b}.
Planet engulfment happens when the separation between a planet and its star goes below the planet Roche limit, leading to its disruption and stellar pollution. It is believed that chemical inhomogeneities between the two stellar components of a binary system are best explained by planet engulfment \citep{pinsonneaultMassConvectiveZone2001}. 
The typical amount of material needed to account for the observed inhomogeneities corresponds to several Earth masses per system.

The Roche limit of an Earth-like planet around a solar-type star is less than the Sun's radius. We thus set the disruption radius to $r_{\rm dis} = 5\times 10^{-3}~a_{\rm p}$, assuming the planet is initially located at $a_{\rm p} \simeq 1$ au. 
According to our results in Section \ref{sec:qmin},
the scattering between a Jupiter-mass planet and an Earth-mass planet at $1$ au can lead to the engulfment of the Earth if its eccentricity is greater than $0.6$ ($T \lesssim 2.6$) or if the eccentricity of the giant planet is greater than $0.1$. On the other hand, the scattering between two nearly equal-mass planets could also lead to  engulfment, assuming the eccentricity of one of the body is greater than $0.6$. 

\section{Summary}
\label{sec:conclusion}

\begin{figure*}
    \centering
    \includegraphics[width=0.6\linewidth]{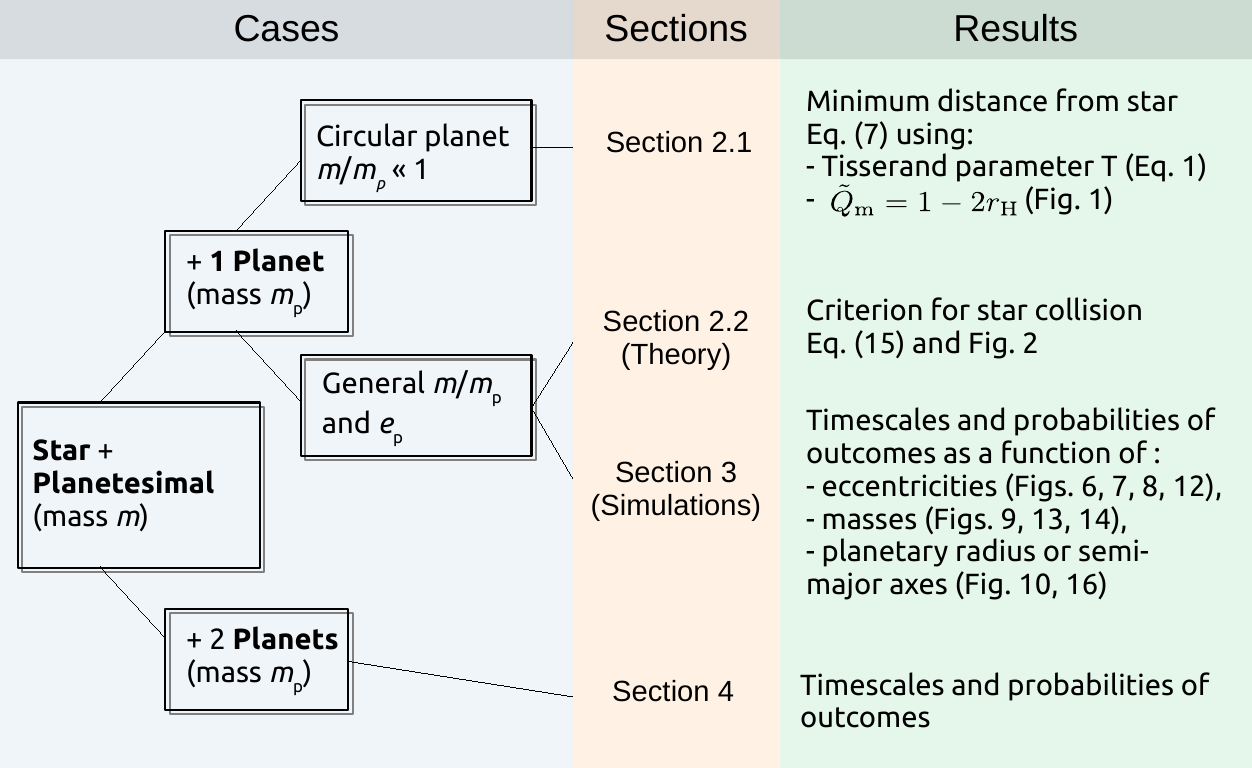}
    \caption{Summary flowchart of our findings for planet-planetesimal
    scatterings.}
    \label{fig:flowchart}
\end{figure*}

In this paper, we have investigated the gravitational scatterings between planet(s) and planetesimals and explore the possibility of disruption or consumption of these small bodies by their host stars following close encounters with the planets. While previous research has examined these scatterings in different contexts, our work aims to provide a more comprehensive analysis that covers a broad range of parameter spaces and yields relevant analytical relations that can be applied to various astrophysical or planetary situations. Figure~\ref{fig:flowchart} summarizes the relevant equations and figures for different cases.
Our key findings are as follows:


1. We first examine the theoretical constraints on the outcomes of planetesimal scatterings by a single planet, and derive a new analytical expression for the minimum stellar approach $q_{\rm min}$ of a planetesimal
initially in the chaotic zone around the planet's 
orbit (Section \ref{sec:qmin}). Our analysis goes beyond previous works by considering general planet masses and eccentricities, as well as finite planetesimal-to-planet mass ratio.
Our analytical results highlight the importance of the planet's initial eccentricity and of the planet-to-star mass ratio. 
In particular, we demonstrate that there is no theoretical barrier that prevents a test particle from plunging into the star after being scattered by a planet, as long as the planet has a finite eccentricity.


2. We carry out extensive simulations of planetesimal (treated as a test mass) scatterings by a planet (Section \ref{sec:fate}).
We identify four dimensionless parameters, which, 
apart from the initial condition of the planetesimal orbit,
completely determine the probabilities of various scattering outcomes (ejection, disruption by the host star, and collision with the planet) and 
the associated timescales:
the planet eccentricity $e_{\rm p}$,  planet-to-star mass ratio $m_{\rm p}/M_\star$, planet radius to semi-major axis ratio $R_{\rm p}/a_{\rm p}$, and the stellar disruption radius to planet semi-major axis ratio $r_{\rm dis}/a_{\rm p}$. 

For planetesimals initially within the chaotic zone around the planet, our results are summarized in Figure~\ref{fig:timescale_vs_ci} and related figures. In particular, the distribution of $q_{\rm min}$ is mainly dependent on $e_{\rm p}$ (Figures~\ref{fig:dmin} and \ref{fig:dmin_log}). 
The probability of planetesimal disruption by the star depends on the value of $r_{\rm dis}$, but for moderate to large planet eccentricity, this probability is significant as long as $r_{\rm dis} \gtrsim 10^{-4}~a_{\rm p}$ (Figure~\ref{fig:proba_vs_e}). For $e_{\rm p} \gtrsim 0.4$, assuming a Jovian planet and a disruption radius of $r_{\rm dis} \simeq 10^{-1}~a_{\rm p}$, this probability becomes greater than the combined probabilities of ejection or collision with the planet (Figure~\ref{fig:branchingratio}). The branching ratios between the different outcomes also vary with $m_{\rm p}/M_\star$ and $R_{\rm p}/a_{\rm p}$. Increasing the mass ratio mostly increases the likelihood of ejection over the other outcomes (Figures~\ref{fig:proba_vs_mp}), while increasing the planet radius increases the likelihood of collision with the planet (Figure~\ref{fig:proba_vs_ap}).

The timescales associated with the different outcomes scale with the planet orbital period $P_{\rm p}$ and span several orders of magnitude. For a Jovian planet, collision typically takes $10^2$ to $10^4~P_{\rm p}$, ejection takes $10^3$ to $10^5~P_{\rm p}$, and disruption by the star takes $10^2$ to $10^6~P_{\rm p}$ (Figure~\ref{fig:timescales}). All timescales depend strongly on the planet to star mass ratio [with the scaling $t \propto (m_{\rm p}/M_\star)^{-1.5}$; see Figure~\ref{fig:timescales_vs_mu}]. Moreover, unsurprisingly, the timescale for disruption depends on the disruption radius (Figure~\ref{fig:timescalesthreshold}) and the timescale for collision (with the planet) depends on the planet radius (Figure~\ref{fig:timescales_vs_rp}).

While our numerical study focuses on the case of a single planet and no replenishment of planetesimals in the chaotic zone,  our results can be adapted to the situations with general planetesimal replenishment profile (see Figure~\ref{fig:spiked}). 

3. We perform more limited numerical  simulations with a two-planet configuration, inspired by the $\beta$ Pictoris system, and compare the results to the single planet case (Figures~\ref{fig:timescale_vs_ci_betapic} and \ref{fig:branchingratio_vs_ci_betapic}). Qualitatively, we find that the proportion of scattered planetesimals is controlled by the inner planet, and the proportion of planetesimals that collide with either planet is controlled by the more eccentric one. The timescales involved are greater than the expectations from either of the planets for all possible outcomes.

4. We apply our general results to several topics of current interest, including  the production of exocomets in young stellar systems, white dwarf pollution via planetesimal disruption, and planet engulfment by main-sequence stars (see Section~\ref{sec:discussion}). The timescales that we derived are not compatible with the sustained production of exocomets to the extent observed in the $\beta$ Pictoris system, suggesting either an exceptional recent influx of planetesimals or a different injection  mechanism with longer timescale. This conclusion also applies to the problem of white-dwarf pollution, since the ages of the polluted white dwarfs are much longer than the time it takes for a planet to deplete its chaotic zone. On the other hand, planet engulfment appears to be a likely outcome of planet-planet scattering assuming a large mass ratio between the two planets and a moderate eccentricity of the more massive planet, although the engulfment rate depends on the occurrence rate of two-planet systems and their orbital parameters.

\section*{Acknowledgements}

We thank the referee for their constructive comments that improved the quality of the manuscript. This work has been supported in part by the NSF grant AST-2107796. We made use of the \textsc{python} libraries \textsc{NumPy} \citep{harrisArrayProgrammingNumPy2020}, \textsc{SciPy} \citep{virtanenSciPyFundamentalAlgorithms2020}, and \textsc{PyQt-Fit}, and the figures were made with \textsc{Matplotlib} \citep{hunterMatplotlib2DGraphics2007}.

\section*{Data Availability}

Our results can be reproduced from the calculations laid down in the appendix, as well as the outcome of the simulations for 31 different parameters. The product files of the simulations have been uploaded on GitHub and can be found at \url{https://github.com/LaRodet/ScatteringIntoAStar.git}.



\bibliographystyle{mnras}
\bibliography{Biblio} 

\begin{thebibliography}{}
\makeatletter
\relax
\def\mn@urlcharsother{\let\do\@makeother \do\$\do\&\do\#\do\^\do\_\do\%\do\~}
\def\mn@doi{\begingroup\mn@urlcharsother \@ifnextchar [ {\mn@doi@}
  {\mn@doi@[]}}
\def\mn@doi@[#1]#2{\def\@tempa{#1}\ifx\@tempa\@empty \href
  {http://dx.doi.org/#2} {doi:#2}\else \href {http://dx.doi.org/#2} {#1}\fi
  \endgroup}
\def\mn@eprint#1#2{\mn@eprint@#1:#2::\@nil}
\def\mn@eprint@arXiv#1{\href {http://arxiv.org/abs/#1} {{\tt arXiv:#1}}}
\def\mn@eprint@dblp#1{\href {http://dblp.uni-trier.de/rec/bibtex/#1.xml}
  {dblp:#1}}
\def\mn@eprint@#1:#2:#3:#4\@nil{\def\@tempa {#1}\def\@tempb {#2}\def\@tempc
  {#3}\ifx \@tempc \@empty \let \@tempc \@tempb \let \@tempb \@tempa \fi \ifx
  \@tempb \@empty \def\@tempb {arXiv}\fi \@ifundefined
  {mn@eprint@\@tempb}{\@tempb:\@tempc}{\expandafter \expandafter \csname
  mn@eprint@\@tempb\endcsname \expandafter{\@tempc}}}

\bibitem[\protect\citeauthoryear{Anderson, Lai  \& Pu}{Anderson
  et~al.}{2020}]{andersonInSituScatteringWarm2020}
Anderson K.~R.,  Lai D.,   Pu B.,  2020, \mn@doi [\mnras]
  {10.1093/mnras/stz3119}, 491, 1369

\bibitem[\protect\citeauthoryear{Antoniadou \& Veras}{Antoniadou \&
  Veras}{2016}]{antoniadouLinkingLongtermPlanetary2016}
Antoniadou K.~I.,  Veras D.,  2016, \mn@doi [\mnras] {10.1093/mnras/stw2264},
  463, 4108

\bibitem[\protect\citeauthoryear{Artymowicz}{Artymowicz}{1997}]{artymowiczBetaPictorisEarly1997}
Artymowicz P.,  1997, \mn@doi [Annual Review of Earth and Planetary Sciences]
  {10.1146/annurev.earth.25.1.175}, 25, 175

\bibitem[\protect\citeauthoryear{Barstow, Barstow, Casewell, Holberg  \&
  Hubeny}{Barstow et~al.}{2014}]{barstowEvidenceExternalOrigin2014}
Barstow M.~A.,  Barstow J.~K.,  Casewell S.~L.,  Holberg J.~B.,   Hubeny I.,
  2014, \mn@doi [\mnras] {10.1093/mnras/stu216}, 440, 1607

\bibitem[\protect\citeauthoryear{{Behmard}, {Sevilla}  \& {Fuller}}{{Behmard}
  et~al.}{2023a}]{Behmard2023a}
{Behmard} A.,  {Sevilla} J.,   {Fuller} J.,  2023a, \mn@doi [\mnras]
  {10.1093/mnras/stac3435}, \href
  {https://ui.adsabs.harvard.edu/abs/2023MNRAS.518.5465B} {518, 5465}

\bibitem[\protect\citeauthoryear{{Behmard}, {Dai}, {Brewer}, {Berger}  \&
  {Howard}}{{Behmard} et~al.}{2023b}]{Behmard2023}
{Behmard} A.,  {Dai} F.,  {Brewer} J.~M.,  {Berger} T.~A.,   {Howard} A.~W.,
  2023b, \mn@doi [\mnras] {10.1093/mnras/stad745}, \href
  {https://ui.adsabs.harvard.edu/abs/2023MNRAS.521.2969B} {521, 2969}

\bibitem[\protect\citeauthoryear{{Behmard}, {Dai}, {Brewer}, {Berger}  \&
  {Howard}}{{Behmard} et~al.}{2023c}]{Behmard2023b}
{Behmard} A.,  {Dai} F.,  {Brewer} J.~M.,  {Berger} T.~A.,   {Howard} A.~W.,
  2023c, \mn@doi [\mnras] {10.1093/mnras/stad745}, \href
  {https://ui.adsabs.harvard.edu/abs/2023MNRAS.521.2969B} {521, 2969}

\bibitem[\protect\citeauthoryear{Beust \& Morbidelli}{Beust \&
  Morbidelli}{1996}]{beustMeanMotionResonancesSource1996}
Beust H.,  Morbidelli A.,  1996, \mn@doi [Icarus] {10.1006/icar.1996.0056},
  120, 358

\bibitem[\protect\citeauthoryear{Beust \& Morbidelli}{Beust \&
  Morbidelli}{2000}]{beustFallingEvaporatingBodies2000}
Beust H.,  Morbidelli A.,  2000, \mn@doi [Icarus] {10.1006/icar.1999.6238},
  143, 170

\bibitem[\protect\citeauthoryear{Beust, {Vidal-Madjar}  \& {Ferlet}}{Beust
  et~al.}{1991}]{beustBetaPectorisProtoplanetary1991}
Beust H.,  {Vidal-Madjar} A.,   {Ferlet} R.,  1991, Astronomy and Astrophysics,
  247, 505

\bibitem[\protect\citeauthoryear{Beust, Lagrange, Plazy  \& Mouillet}{Beust
  et~al.}{1996}]{beustPictorisCircumstellarDisk1996}
Beust H.,  Lagrange A.-M.,  Plazy F.,   Mouillet D.,  1996, Astronomy and
  Astrophysics, 310, 181

\bibitem[\protect\citeauthoryear{Beust et~al.,}{Beust
  et~al.}{2023}]{beustDynamicsBetaPictoris2023}
Beust H.,  et~al., 2023, To be submitted to Astronomy and Astrophysics

\bibitem[\protect\citeauthoryear{Blouin \& Xu}{Blouin \&
  Xu}{2022}]{blouinNoEvidenceStrong2022}
Blouin S.,  Xu S.,  2022, \mn@doi [\mnras] {10.1093/mnras/stab3446}, 510, 1059

\bibitem[\protect\citeauthoryear{Bonsor \& Wyatt}{Bonsor \&
  Wyatt}{2012}]{bonsorScatteringSmallBodies2012}
Bonsor A.,  Wyatt M.~C.,  2012, \mn@doi [\mnras]
  {10.1111/j.1365-2966.2011.20156.x}, 420, 2990

\bibitem[\protect\citeauthoryear{Bonsor, Mustill  \& Wyatt}{Bonsor
  et~al.}{2011}]{bonsorDynamicalEffectsStellar2011}
Bonsor A.,  Mustill A.~J.,   Wyatt M.~C.,  2011, \mn@doi [\mnras]
  {10.1111/j.1365-2966.2011.18524.x}, 414, 930

\bibitem[\protect\citeauthoryear{Chambers, Wetherill  \& Boss}{Chambers
  et~al.}{1996}]{chambersStabilityMultiPlanetSystems1996}
Chambers J.~E.,  Wetherill G.~W.,   Boss A.~P.,  1996, \mn@doi [Icarus]
  {10.1006/icar.1996.0019}, 119, 261

\bibitem[\protect\citeauthoryear{Chauvin et~al.,}{Chauvin
  et~al.}{2012}]{chauvinOrbitalCharacterizationBeta2012}
Chauvin G.,  et~al., 2012, \mn@doi [Astronomy \& Astrophysics]
  {10.1051/0004-6361/201118346}, 542, A41

\bibitem[\protect\citeauthoryear{{De} et~al.,}{{De} et~al.}{2023}]{De2023}
{De} K.,  et~al., 2023, \mn@doi [\nat] {10.1038/s41586-023-05842-x}, \href
  {https://ui.adsabs.harvard.edu/abs/2023Natur.617...55D} {617, 55}

\bibitem[\protect\citeauthoryear{Debes, Walsh  \& Stark}{Debes
  et~al.}{2012}]{debesLinkPlanetarySystems2012}
Debes J.~H.,  Walsh K.~J.,   Stark C.,  2012, \mn@doi [The Astrophysical
  Journal] {10.1088/0004-637X/747/2/148}, 747, 148

\bibitem[\protect\citeauthoryear{Debes et~al.,}{Debes
  et~al.}{2019}]{debesGyrWhiteDwarf2019}
Debes J.~H.,  et~al., 2019, \mn@doi [The Astrophysical Journal]
  {10.3847/2041-8213/ab0426}, 872, L25

\bibitem[\protect\citeauthoryear{Doyle, Desch  \& Young}{Doyle
  et~al.}{2021}]{doyleIcyExomoonsEvidenced2021}
Doyle A.~E.,  Desch S.~J.,   Young E.~D.,  2021, \mn@doi [The Astrophysical
  Journal] {10.3847/2041-8213/abd9ba}, 907, L35

\bibitem[\protect\citeauthoryear{Farihi}{Farihi}{2016}]{farihiCircumstellarDebrisPollution2016}
Farihi J.,  2016, \mn@doi [New Astronomy Reviews]
  {10.1016/j.newar.2016.03.001}, 71, 9

\bibitem[\protect\citeauthoryear{Ford, Havlickova  \& Rasio}{Ford
  et~al.}{2001}]{fordDynamicalInstabilitiesExtrasolar2001}
Ford E.~B.,  Havlickova M.,   Rasio F.~A.,  2001, \mn@doi [Icarus]
  {10.1006/icar.2001.6588}, 150, 303

\bibitem[\protect\citeauthoryear{Frewen \& Hansen}{Frewen \&
  Hansen}{2014}]{frewenEccentricPlanetsStellar2014}
Frewen S. F.~N.,  Hansen B. M.~S.,  2014, \mn@doi [\mnras]
  {10.1093/mnras/stu097}, 439, 2442

\bibitem[\protect\citeauthoryear{Girven, Brinkworth, Farihi, G{\"a}nsicke,
  Hoard, Marsh  \& Koester}{Girven
  et~al.}{2012}]{girvenConstraintsLifetimesDisks2012}
Girven J.,  Brinkworth C.~S.,  Farihi J.,  G{\"a}nsicke B.~T.,  Hoard D.~W.,
  Marsh T.~R.,   Koester D.,  2012, \mn@doi [The Astrophysical Journal]
  {10.1088/0004-637X/749/2/154}, 749, 154

\bibitem[\protect\citeauthoryear{Gladman}{Gladman}{1993}]{gladmanDynamicsSystemsTwo1993}
Gladman B.,  1993, \mn@doi [Icarus] {10.1006/icar.1993.1169}, 106, 247

\bibitem[\protect\citeauthoryear{Harris et~al.,}{Harris
  et~al.}{2020}]{harrisArrayProgrammingNumPy2020}
Harris C.~R.,  et~al., 2020, \mn@doi [\nat] {10.1038/s41586-020-2649-2}, 585,
  357

\bibitem[\protect\citeauthoryear{Helled \& Morbidelli}{Helled \&
  Morbidelli}{2021}]{helledPlanetFormation2021}
Helled R.,  Morbidelli A.,  2021, Planet {{Formation}}.
IOP Publishing, \mn@doi{10.1088/2514-3433/abfa8fch12}

\bibitem[\protect\citeauthoryear{Hollands, G{\"a}nsicke  \& Koester}{Hollands
  et~al.}{2018}]{hollandsCoolDZWhite2018}
Hollands M.~A.,  G{\"a}nsicke B.~T.,   Koester D.,  2018, \mn@doi [\mnras]
  {10.1093/mnras/sty592}, 477, 93

\bibitem[\protect\citeauthoryear{Hunter}{Hunter}{2007}]{hunterMatplotlib2DGraphics2007}
Hunter J.~D.,  2007, \mn@doi [Comput Sci Eng] {10.1109/MCSE.2007.55}, 9, 90

\bibitem[\protect\citeauthoryear{Koester, G{\"a}nsicke  \& Farihi}{Koester
  et~al.}{2014}]{koesterFrequencyPlanetaryDebris2014}
Koester D.,  G{\"a}nsicke B.~T.,   Farihi J.,  2014, \mn@doi [Astronomy \&
  Astrophysics] {10.1051/0004-6361/201423691}, 566, A34

\bibitem[\protect\citeauthoryear{Lacour et~al.,}{Lacour
  et~al.}{2021}]{lacourMassPictorisPictoris2021}
Lacour S.,  et~al., 2021, \mn@doi [Astronomy & Astrophysics]
  {10.1051/0004-6361/202141889}, 654, L2

\bibitem[\protect\citeauthoryear{Lagrange et~al.,}{Lagrange
  et~al.}{2010}]{lagrangeGiantPlanetImaged2010}
Lagrange A.-M.,  et~al., 2010, \mn@doi [Science] {10.1126/science.1187187},
  329, 57

\bibitem[\protect\citeauthoryear{Lagrange et~al.,}{Lagrange
  et~al.}{2019}]{lagrangeEvidenceAdditionalPlanet2019}
Lagrange A.~M.,  et~al., 2019, \mn@doi [Nature Astronomy]
  {10.1038/s41550-019-0857-1}, 3, 1135

\bibitem[\protect\citeauthoryear{Li, Lai, Anderson  \& Pu}{Li
  et~al.}{2021}]{liGiantPlanetScatterings2021}
Li J.,  Lai D.,  Anderson K.~R.,   Pu B.,  2021, \mn@doi [\mnras]
  {10.1093/mnras/staa3779}, 501, 1621

\bibitem[\protect\citeauthoryear{Malmberg, Davies  \& Heggie}{Malmberg
  et~al.}{2011}]{malmbergEffectsFlybysPlanetary2011}
Malmberg D.,  Davies M.~B.,   Heggie D.~C.,  2011, \mn@doi [\mnras]
  {10.1111/j.1365-2966.2010.17730.x}, 411, 859

\bibitem[\protect\citeauthoryear{Marchal \& Saari}{Marchal \&
  Saari}{1975}]{marchalHillRegionsGeneral1975}
Marchal C.,  Saari D.~G.,  1975, \mn@doi [Celestial Mechanics]
  {10.1007/BF01230206}, 12, 115

\bibitem[\protect\citeauthoryear{Murray \& Dermott}{Murray \&
  Dermott}{2000}]{murraySolarSystemDynamics2000}
Murray C.~D.,  Dermott S.~F.,  2000, Solar {{System Dynamics}}.
{Cambridge University Press}

\bibitem[\protect\citeauthoryear{Mustill, Davies  \& Johansen}{Mustill
  et~al.}{2015}]{mustillDestructionInnerPlanetary2015}
Mustill A.~J.,  Davies M.~B.,   Johansen A.,  2015, \mn@doi [The Astrophysical
  Journal] {10.1088/0004-637X/808/1/14}, 808, 14

\bibitem[\protect\citeauthoryear{Mustill, Villaver, Veras, G{\"a}nsicke  \&
  Bonsor}{Mustill et~al.}{2018}]{mustillUnstableLowmassPlanetary2018}
Mustill A.~J.,  Villaver E.,  Veras D.,  G{\"a}nsicke B.~T.,   Bonsor A.,
  2018, \mn@doi [\mnras] {10.1093/mnras/sty446}, 476, 3939

\bibitem[\protect\citeauthoryear{Nowak et~al.,}{Nowak
  et~al.}{2020}]{nowakDirectConfirmationRadialvelocity2020}
Nowak M.,  et~al., 2020, \mn@doi [Astronomy and Astrophysics]
  {10.1051/0004-6361/202039039}, 642, L2

\bibitem[\protect\citeauthoryear{O'Connor, Teyssandier  \& Lai}{O'Connor
  et~al.}{2022a}]{oconnorSecularChaosWhite2022}
O'Connor C.~E.,  Teyssandier J.,   Lai D.,  2022a, \mn@doi [\mnras]
  {10.1093/mnras/stac1189}, 513, 4178

\bibitem[\protect\citeauthoryear{{O'Connor}, {Teyssandier}  \&
  {Lai}}{{O'Connor} et~al.}{2022b}]{O'connor2022}
{O'Connor} C.~E.,  {Teyssandier} J.,   {Lai} D.,  2022b, \mn@doi [\mnras]
  {10.1093/mnras/stac1189}, \href
  {https://ui.adsabs.harvard.edu/abs/2022MNRAS.513.4178O} {513, 4178}

\bibitem[\protect\citeauthoryear{Pavlenko, Kulyk, Shubina, Vasylenko,
  Dobrycheva  \& Korsun}{Pavlenko et~al.}{2022}]{pavlenkoNewExocometsPic2022}
Pavlenko Y.,  Kulyk I.,  Shubina O.,  Vasylenko M.,  Dobrycheva D.,   Korsun
  P.,  2022, \mn@doi [Astronomy & Astrophysics] {10.1051/0004-6361/202142111},
  660, A49

\bibitem[\protect\citeauthoryear{Petrovich}{Petrovich}{2015}]{petrovichStabilityFatesHierarchical2015}
Petrovich C.,  2015, \mn@doi [The Astrophysical Journal]
  {10.1088/0004-637X/808/2/120}, 808, 120

\bibitem[\protect\citeauthoryear{Pinsonneault, DePoy  \& Coffee}{Pinsonneault
  et~al.}{2001}]{pinsonneaultMassConvectiveZone2001}
Pinsonneault M.~H.,  DePoy D.~L.,   Coffee M.,  2001, \mn@doi [The
  Astrophysical Journal] {10.1086/323531}, 556, L59

\bibitem[\protect\citeauthoryear{Pu \& Lai}{Pu \&
  Lai}{2021}]{puStrongScatteringsCold2021}
Pu B.,  Lai D.,  2021, \mn@doi [\mnras] {10.1093/mnras/stab2504}, 508, 597

\bibitem[\protect\citeauthoryear{Raymond \& Morbidelli}{Raymond \&
  Morbidelli}{2022}]{raymondPlanetFormationKey2022}
Raymond S.~N.,  Morbidelli A.,  2022, in Astrophysics and {{Space Science
  Library}}, Vol.~466, Demographics of {{Exoplanetary Systems}}.
{Springer International Publishing}, pp 3--82,
  \mn@doi{10.1007/978-3-030-88124-5_1}

\bibitem[\protect\citeauthoryear{{Rodet} \& {Lai}}{{Rodet} \&
  {Lai}}{2022}]{Rodet2022}
{Rodet} L.,  {Lai} D.,  2022, \mn@doi [\mnras] {10.1093/mnras/stab3046}, \href
  {https://ui.adsabs.harvard.edu/abs/2022MNRAS.509.1010R} {509, 1010}

\bibitem[\protect\citeauthoryear{{Rodet}, {Su}  \& {Lai}}{{Rodet}
  et~al.}{2021}]{Rodet2021}
{Rodet} L.,  {Su} Y.,   {Lai} D.,  2021, \mn@doi [\apj]
  {10.3847/1538-4357/abf8a7}, \href
  {https://ui.adsabs.harvard.edu/abs/2021ApJ...913..104R} {913, 104}

\bibitem[\protect\citeauthoryear{Safronov}{Safronov}{1972}]{safronovEvolutionProtoplanetaryCloud1972}
Safronov V.~S.,  1972, Evolution of the Protoplanetary Cloud and Formation of
  the Earth and Planets.

\bibitem[\protect\citeauthoryear{Smallwood, Martin, Livio  \& Veras}{Smallwood
  et~al.}{2021}]{smallwoodRoleResonancesPolluting2021}
Smallwood J.~L.,  Martin R.~G.,  Livio M.,   Veras D.,  2021, \mn@doi [\mnras]
  {10.1093/mnras/stab1077}, 504, 3375

\bibitem[\protect\citeauthoryear{Spina, Sharma, Mel{\'e}ndez, Bedell, Casey,
  Carlos, Franciosini  \& Vallenari}{Spina
  et~al.}{2021}]{spinaChemicalEvidencePlanetary2021}
Spina L.,  Sharma P.,  Mel{\'e}ndez J.,  Bedell M.,  Casey A.~R.,  Carlos M.,
  Franciosini E.,   Vallenari A.,  2021, \mn@doi [Nature Astronomy]
  {10.1038/s41550-021-01451-8}

\bibitem[\protect\citeauthoryear{Steckloff, Debes, Steele, Johnson, Adams,
  Jacobson  \& Springmann}{Steckloff
  et~al.}{2021}]{steckloffHowSublimationDelays2021}
Steckloff J.~K.,  Debes J.,  Steele A.,  Johnson B.,  Adams E.~R.,  Jacobson
  S.~A.,   Springmann A.,  2021, \mn@doi [The Astrophysical Journal Letters]
  {10.3847/2041-8213/abfd39}, 913, L31

\bibitem[\protect\citeauthoryear{{Teyssandier}, {Lai}  \& {Vick}}{{Teyssandier}
  et~al.}{2019}]{Teyssandier2019}
{Teyssandier} J.,  {Lai} D.,   {Vick} M.,  2019, \mn@doi [\mnras]
  {10.1093/mnras/stz1011}, \href
  {https://ui.adsabs.harvard.edu/abs/2019MNRAS.486.2265T} {486, 2265}

\bibitem[\protect\citeauthoryear{Th{\'e}bault, Augereau  \& Beust}{Th{\'e}bault
  et~al.}{2003}]{thebaultDustProductionCollisions2003}
Th{\'e}bault P.,  Augereau J.~C.,   Beust H.,  2003, \mn@doi [Astronomy and
  Astrophysics] {10.1051/0004-6361:20031017}, 408, 775

\bibitem[\protect\citeauthoryear{Tremaine}{Tremaine}{1993}]{tremaineDistributionCometsStars1993}
Tremaine S.,  1993, in Planets around Pulsars; Proceedings of the
  {{Conference}}. {California Inst. of Technology, Pasadena}, pp 335--344

\bibitem[\protect\citeauthoryear{Veras}{Veras}{2016}]{verasPostmainsequencePlanetarySystem2016}
Veras D.,  2016, \mn@doi [Royal Society Open Science] {10.1098/rsos.150571}, 3,
  150571

\bibitem[\protect\citeauthoryear{Veras}{Veras}{2021}]{verasPlanetarySystemsWhite2021}
Veras D.,  2021, Planetary {{Systems Around White Dwarfs}},
  \mn@doi{10.1093/acrefore/9780190647926.013.238.
}

\bibitem[\protect\citeauthoryear{Veras \& Heng}{Veras \&
  Heng}{2020}]{verasLifetimesPlanetaryDebris2020}
Veras D.,  Heng K.,  2020, \mn@doi [\mnras] {10.1093/mnras/staa1632}, 496, 2292

\bibitem[\protect\citeauthoryear{Veras \& Kurosawa}{Veras \&
  Kurosawa}{2020}]{verasGeneratingMetalpollutingDebris2020}
Veras D.,  Kurosawa K.,  2020, \mn@doi [\mnras] {10.1093/mnras/staa621}, 494,
  442

\bibitem[\protect\citeauthoryear{Veras \& Rosengren}{Veras \&
  Rosengren}{2023}]{verasSmallestPlanetaryDrivers2023}
Veras D.,  Rosengren A.~J.,  2023, \mn@doi [\mnras] {10.1093/mnras/stad130},
  519, 6257

\bibitem[\protect\citeauthoryear{Veras \& Scheeres}{Veras \&
  Scheeres}{2020}]{verasPostmainsequenceDebrisRotationinduced2020}
Veras D.,  Scheeres D.~J.,  2020, \mn@doi [\mnras] {10.1093/mnras/stz3565},
  492, 2437

\bibitem[\protect\citeauthoryear{Veras, Jacobson  \& G{\"a}nsicke}{Veras
  et~al.}{2014}]{verasPostmainsequenceDebrisRotationinduced2014}
Veras D.,  Jacobson S.~A.,   G{\"a}nsicke B.~T.,  2014, \mn@doi [\mnras]
  {10.1093/mnras/stu1926}, 445, 2794

\bibitem[\protect\citeauthoryear{Veras, Georgakarakos, Mustill, Malamud,
  Cunningham  \& {Dobbs-Dixon}}{Veras
  et~al.}{2021}]{verasEntryGeometryVelocity2021}
Veras D.,  Georgakarakos N.,  Mustill A.~J.,  Malamud U.,  Cunningham T.,
  {Dobbs-Dixon} I.,  2021, \mn@doi [\mnras] {10.1093/mnras/stab1667}, 506, 1148

\bibitem[\protect\citeauthoryear{Veras, Birader  \& Zaman}{Veras
  et~al.}{2022}]{verasOrbitDecay21002022}
Veras D.,  Birader Y.,   Zaman U.,  2022, \mn@doi [\mnras]
  {10.1093/mnras/stab3490}, 510, 3379

\bibitem[\protect\citeauthoryear{Veras, Georgakarakos  \& {Dobbs-Dixon}}{Veras
  et~al.}{2023}]{verasHighresolutionResonantPortraits2023}
Veras D.,  Georgakarakos N.,   {Dobbs-Dixon} I.,  2023, \mn@doi [\mnras]
  {10.1093/mnras/stac3274}, 518, 4537

\bibitem[\protect\citeauthoryear{Virtanen et~al.,}{Virtanen
  et~al.}{2020}]{virtanenSciPyFundamentalAlgorithms2020}
Virtanen P.,  et~al., 2020, \mn@doi [\nat Methods] {10.1038/s41592-019-0686-2},
  17, 261

\bibitem[\protect\citeauthoryear{Zuckerman, Koester, Reid  \&
  H{\"u}nsch}{Zuckerman et~al.}{2003}]{zuckermanMetalLinesWhite2003}
Zuckerman B.,  Koester D.,  Reid I.~N.,   H{\"u}nsch M.,  2003, \mn@doi [The
  Astrophysical Journal] {10.1086/377492}, 596, 477

\bibitem[\protect\citeauthoryear{Zuckerman, Melis, Klein, Koester  \&
  Jura}{Zuckerman et~al.}{2010}]{zuckermanAncientPlanetarySystems2010}
Zuckerman B.,  Melis C.,  Klein B.,  Koester D.,   Jura M.,  2010, \mn@doi [The
  Astrophysical Journal] {10.1088/0004-637X/722/1/725}, 722, 725

\makeatother
\end{thebibliography}




\appendix

\onecolumn

\section{Derivation of the minimum periastron from the Tisserand parameter}

\label{sec:appendixtisserand}

\begin{figure}
	\centering
	\includegraphics[width=0.45\linewidth]{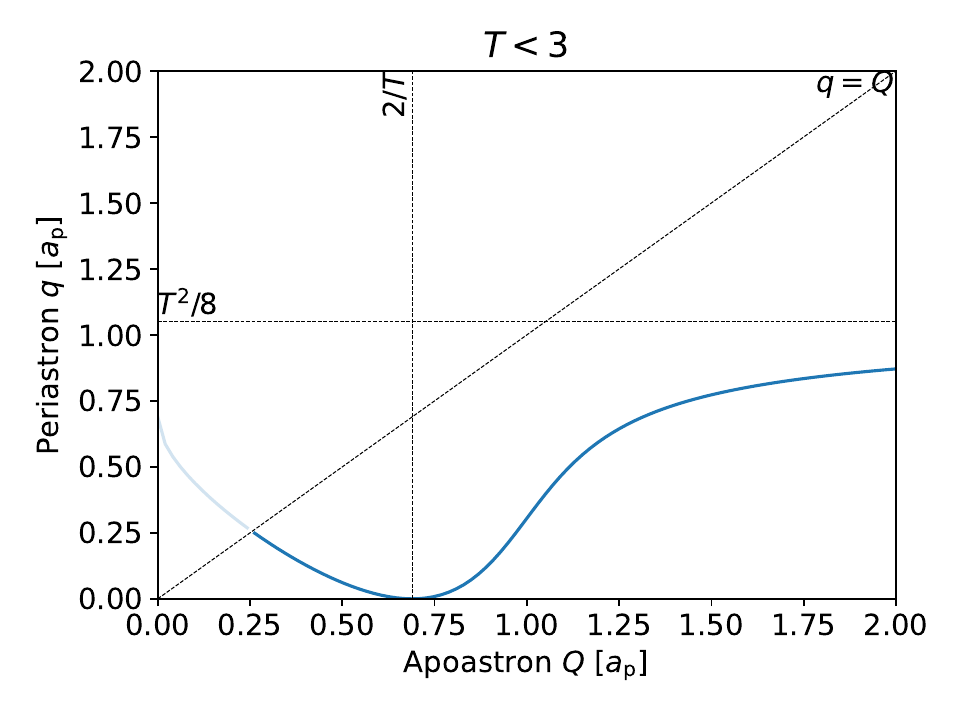}
	\includegraphics[width=0.45\linewidth]{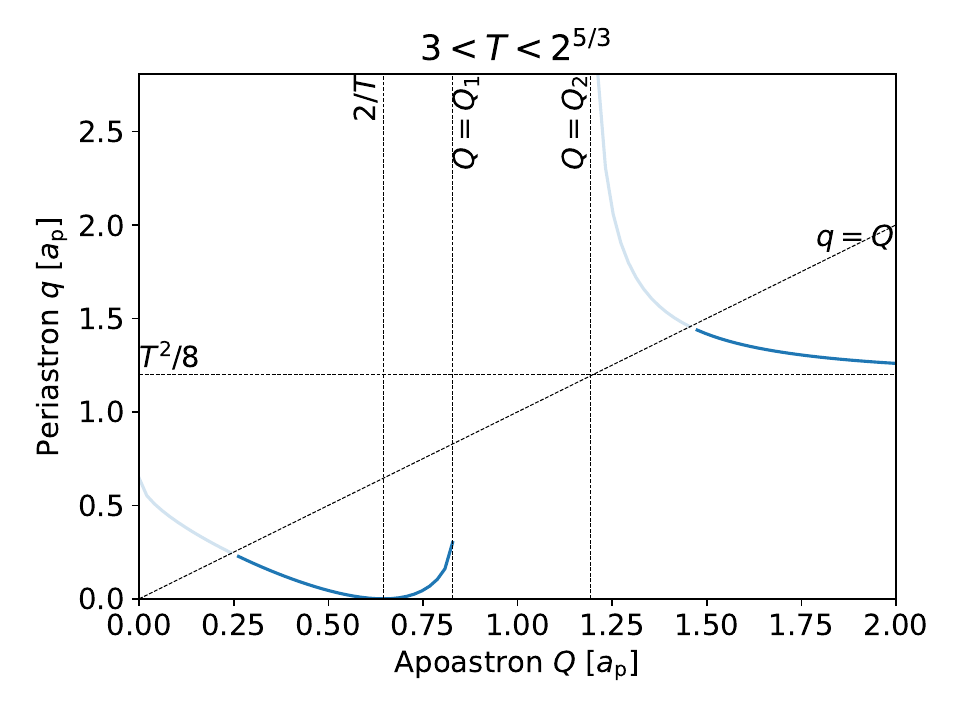}
	\includegraphics[width=0.45\linewidth]{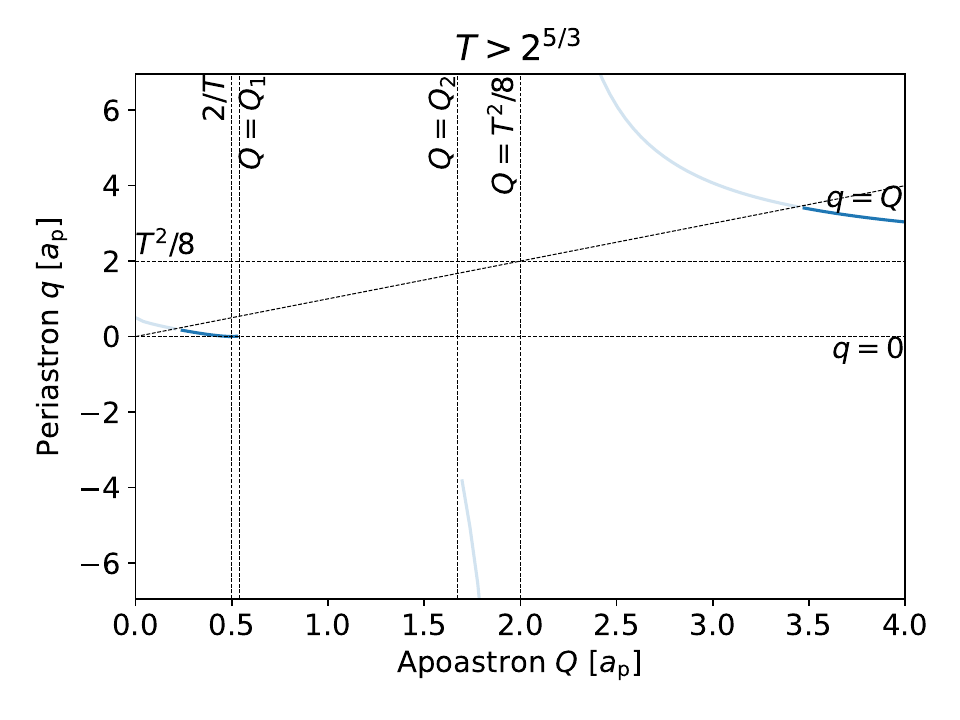}
	\caption{Periastron of the test mass as a function of its apoastron after a close encounter with a giant planet, for given Tisserand parameters (equation~\ref{eq:q}). The light blue line corresponds to values such that $q > Q$ or $q < 0$, which are forbidden. When $T > 3$, equation~\ref{eq:q} has complex solutions for $Q$s ranging from $Q_1$ to $Q_2$, which are given in equations~\eqref{eq:Q1} and \eqref{eq:Q2}.}\label{fig:Tisserand}
\end{figure}

We analyse here the behaviour of the function $\tilde{q}(\tilde{Q},T)$ given by equation~\eqref{eq:q}. We recall that $\tilde{q}$ and $\tilde{Q}$ should be positive with $\tilde{q} < \tilde{Q}$. Our goal is, for a given $T$, to find the minimum $\tilde{q}(\tilde{Q},T)$ that satisfies equation~\eqref{eq:condition} ($\tilde{Q}>\tilde{Q}_{\rm m}$).

First, we note that equation~\eqref{eq:condition} restricts the range of possible Tisserand parameter. After some algebra, we find that $T$ is minimum when $q = 0$ and $Q\to \infty$ (parabolic orbit), where $T = 0$. On the other hand, $T$ has a finite maximum value when $\tilde{Q} = \tilde{Q}_{\rm m}$ and $\tilde{q} = \tilde{Q}_{\rm m}^4/(2-\tilde{Q}_{\rm m}^3)$, given by
\begin{equation}
	T_{\rm max} = \frac{2}{\tilde{Q}_{\rm m}} + \tilde{Q}_{\rm m}^2. \label{eq:Tmax}
\end{equation}
Typically, for a Jupiter-Sun mass ratio and assuming $\beta = 3$, we get $T_{\rm max} \approx 3.15$.

We identify three cases, depending on the value of $T$, as depicted in Figure~\ref{fig:Tisserand}.

\subsection{$T < 3$}

When $T < 3$, the function $\tilde{q}$ is well-defined and continuous for the entire range of $\tilde{Q}$ from $0$ to $+ \infty$: $\tilde{q}$ decreases from $\tilde{Q} = 0$ to $2/T$, where it reaches its minimum $\tilde{q} = 0$. For $\tilde{Q}>2/T$, it increases with an asymptote at $\tilde{q} = T^2/8$.

In this case, there are two possibilities, as given by equation~\eqref{eq:qmin}:
\begin{itemize}
	\item $\tilde{Q}_{\rm m} < 2/T$: in this case, the apoastron $\tilde{Q} = 2/T$ is allowed and the minimum periastron is $\tilde{q}_{\rm min} = 0$.
	\item $\tilde{Q}_{\rm m} > 2/T$: in this case, since $\tilde{q}(\tilde{Q}, T)$ is an increasing function of $\tilde{Q}$ for $\tilde{Q} > 2/T$, the minimum periastron is  $\tilde{q}_{\rm min} = \tilde{q}(\tilde{Q}_{\rm m}, T)$, which is a non-trivial function of $\tilde{Q}_{\rm m}$ and $T$.
\end{itemize}

\subsection{$3 < T < 2^{5/3}$}

When $3 < T < 2^{5/3} \approx 3.2$, the function $\tilde{q}$ has complex values for $\tilde{Q}$ between $\tilde{Q}_1$ and $\tilde{Q}_2$, given by:
\begin{align}
	&\tilde{Q}_1 = 2\sqrt{\frac{T}{3}} \cos(\frac{\pi}{3}+\theta)\\ \label{eq:Q1}
	&\tilde{Q}_2 = 2\sqrt{\frac{T}{3}} \cos(\frac{\pi}{3}-\theta)\\ \label{eq:Q2}
	&\text{with }\theta = \frac{1}{3}\arctan\left(\sqrt{\left(\frac{T}{3}\right)^3 -1 }\right).
\end{align}
The function $\tilde{q}$ first decreases from $\tilde{Q} = 0$ to $\tilde{Q} = 2/T$, where it reaches its minimal value $\tilde{q} = 0$. It then increases from $\tilde{Q} = 2/T$ to $\tilde{Q} = \tilde{Q}_1$. Finally, $\tilde{q}$ decreases from $\tilde{Q} = \tilde{Q}_2$ to $+\infty$, towards the asymptote $\tilde{q} = T^2/8$.

If the test particle is initially in the chaotic zone of around the planet ($\tilde{q} < 1 + \beta r_{\rm H}$ and  $\tilde{Q} > 1 - \beta r_{\rm H}$), it can be shown that $2/T < \tilde{Q}_1$. Moreover,  $T^2/8 > q(\tilde{Q}= \tilde{Q}_1, T)$, so that the minimum $q$ always belongs to the $\tilde{Q} < \tilde{Q}_1$ zone. This case thus boils down to the $T < 3$ case:
\begin{itemize}
	\item $\tilde{Q}_{\rm m} < 2/T$: in this case, $\tilde{Q} = 2/T$ is authorized and the minimum periastron is $\tilde{q}_{\rm min} = 0$.
	\item $\tilde{Q}_{\rm m} > 2/T$: in this case, since $\tilde{q}(\tilde{Q}, T)$ is an increasing function of $\tilde{Q}$ for $2/T < \tilde{Q} < \tilde{Q}_1$,  the minimum periastron is  $\tilde{q}_{\rm min} = \tilde{q}(\tilde{Q}_{\rm m}, T)$, which is a non-trivial function of $\tilde{Q}_{\rm m}$ and $T$.
\end{itemize}

\subsection{$T > 2^{5/3}$}

When $T > 2^{5/3} \approx 3.2$, the function $\tilde{q}$ not only has complex values for $\tilde{Q}$ between $\tilde{Q}_1$ and $\tilde{Q}_2$, but also diverges in $\tilde{Q} = T^2/8 > \tilde{Q}_2$. It also has negative values from $\tilde{Q} = \tilde{Q}_2$ to $T^2/8$.

The function $\tilde{q}$ first decreases from $\tilde{Q} = 0$ to $\tilde{Q} = 2/T$, where it reaches its minimal value $\tilde{q} = 0$. It then increases from $\tilde{Q} = 2/T$ to $\tilde{Q} = \tilde{Q}_1$. It continuously decreases from $\tilde{Q} = T^2/8$ to $+\infty$, towards the asymptote $\tilde{q} = T^2/8$. Again, if the test particle is initially in the chaotic zone of around the planet ($\tilde{q} < 1 + \beta r_{\rm H}$ and  $\tilde{Q} > 1 - \beta r_{\rm H}$), it can be shown that $2/T < \tilde{Q}_1$. Moreover,  $T^2/8 > q(\tilde{Q}= \tilde{Q}_1, T)$, so that the minimum $q$ always belongs to the $\tilde{Q} < \tilde{Q}_1$ zone. Once again, this case is comparable to the previous one:
\begin{itemize}
	\item $\tilde{Q}_{\rm m} < 2/T$: in this case, $\tilde{Q} = 2/T$ is authorized and the minimum periastron is $\tilde{q}_{\rm min} = 0$.
	\item $\tilde{Q}_{\rm m} > 2/T$: in this case, since $\tilde{q}(\tilde{Q}, T)$ is an increasing function of $\tilde{Q}$ for $2/T < \tilde{Q} < \tilde{Q}_1$,  the minimum periastron is  $\tilde{q}_{\rm min} = \tilde{q}(\tilde{Q}_{\rm m}, T)$, which is a non-trivial function of $\tilde{Q}_{\rm m}$ and $T$.
\end{itemize}

\begin{figure}
	\centering
	\includegraphics[width=0.5\linewidth]{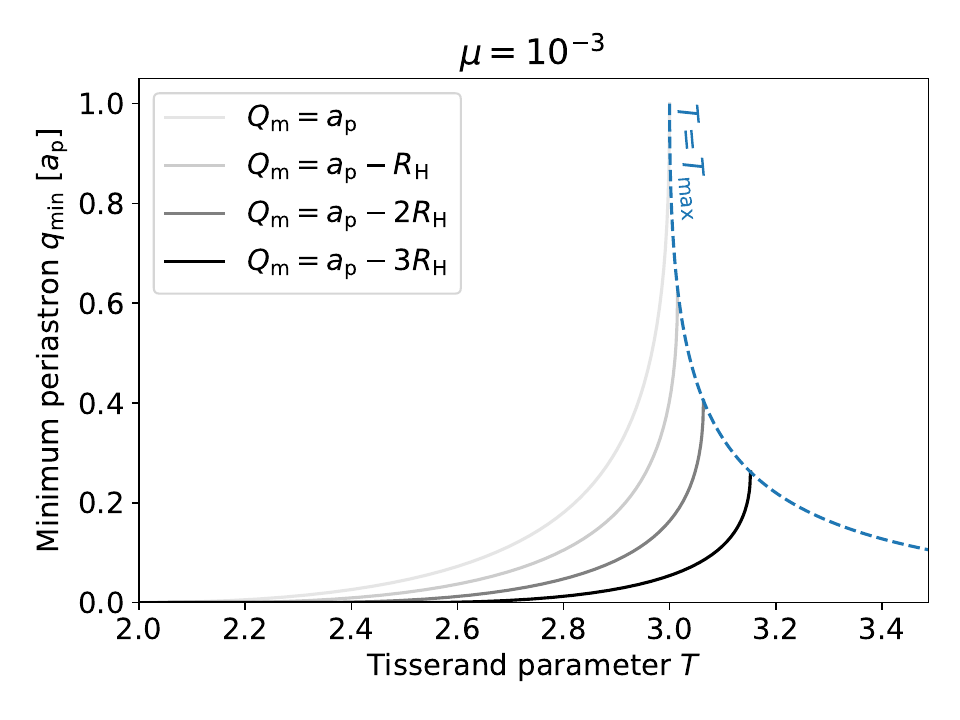}
	\caption{Minimal test mass periastron as a function of the Tisserand parameter and the minimum allowed apoastron $Q_{\rm m} = \tilde{Q}_{\rm m} a_{\rm p}$, for a fixed planet mass ratio $\mu = 10^{-3}$. The blue dotted line corresponds to $T = T_{\rm max}$ (equation~\ref{eq:Tmax}) for $Q_{\rm m}$ between $a_{\rm p}$ and $a_{\rm p} - 5 R_{\rm H}$. The lightest line ($Q_{\rm m} = a_{\rm p}$) corresponds to the result of Bonsor et Wyatt (2012). Our best estimate from the $N$-body simulations is much smaller ($Q_{\rm m} = a_{\rm p} - 2R_{\rm H}$).} \label{fig:qminvsT}
\end{figure}

\section{Derivation of the minimum periastron from the conservation laws}
\label{sec:appendixconservation}

\begin{figure}
	\centering
	\includegraphics[width=0.6\linewidth]{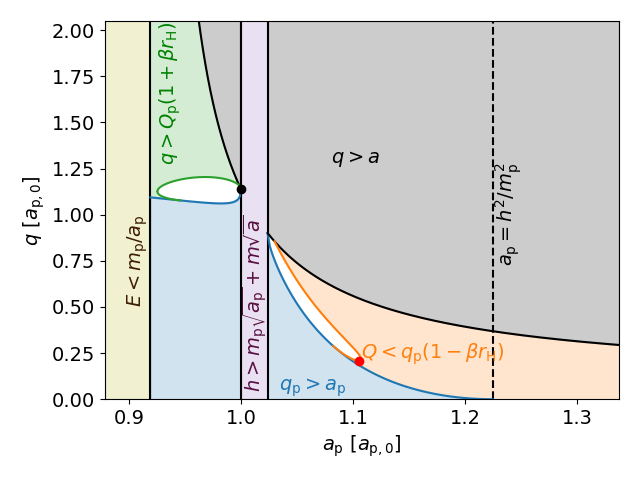}
	\caption{Possible parameter space satisfying equations~\eqref{eq:E}, \eqref{eq:h}, \eqref{eq:rh} and \eqref{eq:rh2} for the variables $q$ (planetesimal periastron) and $a_{\rm p}$ (semi-major axis of the planet). The initial conditions are $a_{\rm 0} = a_{\rm p, 0}(1+\beta r_{\rm H})$ (with $\beta = 2$) and $e_{\rm p, 0} = e_{\rm 0} = 0$. Each coloured zone corresponds to a different criterion derived from the set of equations or from the definition of the quantities. The black dot corresponds to the initial condition, and the red dot to the minimal possible periastron $q$. The entire white zone is accessible from the black dot, since this diagram is valid before and after the scattering but not during the close encounter.} \label{fig:q_vs_ap}
\end{figure}

We derive here the minimum periastron $q_{\rm min}$ that the planetesimal $m$ can reach after scattering with $m_{\rm p}$, using the energy conservation laws (equations~\ref{eq:E} and \ref{eq:h})  and the stability (or rather instability) constraints (equations~\ref{eq:rh} and \ref{eq:rh2}). For given initial conditions, only a subset of the 4-dimensional space of the post-scattering quantities $a$, $a_{\rm p}$, $q$ and $q_{\rm  p}$ is possible, which is represented on Figure~\ref{fig:q_vs_ap}.

\subsection{Formal solution}
\label{sec:formal}

We rewrite equation~\eqref{eq:h} to express the angular momentum $h$ as a function of $q$ and $q_{\rm p}$ instead of $e$ and $e_{\rm p}$:
\begin{equation}
	h = m_{\rm p}\sqrt{q_{\rm p}\left(2-\frac{q_{\rm p}}{a_{\rm p}}\right)} + m\sqrt{q\left(2-\frac{q}{a}\right)}. \label{eq:hbis}
\end{equation}
Since $h$ is constant and $x \mapsto \sqrt{x (2-x/y)}$ is a monotonically increasing function for $x$ between $0$ and $y$, it follows that $q$ is a decreasing function of $q_{\rm p}$ for fixed $a$ and $a_{\rm p}$. The minimum $q_{\rm min}$ is thus reached when $q_{\rm p}$ reaches its maximum value $q_{\rm p, max}$. The value of $q_{\rm p, max}$ depends on the value of $a_{\rm p}$ and $a$. It stems from three constraints:
\begin{align}
	&q_{\rm p}  \leq a_{\rm p} \quad \text{(definition of the periastron)}, \label{eq:qpmax1}\\
	&m_{\rm p}\sqrt{q_{\rm p}(2-\frac{q_{\rm p}}{a_{\rm p}})} \leq h \quad \text{(from equation~\ref{eq:hbis})}\label{eq:qpmax2},\\
	&q_{\rm p} \leq \frac{Q}{1-\beta r_{\rm H}} = \frac{2a-q}{1-\beta r_{\rm H}} \quad \text{(from equation~\ref{eq:rh2})}\label{eq:qpmax3}.
\end{align}
We note that equation~\eqref{eq:qpmax2} is automatically satisfied when $a_{\rm p} \leq h^2/m_{\rm p}^2$ given equation~\eqref{eq:qpmax1}. Conversely, when $a_{\rm p} > h^2/m_{\rm p}^2$, equations~\eqref{eq:qpmax1} and \eqref{eq:qpmax2} reduce to
\begin{equation}
	q_{\rm p} \leq a_{\rm p}\left(1-\sqrt{1-\frac{h^2}{m_{\rm p}^2 a_{\rm p}}}\right). \label{eq:qpmax2bis}
\end{equation}
Thus, we have
\begin{equation}
	q_{\rm p, max} = \begin{cases}
		\min\left[a_{\rm p}, \frac{2a-q}{1-\beta r_{\rm H}}\right] &\text{if } a_{\rm p} \leq h^2/m_{\rm p}^2,\\
		\min\left[a_{\rm p}\left(1-\sqrt{1-\frac{h^2}{m_{\rm p}^2 a_{\rm p}}}\right), \frac{2a-q}{1-\beta r_{\rm H}}\right] &\text{if } a_{\rm p} \geq h^2/m_{\rm p}^2.
	\end{cases}\label{eq:qpmax}
\end{equation}
Furthermore, equation~\eqref{eq:E} gives $a$ as a function of $a_{\rm p}$: 
\begin{equation}
	a = \frac{m}{E-\frac{m_{\rm p}}{a_{\rm p}}}. \label{eq:a}
\end{equation}
Combining equations~\eqref{eq:qpmax} and \eqref{eq:a} into equation~\eqref{eq:hbis}, we obtain an expression linking $q$ and $a_{\rm p}$, of the form
\begin{equation}
	f(q, a_{\rm p}) = 0. \label{eq:f}
\end{equation}
This equation has no analytic solution, but can be solved numerically. It corresponds to a contour in the $(a_{\rm p}, q)$ plane (that of the lower white zone in Figure~\ref{fig:q_vs_ap}). At minimum periastron, the following condition is satisfied:
\begin{equation}
	\pdv{f}{a_{\rm p}} \left(q_{\rm min}, a_{\rm p}\right) = 0. \label{eq:df}
\end{equation}
This equation along with equation~\eqref{eq:f} can be used to find the minimum periastron $q_{\rm min}$ by using a $2D$ root finding algorithm. It so happens that there is only one solution that satisfies both equations~\eqref{eq:f} and \eqref{eq:df} for the cases we studied, represented by the red dot in Figure~\ref{fig:q_vs_ap}. This method was used to plot the minimum periastron as a function of the mass ratio and initial planet eccentricity in Fig.~\ref{fig:qmin_vs_m}.
We study below the conditions necessary for $E$, $h$ and $\beta r_{\rm H}$ to allow $q_{\rm min} = 0$ while satisfying all the necessary equations~\eqref{eq:E}, \eqref{eq:h}, \eqref{eq:rh} and \eqref{eq:rh2}.

\subsection{Conditions for $q_{\rm min} = 0$}
\label{sec:qmin0}

If $q = 0$ is a possible solution of the post-scattering problem, the periastron and semi-major axis of the planet $q_{\rm p}$ and $a_{\rm p}$ must satisfy equations~\eqref{eq:h_q=0} and \eqref{eq:qpsolmax}.
Solving equation~\eqref{eq:h_q=0} for $q_{\rm p} < a_{\rm p}$ gives
\begin{equation}
	q_{\rm p} = a_{\rm p} \left(1 - \sqrt{1-\frac{h^2}{m_{\rm p}^2 a_{\rm p}}}\right). \label{eq:qpsol}
\end{equation}
This solution exists only if $a_{\rm p} \geq {h^2}/{m_{\rm p}^2}$. Let us find the conditions required for the existence of a solution $a_{\rm p}$ compatible with $q = 0$ and equations~\eqref{eq:qpsolmax} and \eqref{eq:qpsol}.
Combining these equations, we get
\begin{equation}
	a_{\rm p} \left(1 - \sqrt{1-\frac{h^2}{m_{\rm p}^2 a_{\rm p}}}\right) < \frac{2m}{(1-\beta r_{\rm H})\left(E-\frac{m_{\rm p}}{a_{\rm p}}\right)}
	\iff  1-\sqrt{1-\left(\frac{m_{\rm p}^2 a_{\rm p}}{h^2}\right)^{-1}} < \frac{2m m_{\rm p}^2}{(1-\beta r_{\rm H})Eh^2} \left(\frac{m_{\rm p}^2a_{\rm p}}{h^2}-\frac{m_{\rm p}^3}{Eh^2}\right)^{-1}. \label{eq:eqcomplique}
\end{equation}
Let us define
\begin{align}
	&x = \frac{a_{\rm p} m_{\rm p}^2}{h^2} > 1\\
	&\alpha_1 = \frac{2m m_{\rm p}^2}{(1-\beta r_{\rm H})Eh^2}\\
	&\alpha_2 = \frac{m_{\rm p}^3}{Eh^2}.
\end{align}
We note that, according to equation~\eqref{eq:E}, $x > \alpha_2$. Simple manipulations then lead to:
\begin{equation}
	\frac{x - (\alpha_1+\alpha_2)}{x-\alpha_2} < \sqrt{1-x^{-1}}.\label{eq:alphabeta}
\end{equation}
If $\alpha_1 + \alpha_2 >1$, equation~\eqref{eq:alphabeta} always has a solution. This is because one can find $x$ such that $1 < x < \alpha_1 + \alpha_2$, in which case
\begin{equation}
	\frac{x - (\alpha_1+\alpha_2)}{x-\alpha_2} < 0 < \sqrt{1-x^{-1}}
\end{equation}
The condition $\alpha_1 + \alpha_2 >1$ corresponds to
\begin{align}
	\alpha_1 + \alpha_2 > 1 \iff \frac{h^2E}{m_{\rm p}^3} < \frac{2m}{m_{\rm p}(1-\beta r_{\rm H})} + 1. \label{eq:condition1}
\end{align}
On the other hand, if $\alpha_1+\alpha_2 < 1$, then condition~\eqref{eq:alphabeta} becomes
\begin{align}
	&\left(\frac{x-\alpha_1-\alpha_2}{x-\alpha_2}\right)^2 < 1-x^{-1}\\
	\iff & x\left(x-\alpha_1-\alpha_2\right)^2 < (x-1) (x-\alpha_2)^2\\
	\iff & (1-2\alpha_1)x^2 + \left[\alpha_1^2 + 2\alpha_2(\alpha_1-1)\right]x + \alpha_2^2 < 0.
\end{align}
The left-hand side of the last inequality is a quadratic function of $x$, so the condition can be solved analytically. We have two cases: either
\begin{equation}
	1-2\alpha_1 < 0,
\end{equation}
or 
\begin{align}
	&\Delta = \alpha_1^2\left[(\alpha_1+2\alpha_2)^2 - 4 \alpha_2\right] > 0, \label{eq:condition2}\\
	\text{and }& x_{\rm root} = \frac{\alpha_1^2 + 2\alpha_1\alpha_2 - 2\alpha_2 + \sqrt{\Delta}}{2(2\alpha_1-1)} > 1.
\end{align}

\subsection{Small mass limit}

When the mass $m$ is small compared to the planet mass $m_{\rm p}$, then the results can be simplified and expressed in terms of the Tisserand parameter $T$ (equation~\ref{eq:T}), the initial planetary eccentricity $e_{\rm p,0}$ and the mass ratio $\varepsilon = m/m_{\rm p} \ll 1$. We have
\begin{align}
	&E  = \frac{m_{\rm p}}{a_{\rm p}} + \frac{m}{a} = \frac{m_{\rm p}}{a_{\rm p, 0}}\left(1 + \varepsilon \frac{a_{\rm p,0}}{a_0}\right), \label{eq:ET}\\
	&h = m_{\rm p}\sqrt{a_{\rm p}(1-e_{\rm p}^2)} + m\sqrt{a(1-e^2)} = m_{\rm p}\sqrt{a_{\rm p,0}}\left[\sqrt{1-e_{\rm p,0}^2} + \frac{\varepsilon}{2}\left(T - \frac{a_{\rm p,0}}{a_0}\right)\right], \label{eq:hT}
\end{align}
where $a_{\rm p, 0}$ and $a_0$ are the initial semi-major axes of the planet and the planetesimal respectively. 

\subsubsection{Circular case: retrieving the result from the Tisserand parameter's approach}

\begin{figure}
	\centering
	\includegraphics[width=0.6\linewidth]{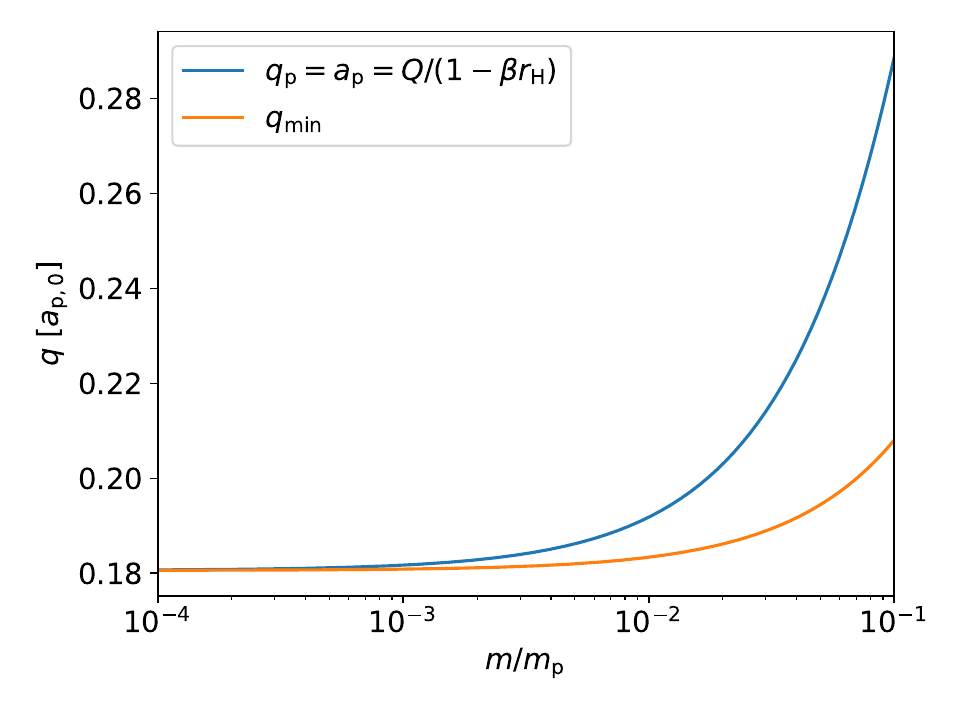}
	\caption{planetesimal periastron as a function of the mass ratio $m/m_{\rm p}$. We assume that $a_{\rm 0} = a_{\rm p, 0}(1+2\beta r_{\rm H})$ (with $\beta = 2$) and $e_{\rm p, 0} = e_{\rm 0} = 0$. The $q_{\rm min}$ line is the true minimum, derived numerically from equations~\eqref{eq:f} and \eqref{eq:df}. The $q_{\rm p} = a_{\rm p}= Q/(1-\beta r_{\rm H})$ line corresponds to the intersection between the blue and orange line in Figure~\ref{fig:q_vs_ap}, and is a good proxy for the minimum $q_{\rm min}$ at small masses.} \label{fig:qmin_vs_m_circular}
\end{figure}

Our goal here is to derive an analytical estimate for $q_{\rm min}$ when the mass $m \ll m_{\rm p}$ and $e_{\rm p,0} = 0$. Using the results of Section~\ref{sec:formal}, we plot $q_{\rm min}$ as a function of the mass ratio $m/m_{\rm p}$ in Figure~\ref{fig:qmin_vs_m}. We compare that value to the intersection between the $q_{\rm p} = a_{\rm p}$ (or equivalently $e_{\rm p} = 0$) and $q_{\rm p} = Q/(1-\beta r_{\rm H})$ branches, which seems somewhat close to $q_{\rm min}$ in Figure~\ref{fig:q_vs_ap}. Figure~\ref{fig:qmin_vs_m_circular} corroborates this observation. Writing equation~\eqref{eq:hT} with $q_{\rm p} = a_{\rm p} = Q/(1-\beta r_{\rm H})$, we get
\begin{align}
	&h = m_{\rm p} \sqrt{a_{\rm p}} + m\sqrt{(2-\frac{Q}{a})Q}  \\
	\iff &1 + \frac{\varepsilon}{2}\left(T - \frac{a_{\rm p,0}}{a_0} \right) = \sqrt{\frac{a_{\rm p}}{a_{p,0}}} + \varepsilon\sqrt{\left(2 - \frac{a_{\rm p}}{a}(1-\beta r_{\rm H})\right)\frac{a_{\rm p}}{a_{p,0}}(1-\beta r_{\rm H})}.
\end{align}
We define $\delta = (a_{\rm p}-a_{\rm p, 0})/a_{\rm p, 0}$. From the conservation of energy (equation~\ref{eq:E}), we know that $\delta \propto \varepsilon$, so that we define $\gamma = \delta/\varepsilon$. From equation~\eqref{eq:ET}, we derive $a_{\rm p}/a$ as a function of $\delta$ and $\gamma$:
\begin{equation}
	\frac{a_{\rm p}}{a} = \frac{a_{\rm p, 0}}{a_0}(1+\delta) + \gamma
\end{equation}
Finally, we have:
\begin{equation}
		1 + \frac{\varepsilon}{2}\left(T - \frac{a_{\rm p,0}}{a_0} \right) = \sqrt{1 + \delta} + \varepsilon\sqrt{\left[2 - \left(\frac{a_{\rm p, 0}}{a_0}(1+\delta) + \gamma\right)(1-\beta r_{\rm H})\right](1+\delta)(1-\beta r_{\rm H})}
\end{equation}
For $\delta, \varepsilon \ll 1$, it yields
 \begin{align}
 	&\frac{\varepsilon}{2}\left(T - \frac{a_{\rm p,0}}{a_0} \right) \simeq \frac{\delta}{2} + \varepsilon\sqrt{\left[2 - \left(\frac{a_{\rm p, 0}}{a_0} + \gamma\right)(1-\beta r_{\rm H})\right](1-\beta r_{\rm H})}\\
 	\implies & \frac{1}{4}\left(T - \frac{a_{\rm p,0}}{a_0}  - \gamma\right)^2 \simeq \left[2 - \left(\frac{a_{\rm p, 0}}{a_0} + \gamma\right)(1-\beta r_{\rm H})\right](1-\beta r_{\rm H}),
 \end{align}
which can be solved analytically for $\gamma$ (we choose the largest solution, corresponding to the maximum $q_{\rm p}$):
\begin{equation}
	\gamma = T - \frac{a_{\rm p,0}}{a_0} +2(1-\beta r_{\rm H})^2 \left(\sqrt{1-\frac{T}{(1-\beta r_{\rm H})^2} + \frac{2}{(1 - \beta  r_{\rm H})^3}}-1\right).
\end{equation}
From there, we can derive the corresponding semi-major axes and periastrons of the planet and planetesimal:
\begin{align}
	&a_{\rm p} = q_{\rm p} = a_{\rm p,0}(1+\gamma \frac{m}{m_{\rm p}}),\\
	&a = \frac{m}{E-\frac{m_{\rm p}}{a_{\rm p}}} \simeq \frac{a_{\rm p, 0}}{\frac{a_{\rm p, 0}}{a_0} + \gamma},\\
	&q_{\rm min} = 2a-Q \simeq \frac{2a_{\rm p, 0}}{\frac{a_{\rm p, 0}}{a_0} + \gamma} - a_{\rm p,0}(1-\beta r_{\rm H}).
\end{align}
After some algebra, we find that $q_{\rm min}$ has the exact same expression as in equation~\eqref{eq:qmin}, derived in Section~\ref{sec:tisserand} for the restricted three-body problem. This confirms the validity of our approximations for $m \ll m_{\rm p}$.

\subsubsection{Condition for $q_{\rm min} = 0$}
We use here the results from Section~\ref{sec:qmin0} to determine a simpler approximate condition for the planetesimal to reach $q_{\rm min} = 0$ in the small mass limit $m \ll m_{\rm p}$. Equation~\eqref{eq:condition1} can be rewritten as
\begin{equation}
	\left[\sqrt{1-e_{\rm p,0}^2} + \frac{\varepsilon}{2}\left(T - \frac{a_{\rm p,0}}{a_0}\right)\right]^2 \left(1 + \varepsilon \frac{a_{\rm p,0}}{a_0}\right) < \frac{2\varepsilon}{1-\beta r_{\rm H}} +1.
\end{equation}
To first order in $\varepsilon$, this is equivalent to
\begin{equation}
	\varepsilon\left(T\sqrt{1-e_{\rm p,0}^2} - \frac{2}{1-\beta r_{\rm H}}\right) < e_{\rm p, 0}^2.
\end{equation}
This condition is satisfied for very high eccentricities
\begin{equation}
	T\sqrt{1-e_{\rm p,0}^2} < {2}/{(1-\beta r_{\rm H})},
\end{equation}
or for a small mass ratio
\begin{equation}
	\varepsilon < \frac{e_{\rm p,0}^2}{T\sqrt{1-e_{\rm p,0}^2} - \frac{2}{1-\beta r_{\rm H}}}. \label{eq:conditionsmallm}
\end{equation}
Equation~\eqref{eq:condition2} reduces to equation~\eqref{eq:conditionsmallm} to first order in $\varepsilon$, so that equation~\eqref{eq:conditionsmallm} is a sufficient and necessary condition for $q_{\rm min} = 0$ to be a valid solution.

\section{Probability of different outcomes as a function of the Safronov parameter}

\begin{figure}
	\centering
	\includegraphics[width=0.6\linewidth]{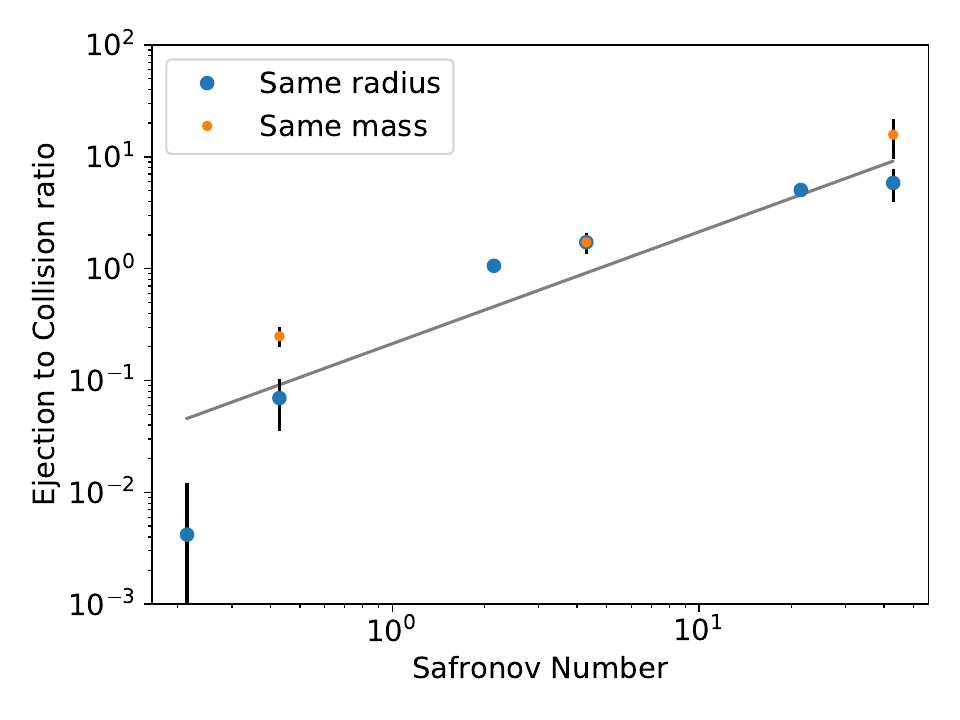}
	\caption{Ejection to collision ratio as a function of the Safronov number (equation~\ref{eq:safronov}) using the simulations from Figures~\ref{fig:proba_vs_mp} and \ref{fig:proba_vs_ap}, for $e_{\rm p} = 0.1$ and $r_{\rm dis} = 0$. The blue dots have $R_{\rm p} = 5\times 10^{-4}~a_{\rm p}$, the orange dots have $m_{\rm p} = 10^{-3}~M_\star$.}\label{fig:safronov}
\end{figure}

\begin{figure}
	\centering
	\includegraphics[width=0.6\linewidth]{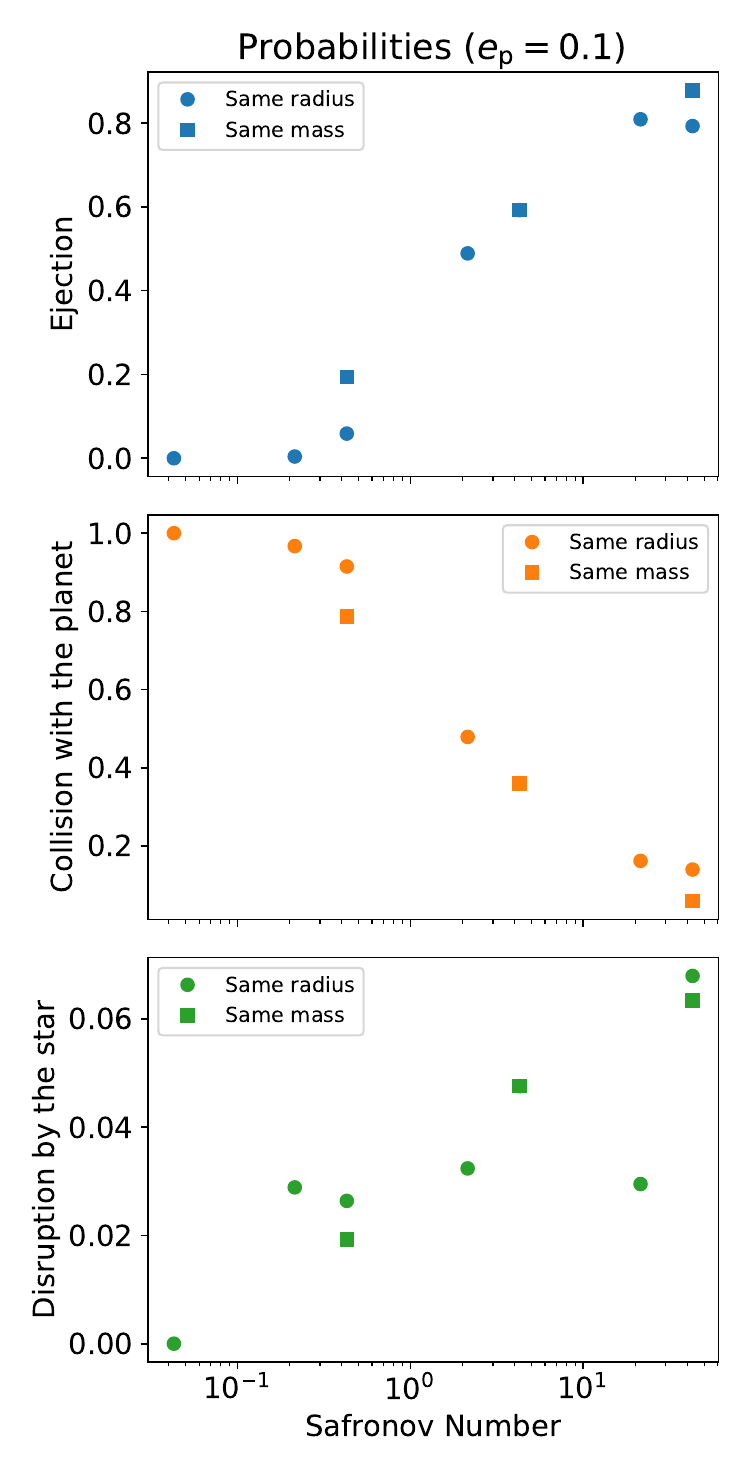}
	\caption{Same as Figure~\ref{fig:safronov} for the probabilities of each outcome, assuming $r_{\rm dis} = 0.1~a_{\rm p}$. The circles have $R_{\rm p} = 5 \times 10^{-4}~a_{\rm p}$, the squares have $m_{\rm p} = 10^{-3}~M_\star$.}\label{fig:proba_safronov}
\end{figure}

Figures~\ref{fig:safronov} and \ref{fig:proba_safronov} show the probabilities of different outcomes as a function of the Safronov parameter (Equation~\ref{eq:safronov}).

\bsp	
\label{lastpage}
\end{document}